\documentclass[a4paper,oneside,final,notitlepage,onecolumn,12pt]{article}

\usepackage[T1]{fontenc}
\usepackage[utf8]{inputenc}

\usepackage{hyperref}
\hypersetup
{
    colorlinks=true,
    linktocpage=true,
    linkcolor={red},
    anchorcolor={black},
    citecolor={green},
    menucolor={red},
    unicode=true,
    pdftitle={Superradiance with TME},
    pdfauthor={K. Z. Csukás, L. András, I. Rácz and G.Zs. Tóth }
}

\usepackage{amsmath}
\usepackage{amssymb}
\usepackage{authblk}
\usepackage{bbm}
\usepackage{accents}

\usepackage{graphicx}
\usepackage{ifpdf,epsfig,array,psfrag}

\usepackage{subcaption}

\usepackage[abs]{overpic}

\newcommand{\ld}{D}
\newcommand{\nd}{\Delta}
\newcommand{\md}{\delta}
\newcommand{\mvd}{\overline{\delta}}

\newcommand{\npt}{\tau}

\newcommand{\npp}{\varpi}
\newcommand{\npa}{\alpha}
\newcommand{\npb}{\beta}
\newcommand{\npg}{\gamma}
\newcommand{\npe}{\epsilon}

\newcommand{\npm}{\mu}

\newcommand{\npr}{\rho}

\newcommand{\nptv}{\overline{\tau}}

\newcommand{\nppv}{\overline{\varpi}}
\newcommand{\npav}{\overline{\alpha}}
\newcommand{\npbv}{\overline{\beta}}

\newcommand{\npev}{\overline{\epsilon}}

\newcommand{\npmv}{\overline{\mu}}

\newcommand{\nprv}{\overline{\rho}}

\def\tf{\textit{\texttt{f}}}

\usepackage[normalem]{ulem}
\usepackage{color}

\newcounter{mnotecount}

\newcommand{\mnotex}[1]
{\protect{\stepcounter{mnotecount}}$^{\mbox{\footnotesize $\bullet$\themnotecount}}$
	\marginpar{\color{red}
		\raggedright\tiny\em
		$\!\!\!\!\!\!\,\bullet$\themnotecount: #1} }

\usepackage[normalem]{ulem}
\usepackage{color}
\definecolor{blue}{rgb}{0,0,1}
\definecolor{red}{rgb}{1,0,0}

\DeclareFontFamily{OT1}{rsfs}{} \DeclareFontShape{OT1}{rsfs}{m}{n}{
	<-7> rsfs5 <7-10> rsfs7 <10-> rsfs10}{}
\DeclareMathAlphabet{\mathscr}{OT1}{rsfs}{m}{n}

\def\scri{{\mathscr I}}
\def\scrip{\scri^{+}}%
\def\scrim{\scri^{-}}%

\psfrag{I+}{$\scrip$}
\psfrag{I-}{$\scrim$}
\psfrag{H+}{$\mathcal H_+$}
\psfrag{H-}{$\mathcal H_-$}
\psfrag{J^-[D]}{$J^-[D]$}
\psfrag{i+}{$i^+$}
\psfrag{i-}{$i^-$}
\psfrag{i0}{$i^0$}

\newtheorem{theorem}{Theorem}[section]

\renewcommand{\theequation}{\thesection.\arabic{equation}} 

\title{Numerical investigation of the dynamics of linear spin $s$ fields on a Kerr background II: \\ Superradiant scattering}
\author{K\'aroly 
	Csuk\'as\footnote{E-mail address:{\tt csukas.karoly@wigner.hu}} }
\author{Istv\'an R\'acz\footnote{E-mail address:{\tt racz.istvan@wigner.hu}}}
\affil{Wigner RCP, H-1121 Budapest, Konkoly Thege Mikl\'{o}s \'{u}t  29-33, Hungary}

\begin{document}
    \maketitle
    \begin{abstract}
    	Superradiant scattering of linear spin $s=0,\pm 1,\pm 2$ fields on Kerr black hole background is investigated in the time domain by integrating numerically the homogeneous Teukolsky master equation. The applied numerical setup has already been used in studying long time evolution and tail behavior of electromagnetic and metric perturbations on rotating black hole background \cite{cskritgzs-2019}. To have a clear setup the initial data is chosen to be of the compact support, while to optimize superradiance the frequency of the initial data is fine tuned. Our most important finding is that the rate of superradiance strongly depends on the relative position of the (compact) support of the initial data and the ergoregion. When they are well-separated then only a modest---in case of $s=0$ scalar fields negligible---superradiance occurs, whereas it can get to be amplified significantly whenever the support of the initial data and the ergoregion overlap. 
	\end{abstract}

\section{Introduction}	
\setcounter{equation}{0}

Superradiant scattering is one of the most challenging phenomenon in black hole physics ever since it was pointed out by Misner, Zel'dovich and Starobinskii \cite{misner,zeld,staro,staro-churil} that more energy return to infinity than sent in originally. Superradiance is expected to occur in the scattering of a suitably frequency tuned scalar, electromagnetic or gravitational wave packet on a rotating black hole background. 

Notably, while studying the superradiant scattering of test fields on black hole background, most of the works have been done by applying frequency-domain analyses (see, e.g.~\cite{cardoso} and references in it). As opposed to this, time-domain studies are much rarer (see, e.g., \cite{laguna-1998,Csizmadia:2012kq,Andras:2014kq,Pretorius-2014,oliver-2016}) even though some of the physically relevant processes cannot be investigated via frequency-domain analyses.
The most natural mathematical framework to investigate scattering of scalar, electromagnetic and gravitational perturbations in the time-domain is provided by the homogeneous Teukolsky equation \cite{Teukolsky-I}. The conventional argument rests on the completely plausible assumptions that the ``transmitted wave'' is absorbed by the black hole while the ``reflected wave'' escapes to infinity. It is also important to be emphasized here that, although the occurrence of  superradiance was found to be completely consistent with the laws of black hole thermodynamics \cite{bekenstein,Teukolsky-III}, most of the arguments supporting energy extraction boil down to the investigation of certain individual, thereby, preferred eigenstate modes \cite{Teukolsky-III}.

\medskip

This paper is to report on our numerical findings on superradiant scattering of scalar waves and electromagnetic and gravitational perturbations on a rotating Kerr black hole background. In doing so we shall use the numerical method we developed to investigate the time evolution of linear fields of spin $s=0,\pm 1,\pm 2$ on Kerr black hole spacetimes by solving the homogeneous Teukolsky equation numerically \cite{cskritgzs-2019}. Accordingly, as in \cite{cskritgzs-2019}, the numerical setup we shall apply incorporates both the elements of conformal compactification and those of the hyperbolic initial value problem. Based on the fact that the fixed Kerr background is smoothly foliated by a two-parameter family of topological two-spheres---that are the Boyer-Lindquist $t=const$ and $r=const$ ellipsoids---the evolved fields are expanded in terms of spin-weighted spherical harmonics.
This allows us to evaluate all the angular derivatives analytically as they can be expressed either in terms of the ``eth'' and ``ethbar'', $\eth$ and $\overline{\eth}$, operators or by Lie-derivatives with respect to the axial Killing vector of the Kerr background. The expansion coefficients are then evolved, in the time-radial section, by applying the method of lines. 
The corresponding finite difference part of the code was verified to be of fourth order accurate \cite{cskritgzs-2019}. Here it is verified that in the angular directions the convergence rate is, as expected, indeed exponential. In principle, the full spectra of nonaxisymmetric configurations can be investigated, though, in practice, our attention will be restricted to the values $m=\pm1,\pm2$ of the azimuthal parameter. 
In any numerical investigation it is critical to verify that the numerical findings are not simple artifacts of the applied numerical method but they are rooted in true physical nature of the investigated fields. Therefore---in addition to the conventional checks on the convergence rates---new type of ``energy'' and ``angular momentum'' balance relations were also used to verify both the proper implementation of the underlying mathematical model and the  reliability and robustness of the developed numerical schema.

\medskip  

The analysis of individual back-scattered modes---see subsection \ref{subsec: mode-analyse} for a brief discussion on some of the related issues---shows that in order to ensure a superradiant scattering to occur these modes have to be both frequency tuned and also guaranteed to submerge deeply into the ergoregion. To demonstrate that the second assumption is indeed of critical importance three different types of initial data are applied in our numerical investigations. Each of these is of compact support. The most significant difference between them is that for type-I data the compact support is clearly separated from the ergoregion, for type-II data the support is concentrated to the ergoregion, whereas for type-III data the support inter-bridges the domains of type-I and type-II data, i.e.~the radial domain of type-III data significantly overlaps with the ergoregion and also it has a non-negligible part outside of the ergoregion.

\medskip

One of our most important findings is that if the initial data is of type-I, regardless of how well it is frequency-tuned to be superradiant, considerably large portion of the incident pulse goes through of an almost perfect reflection before reaching the ergoregion. Most cases this reduces the transmission coefficient, and, thereby, the effect of superradiance is suppressed significantly. As opposed to this, in some cases, for type-II and type-III initial data considerably large superradiance can be observed. 

\medskip  

In all circumstances the energy and angular momentum balance relations are used to give a proper measure of the energy and angular momentum gains. In doing so the handy conserved currents introduced in \cite{Toth:2018ybm} are applied which, however, require the use of a pair of spin $s$ and $-s$ solutions. This, in particular, means that whenever we intend to characterize the superradiance of a spin $s$ solution we need to construct an adequate spin $-s$ counterpart solution to the Teukolsky master equation. Exactly this type of construction is offered for electromagnetic perturbations by Cohen-Kegeles in  \cite{Cohen:1974cm},  and both for electromagnetic and metric perturbations by Chrzanowski in \cite{Chrzanowski:1975wv}. All the related heuristic aspects of the constructions in \cite{Cohen:1974cm,Chrzanowski:1975wv} were explained by Wald in \cite{Wald:1978vm}. As we need to apply these transformations a short review on the most important elements is provided in the appendix of the present paper. 

\medskip  

The rest of this paper is organized as follows. In section \ref{sec: prelim} some of the basic notions and notations are recalled including the Teukolsky equation, the mode analysis of superradiance and our guiding principles applied in selecting the initial data to get superradiant solutions. Section \ref{sec: initialization} is to discuss the regularization of the basic variables, the new type of energy and angular momentum balance relations, the applied multipole expansions and the actual initialization of time evolution. Section \ref{sec: numerical-results} is to present our numerical findings. We start, in subsection \ref{subsec: ID-separated}, by investigating configurations with type-I initial data, i.e.~the compact support of the data is separated from the ergoregion. Then, in subsection \ref{subsec: ID-overlapping}, results relevant for configurations with initial data concentrated to the ergoregion are presented. Subsection \ref{subsec: ID-III} includes investigations of type-III configurations the radial support of the initial data of which inter-bridges the domains of types I and II data.  The discussions on our findings are closed by final remarks in section \ref{sec: final-remarks}. The paper completed by several appendices providing useful insights on the analytic backgrounds we used to generate adequate spin $-s$ counterpart to spin $s$ solutions of the Teukolsky master equation.  

\section{Preliminaries}\label{sec: prelim}	
\setcounter{equation}{0}

This section is to recall Teukolsky's master equation, the mode analyses of superradiance and also to motivate the choice of the initial data to get superradiant solutions.

\subsection{Teukolsky's master equation}

The line element of the Kerr background in Boyer-Lindquist coordinates $(t,r,\vartheta,\phi)$ is
\begin{multline}
(d s)^2=\left(1-\frac{2Mr}{\Sigma}\right)(d t)^2+\frac{4arM}{\Sigma}\sin^2\vartheta\,d t d\phi\\-\frac{\Sigma}{\Delta}\,(d r)^2-\Sigma\,(d \vartheta)^2-\frac{(r^2+a^2)^2-a^2\Delta\sin^2\vartheta}{\Sigma}\sin^2\vartheta\,(d\phi)^2,
\end{multline}
where $\Sigma=r^2+a^2\cos^2\vartheta$ and $\Delta=r^2-2Mr+a^2$. The parameters $M$ and $a$ denote the ADM mass and the angular momentum per unit mass parameters of the Kerr solution, respectively. 

\medskip  

A very simple, probably the simplest, covariant form of the homogeneous Teukolsky master equation was given in \cite{Bini:2002jx} and it reads as
\begin{equation}
\label{eq:TMEBini}
\left[\,\left(\nabla^a+s\,\Gamma^a\right)\,\left(\nabla_a+s\,\Gamma_a\right)-4\,s^2\,\Psi_2\,\right]\,\psi^{(s)}=0\,,
\end{equation}
where $\psi^{(s)}$ is a linear spin $s$ field, and 
\begin{equation}
\Psi_2=-\frac{M}{(r-\mathrm{i}\, a \cos\vartheta)^3}
\end{equation}
is the only nonvanishing and gauge invariant Weyl-scalar component on Kerr background. The components of the ``connection vector'' $\Gamma^a$ in \eqref{eq:TMEBini} are \cite{Bini:2002jx}
\begin{align}
\Gamma^t&=-\frac{1}{\Sigma}\left[\frac{M(r^2-a^2)}{\Delta}-(r+\mathrm{i}\, a \cos\vartheta)\right],\\
\Gamma^r&=-\frac{1}{\Sigma}\,\big(r-M\big),\\
\Gamma^\vartheta&=0,\\
\Gamma^\phi&=-\frac{1}{\Sigma}\left[\frac{a(r-M)}{\Delta}+\mathrm{i}\,\frac{\cos\vartheta}{\sin^2\vartheta}\right]\,.
\end{align}

\subsection{Superradiance in mode analysis}\label{subsec: mode-analyse}

Consider a solution $\psi^{(s)}$  to (\ref{eq:TMEBini}). It was shown by Teukolsky \cite{Teukolsky-I} that it is advantageous to decompose its temporal Fourier transform, 
\begin{equation}
{\mathscr{F}}\hskip-0.5mm\psi^{(s)} =\frac{1}{\sqrt{2\pi}}\int_{-\infty}^{+\infty}\psi^{(s)}\, e^{-\mathrm{i}\,\omega \,t} dt\,,
\end{equation} 
as  
\begin{equation}\label{eq: kerr-representation}
{\mathscr{F}}\hskip-0.5mm\psi^{(s)}(\omega,r,\vartheta,{\phi})=
\sum_{\ell=|s|}^{\infty}\sum_{m=-\ell}^{\ell}{}_sR_{\ell,\omega}^{m}(r)\cdot{}_sS_{\ell,a\omega}^{m}(\vartheta ,{\phi})\,,  
\end{equation}
where $(t,r,\vartheta,\phi)$ stand again for the Boyer-Lindquist coordinates, while $\omega$ is the frequency in the time translation direction. 

The radial functions ${}_s R_{\ell,\omega}^{m}$ in \eqref{eq: kerr-representation}---for any specific value of the angular momentum quantum numbers $\ell,m$ and the frequency $\omega$---are subject to a one-dimensional Schr\"odinger-type equation 
\begin{equation}\label{eq: schro}
\Delta^{-s}\frac{d }{dr}\left( \Delta^{s+1} \,\left(\frac{d} {dr}\,{}_s{R}_{\ell,\omega}^{m}\right) \right) +
\left[\frac{K^2-2\,\mathrm{i}\,s\,(r-M)\,K}{\Delta} +4\,\mathrm{i}\,s\,\omega\,r - {}_s\lambda_{\ell,\omega}^{m}
\right]\,{}_s {R}_{\ell,\omega}^{m}=0\,. 
\end{equation}
Note also that the angular factors  ${}_s S_{\ell,a\omega}^{m}$, in \eqref{eq: kerr-representation}, stand for the spin-weighted (oblate) spheroidal harmonic functions, with oblateness parameter $a\omega$, which are eigenfunctions of a self-adjoint operator (see, e.g., (2.7) in \cite{Teukolsky-III} or (3.74) in \cite{cardoso}), i.e.,~${}_s S_{\ell,a\omega}^{m}$ are subject to
\begin{equation}\label{eq: spheroidal}
\overline{\eth} \eth {}_sS_{\ell,a\omega}^{m} + \left[2\,s\,(a\,\omega\cos\vartheta-1) + a^2\,\omega^2\cos^2\vartheta  - {}_sE_{\ell,a\omega}^{m} \right] {}_sS_{\ell,a\omega}^{m}=0\,.
\end{equation}
The eigenvalues, ${}_s\lambda_{\ell,\omega}^{m}$ and ${}_sE_{\ell,a\omega}^{m}$, are related as ${}_s\lambda_{\ell,\omega}^{m} = {}_sE_{\ell,a\omega}^{m} + a^2\,\omega^2 -2 a m \omega-s(s+1)$ and $K=(r^2+a^2)\,\omega- a\,m$. In \eqref{eq: spheroidal} the $\eth$ and $\overline{\eth}$ operators were used which, when acting on a spin-$s$ field $f$, are defined via the relations
\begin{align}
\eth f&=-\sin^s\vartheta\left(\partial_\vartheta+\frac{\mathrm{i}}{\sin\vartheta}\partial_\phi\right)(\sin^{-s}\!\vartheta \cdot f)\\
\overline{\eth} f&=-\sin^{-s}\vartheta\left(\partial_\vartheta-\frac{\mathrm{i}}{\sin\vartheta}\partial_\phi\right)(\sin^{s}\!\vartheta \cdot f)\,.
\end{align}
In terms of $\eth$ and $\overline{\eth}$ the ``generalized'' two-dimensional Laplace-Beltrami operator reads as
\begin{equation}
\overline{\eth}\eth f=\partial_{\vartheta\vartheta} f+\cot\vartheta\,\partial_\vartheta f+\frac{1}{\sin^2\vartheta}\,\partial_{\phi\phi} f+2\,\mathrm{i}\,s\,\frac{\cot\vartheta}{\sin\vartheta}\,\partial_\phi f+s\,(1-s\cot^2\vartheta)f\,.
\end{equation}

\medskip

Returning to the main line of the argument, recall first that, using the Starobinskii-Teukolsky identity \cite{staro,staro-churil,Teukolsky-III}, along with the tortoise radial coordinate $r_*$, determined by the relation $dr/d r_* =\Delta/(r^2+a^2)$, the ``physical solutions'' to (\ref{eq: schro}) are supposed to possess the asymptotic behavior 
\begin{equation}\label{asympt}
{R}_{\ell,\omega}^{m}\sim 
\begin{cases}
	\mathcal{I}\,e^{-\mathrm{i}\,\omega\, r_{*}} + \mathcal{R} \,e^{+\mathrm{i}\,\omega\, r_{*}} & \mathrm{ as }\;\; r\rightarrow \infty  \\
	\mathcal{T}\,e^{-\mathrm{i}\,(\omega- m\, \Omega_{+})\,r_{*}}        & \mathrm{ as }\;\;r \rightarrow r_+\,,
\end{cases}
\end{equation} 
where $r_+=M + \sqrt{M^2-a^2}$ and $\Omega_{+}=a/(2Mr_+)$ denote the (outer) horizon radius and the angular velocity of the Kerr black hole with respect to the asymptotically stationary observers, respectively. The factors $\mathcal{I}$, $\mathcal{R}$, and $\mathcal{T}$ denote the incident, reflection and transmission coefficients at infinity and at the horizon, respectively. It is important to note and keep in mind that the above boundary condition at the horizon explicitly assumes the existence of a transmitted wave submerging deeply into the ergoregion.

\medskip

In this setup by making use of equations \eqref{eq: schro} and \eqref{eq: spheroidal}, along with the asymptotic behavior as fixed in \eqref{asympt}, one can determine the energy fluxes carried by the fields from the ergoregion toward infinity. The most striking implication (see section IV. in \cite{Teukolsky-III}, and also subsection 3.6.2 in \cite{cardoso} for details) is that, regardless of the value $s=0,\pm1,\pm2$ of the spin of bosonic fields, the total energy flux per unit solid angle at the horizon is always proportional to the product $\omega(\omega- m\, \Omega_{+})\,|\mathcal{T}|^2$, i.e.,
\begin{equation}\label{eq: en-flux}
\left.\frac{d^2\mathscr{E}}{dtd\Omega}\right\vert_{r_+} \sim \omega(\omega- m\, \Omega_{+})\,|\mathcal{T}|^2\,.
\end{equation}
The pertinent positive factors of proportionality---which do not play a primary role in the present argument---can be read off equations (4.26), (4.32) and (4.44) in \cite{Teukolsky-III}.  
In accordance to \eqref{eq: en-flux} whenever $|\mathcal{T}|$ does not vanish and the inequality
\begin{equation}\label{eq: range}
0<\omega<m\,\Omega_{+}
\end{equation}
holds energy can be extracted by back-scattered individual modes of scalar, electromagnetic or gravitational perturbations due to their interaction with the Kerr background.

\subsection{Initialization of to be superradiant solutions}\label{subsec: guess-ID} 

In preparing frequency-tuned incident wave packets we shall adapt the basic ideas applied in \cite{Csizmadia:2012kq, Andras:2014kq}. Accordingly, we assume that in 
the asymptotic region---at least in a sufficiently small neighborhood of the initial data surface---the solution may be approximated as 
\begin{equation}\label{eq: minkowski-solution}
\psi^{(s)}(t,r_{*},\vartheta,\phi) 
\approx A\cdot\exp\left[-\mathrm{i}\,\omega_{0}\,(r_{*}-r_{*0}+t)\right] f(r_{*}-r_{*0}+t)\,{}_{s}Y{}_{\ell}^{m}(\vartheta,\phi)\,.  
\end{equation}
where $f$ is a smooth function of compact support, and where $A$, $\omega_{0}$,  and $r_{*0}$ are suitable real parameters ($r_{*0}$ is simply to provide a shift in the tortoise coordinate). Note that though in \eqref{eq: minkowski-solution} a pure ${}_{s}Y{}_{\ell}^{m}$-mode is specified this does not restrict the generality of the evolving field as a systematic hierarchy of additional $\ell$-modes (thought each of them possessing the same $m$-parameter as the initial pure one) get to be invoked immediately via mode couplings in \eqref{eq:TMEBini}. 
The corresponding initial data reads as 
\begin{align} 
\psi^{(s)}(r_{*},\vartheta,\phi) = {}&  A\cdot\exp\left[-\mathrm{i}\,\omega_{0}\,(r_{*}-r_{*0})\right]
f(r_{*}-r_{*0})\,{}_{s}Y{}_{\ell}^{m}(\vartheta,\phi)\,,\\
\psi^{(s)}_{t}(r_{*},\vartheta,\phi)= {}&
A\cdot\exp\left[-\mathrm{i}\,\omega_{0}\,(r_{*}-r_{*0})\right]\,f'(r_{*}-r_{*0})\,{}_{s}Y{}_{\ell}^{m}(\vartheta,\phi)-\mathrm{i}\,\omega_{0}\,\psi^{(s)}(r_{*},\vartheta,\phi)\,,
\end{align} 
where $f'$ denotes the first derivative of $f$. The Fourier transform, 
${\mathscr{F}}\hskip-0.5mm\psi^{(s)}$, of the approximate solution (\ref{eq: minkowski-solution}) is of the form
\begin{equation}
{\mathscr{F}}\hskip-0.5mm \psi^{(s)}(\omega,r_{*},\vartheta,\phi) 
\approx A\cdot\exp\left[-\mathrm{i}\,\omega\,(r_{*}-r_{*0})\right] {\mathscr{F}}\hskip-1mm 
f(\omega-\omega_{0})\,{}_{s}Y{}_{\ell}{}^{m}(\vartheta,\phi),  
\end{equation}
where $\omega$ is the temporal frequency and ${\mathscr{F}}\hskip-1mm f$ stands for the Fourier-transform of $f$. Accordingly, ${\mathscr{F}}\hskip-1mm f$ plays the role of a frequency profile, which guarantees that whenever ${\mathscr{F}}\hskip-1mm f$ is chosen to be sufficiently narrow the solution (\ref{eq: minkowski-solution}) in the asymptotic region
is well-approximated by an incident monochromatic wave packet, which by choosing  $\omega_0$ from the interval $0< \omega_0 <m\, \Omega_{+}$ gets to be superradiant.   

\section{More on the analytic background}\label{sec: initialization}
\setcounter{equation}{0}

This section is to discuss issues related to the regularization of basic variables, the new type of energy and angular momentum balance relations, the applied multipole expansion and the initialization of the investigated solutions.

\subsection{Regularization}\label{subsec: regularization}

Solutions to the homogeneous Teukolsky master equation \eqref{eq:TMEBini} are known to be singular either in the $\Delta\rightarrow 0$ or $r \rightarrow \infty$ limit. In order to regularize the spin $s$ fields $\psi^{(s)}$ new coordinates were introduced in \cite{cskritgzs-2019} in two steps. First, the Boyer-Lindquist coordinates, $(t,r,\vartheta,\phi)$, were replaced by the horizon penetrating ingoing Kerr coordinates, $(\tau,r,\vartheta,\varphi)$, where $\tau$ and $\varphi$, are determined via
\begin{equation}
\label{eq:tr:ki}
\tau =t-r+\int dr\frac{r^2+a^2}{\Delta} \qquad \mathrm{and} \qquad 
\varphi=\phi+\int dr\frac{a}{\Delta}\,.
\end{equation}

In the second step new coordinates $(T,R)$ were introduced, replacing $(\tau,r)$, via the relations 
\begin{equation}
\label{eq:tr:rt}
\tau =T+\frac{1+R^2}{1-R^2}-4M\log(|1-R^2|) \qquad \mathrm{and} \qquad 
r =\frac{2R}{1-R^2}\,.
\end{equation}
Note that in these new coordinates future null infinity, $\mathscr{I}^+$, is represented by the $R=1$ hypersurface, and also that  each of the $T=const$ hypersurfaces intersect both the horizon and $\mathscr{I}^+$  in regular spherical cuts. 
In our numerical investigations the regularized field variables, $\Phi^{(s)}$, were used which are defined as 
\begin{equation}\label{eq: regularized-variable}
\Phi^{(s)}(T,R,\vartheta,\varphi)=[\,r(R)\cdot\Delta^s(R)\,]\cdot\psi^{(s)}(T,R,\vartheta,\varphi)\,.
\end{equation}
In terms of these regularized variables the homogeneous Teukolsky master equation \eqref{eq:TMEBini} was shown to take the form \cite{cskritgzs-2019} 
\begin{align}
\label{eq:teurt}
&\partial_{TT}\Phi^{(s)}=\frac{1}{\mathscr{A}+\mathscr{B}\cdot Y_2^0}\,\Big[c_{RR}\cdot\partial_{RR}\Phi^{(s)}+c_{TR}\cdot\partial_{TR}\Phi^{(s)}+c_{T\varphi}\cdot\partial_{T\varphi}\Phi^{(s)}+c_{R\varphi}\cdot\partial_{R\varphi}\Phi^{(s)} \nonumber\\
&+c_{\vartheta\vartheta}\cdot\overline{\eth}\eth\,\Phi^{(s)}+c_T\cdot\partial_T\Phi^{(s)}+\mathrm{i}\,c_{Ty} \,Y_1^0\cdot\partial_T\Phi^{(s)}+c_R\cdot\partial_R\Phi^{(s)}+c_\varphi\cdot\partial_\varphi\Phi^{(s)}+c_0\cdot\Phi^{(s)}\Big]\,,
\end{align}
where $Y_1^0$ and $Y_2^0$ are the standard (spin $s=0$) spherical harmonics with $\ell=1,2$ and $m=0$, respectively. The explicit form of the involved $R$-dependent coefficients can be found in Appendix A of \cite{cskritgzs-2019}.

\subsection{The conserved currents}

In case of scalar perturbations the canonical energy and angular momentum are completely satisfactory to describe the studied processes. Yet, if $s\not=0$ we have to resort to more complicated formulas which reduce to the scalar case when s=0. It was shown in \cite{Toth:2018ybm} that meaningful Lagrangian can be associated with an arbitrary pair of spin $s$ and $-s$ fields. More importantly, sensible conserved  energy and angular momentum type currents were also introduced in \cite{Toth:2018ybm}  via the relations 
\begin{align}
\label{eq:curr1}
\mathcal{E}^a&=(\nabla^a-s\Gamma^a)\psi^{(-s)}T^b\partial_b\psi^{(s)}+(\nabla^a+s\Gamma^a)\psi^{(s)}T^b\partial_b\psi^{(-s)}+ T^a\mathcal{L}\,,\\
\label{eq:curr2}
\mathcal{J}^a&=(\nabla^a-s\Gamma^a)\psi^{(-s)}\varphi^b\partial_b\psi^{(s)}+(\nabla^a+s\Gamma^a)\psi^{(s)}\varphi^b\partial_b\psi^{(-s)}+\varphi^a\mathcal{L}\,.
\end{align}
Here $T^a=(\partial_T)^a$ and $\varphi^a=(\partial_\varphi)^a$ are  Killing vector fields generating the underlying one-parameter group of diffeomorphisms. These currents are always conserved when $\psi^{(s)}$ is a solution to the Teukolsky equation with spin-weight $s$ and $\psi^{(-s)}$ is a solution with spin-weight $-s$. The two solutions otherwise may be completely unrelated. In order to attain quantities which can be viewed as if they characterize a single solution we use a solution $\psi^{(s)}$, along with its adjoint, to get a proper pair. The motivation and the process of determining the adjoint solution is explained in details in the Appendix of the present paper.

\medskip 

Since the covariant divergence of these currents vanish the balance relations 
\begin{equation}
\int_\Omega\nabla_a\mathcal{E}^a=\int_{\partial\Omega}n_a\mathcal{E}^a =0  \quad \mathrm{and} \quad \int_\Omega\nabla_a\mathcal{J}^a=\int_{\partial\Omega}n_a\mathcal{J}^a =0\label{eq: balance-E-J}
\end{equation}
hold. In the present paper the domain of integration $\Omega$ will always be the rectangular coordinate domain in $(T,R)$ which is bounded by the initial and final time slices, $T=T_i$ and $T=T_f$,
and also by the null cylinders at $R=R_{+}$ and $R=1$, respectively.

\medskip

The superradiant amplifications can then be characterized by comparing the energy and angular momentum of the incident wave packet
\begin{equation}
E_0=\int_{T=T_i}n_a^{(T)}\mathcal{E}^a\sqrt{|h_T|}\,\,d R\,\mathrm{d}\vartheta\,\mathrm{d}\varphi  \quad \mathrm{and} \quad  J_0=\int_{T=T_i}n_a^{(T)}\mathcal{J}^a\sqrt{|h_T|}\,\,d R\,\mathrm{d}\vartheta\,\mathrm{d}\varphi
\end{equation}
to the energy and angular momentum radiated toward null infinity 
\begin{equation}
E_{out}=\int_{R=1}n_a^{(R)}\mathcal{E}^a\sqrt{|h_R|}\,d T\,\mathrm{d}\vartheta\,\mathrm{d}\varphi  \quad \mathrm{and} \quad J_{out}=\int_{R=1}n_a^{(R)}\mathcal{J}^a\sqrt{|h_R|}\,d T\,\mathrm{d}\vartheta\,\mathrm{d}\varphi\,,
\end{equation}
respectively, where $n_a^{(T)}=(d T)_a/\sqrt{g^{TT}}$ and $n_a^{(R)}=(d R)_a/\sqrt{-g^{RR}}$
are the respective normals of the $T=const$ and $R=const$ hypersurfaces
and $h_T$ and $h_R$ denote the determinant of the metric induced on these hypersurfaces.
Accordingly, if energy and angular momentum gains occur then $E_{out}> E_0$ and $J_{out}> J_0$ and such a situation will always be characterized by the scale independent ratios 
\begin{equation}\label{sup-gains}
\delta E=\frac{E_0 - E_{out}}{E_0} \quad \mathrm{and} \quad \delta J=\frac{J_0 - J_{out}}{J_0}\,,
\end{equation}
respectively. In practice, as $E_0, E_{out}, J_0, J_{out}$ are all complex apart from the scalar field case with $s=0$, therefore we always used the absolute values of $E_0, E_{out}, J_0, J_{out}$ in evaluating \eqref{sup-gains}. 

\medskip

The energy and angular momentum transports do provide important hints on the histories of the investigated configurations. It has been demonstrated in earlier studies (see Figs.~12-14 of   \cite{GF-IR-2008}) that plotting the ``volume normalized'' energy and angular momentum current densities is more informative than the figures showing the energy and angular momentum current densities themselves. In the present case the use of a slightly different notion of volume normalized current densities, $\mathscr{E}=\mathscr{E}(T,R)$ and $\mathscr{J}=\mathscr{J}(T,R)$, turned out to be more appropriate, which are defined by the integrals 
\begin{equation}
\mathscr{E}=\int_{0}^{\pi}\int_{0}^{2\pi}n_a^{(R)}\mathcal{E}^a\sqrt{|h_R|}\,\,\mathrm{d}\vartheta\,\mathrm{d}\varphi\,,  
\quad 
\mathscr{J}=\int_{0}^{\pi}\int_{0}^{2\pi}n_a^{(R)}\mathcal{J}^a\sqrt{|h_R|}\,\,\mathrm{d}\vartheta\,\mathrm{d}\varphi\,.
\end{equation}
Accordingly, the pure $T$-integrals, $\int_R\mathscr{E}(T,R) \,\mathrm{d}T$ and $\int_R\mathscr{J}(T,R) \,\mathrm{d}T$, of these current densities, over specific $R=const$ timelike hypersurfaces give us the energy and angular momentum fluxes through these $R=const$ hypersurfaces, respectively. Recall that, in the general case with $s\not=0$, likewise the conserved currents  $\mathcal{E}^a$ and $\mathcal{J}^a$, the volume normalized energy and angular momentum current densities have real and imaginary parts both. Nevertheless, in what follows, we shall plot only the real parts which, in the considered cases, are always several order larger than the corresponding imaginary parts.

\subsection{Multipole expansion}

In solving \eqref{eq:teurt} the angular derivatives were evaluated analytically by making use of the operators $\partial_\varphi$, $\eth$ and $\overline{\eth}$ by expanding the dynamical field $\Phi^{(s)}$ in terms of spin-weighted spherical harmonics as
\begin{equation}\label{eq: mult-exp}
\Phi^{(s)}(T,R,\vartheta,\varphi) = \sum_{\ell=|s|}^{\ell_{max}}\sum_{m=-\ell}^{\ell}\phi_\ell{}^{m}(T,R)\cdot {}_s{Y_\ell}{}^m(\vartheta,\varphi)\,.
\end{equation}
Then \eqref{eq:teurt} can be replaced by a set of coupled $(1+1)$-dimensional linear wave equations for the expansion coefficients $\phi_\ell{}^{m}(T,R)$. The  value of $\ell_{max}$ in \eqref{eq: mult-exp} was  chosen to be suitably large in order to keep the truncation error at the level of numerical precision.

\medskip

In solving \eqref{eq:teurt} standard order reduction techniques were also applied by introducing 
$(\phi_{T}){}_\ell{}^m=\partial_T\phi_\ell{}^m$ and $(\phi_{R}){}_\ell{}^m=\partial_R\phi_\ell{}^m$ as our new variables. The corresponding first order strongly hyperbolic system---relevant for the vector valued variable $(\phi_\ell{}^{m},(\phi_{R}){}_\ell{}^m,(\phi_{T}){}_\ell{}^m)^T$---was solved numerically by applying the method of lines  in a $4^{th}$ order Runge-Kutta integrator, along with a $6^{th}$ order dissipation term to suppress high frequency spurious solutions \cite{1995tpdm}.

\subsection{The choice made for the initial data}

In solving  \eqref{eq:teurt} it is necessary to specify, on a $T=T_0\,(\in\mathbb{R})$ initial data surface, a pair of functions $(\phi^{(s)},\phi_T^{(s)})$ such that  $\phi^{(s)}=\Phi^{(s)}|_{T=T_0}$ and $\phi_T^{(s)}=(\partial_T\Phi^{(s)})|_{T=T_0}$. 

\medskip

In choosing initial data for the regularized field  $\Phi^{(s)}$ we shall economize the choice we have already discussed in subsection \ref{subsec: guess-ID}. In doing so note first that by combining the coordinate transformations applied in subsection \ref{subsec: regularization} the relations
\begin{equation}\label{eq: relations}
    t+r_*-r_{*0}= \tau + r - r_{*0}=  T + \frac{1+R^2}{1-R^2}-4M\log(|1-R^2|) + \frac{2R}{1-R^2}-R_0
\end{equation}
can be seen to hold, where the $r_{*0}$ shift in the radial tortoise coordinate, in the last step, is simply replaced by $R_0$ that is its suitable correspondent. Denoting the right hand side of \eqref{eq: relations} by $H(T,R;R_0)$ and using the above replacement in \eqref{eq: minkowski-solution}, along with the regularization \eqref{eq: regularized-variable}, an approximate incident wave packet is expected to have the form 
\begin{align}\label{eq: approximate-solution}
\Phi^{(s)}(T,R,\vartheta,\varphi) = {} & \mathcal{A}(R)\cdot \psi^{(s)}(T,R,\vartheta,\varphi) \nonumber \\  \approx {} &  \mathcal{A}(R)\cdot \exp\left[-\mathrm{i}\,\omega_{0}\,H(T,R;R_0)\right] f(H(T,R;R_0))\,{}_{s}Y{}_{\ell}^{m}(\vartheta,\varphi)\,,
\end{align}
where $\mathcal{A}$ denotes the product $A\cdot[\,r(R)\cdot\Delta^s(R)\,]$ of the amplification factors in \eqref{eq: minkowski-solution} and \eqref{eq: regularized-variable}. 
Accordingly, the corresponding initial data $(\phi^{(s)},\phi_T^{(s)})$ read as 
\begin{align}
\phi^{(s)}(R,\vartheta,\varphi) = {} & \mathcal{A}(R)\cdot \exp\left[-\mathrm{i}\,\omega_{0}\,H_0\right] f(H_0)\,{}_{s}Y{}_{\ell}^{m}(\vartheta,\varphi)  \label{eq: Initial-data-TR} \\  
\phi_T^{(s)}(R,\vartheta,\varphi) = {} & \mathcal{A}(R)\cdot\exp\left[-\mathrm{i}\,\omega_{0}\,H_0\right]\,f'(H_0) \,{}_{s}Y{}_{\ell}^{m}(\vartheta,\phi)-\mathrm{i}\,\omega_{0}\,\phi^{(s)}(R,\vartheta,\varphi)\,,\label{eq: Initial-data-TR-T-der} 
\end{align}
where $f'$ denotes the first order derivative of $f$ and $H_0=H_0(R)$ stands for $H(T_0,R;R_0)$.

\medskip

In all of our numerical simulations the radial profile, $f$, was chosen as
\begin{eqnarray}
\label{profilefunction}
\hskip-0.3cm{f(x)=
	\left\{
	\begin{array} {r l}
	\exp\left[4-\left|\frac{w_1}{x+\frac{w_1}{2}}\right|-\left|\frac{w_2}{x-\frac{w_2}{2}}\right|\right]\,, 
	& \mathrm{if} \, x\in[-\frac{w_1}{2},\frac{w_2}{2}]\\
	0\,, & \mathrm{otherwise} \,,
	\end{array}
	\right.}
\end{eqnarray}
which is a smooth real function of the real variable $x$ that is vanishing in the complement of the compact support $[-{w_1}/{2},{w_2}/{2}]$. 

\medskip

\begin{figure}[ht!]
	\begin{centering}
		{\tiny
			\begin{subfigure}{0.49\textwidth}
				\includegraphics[width=\textwidth]{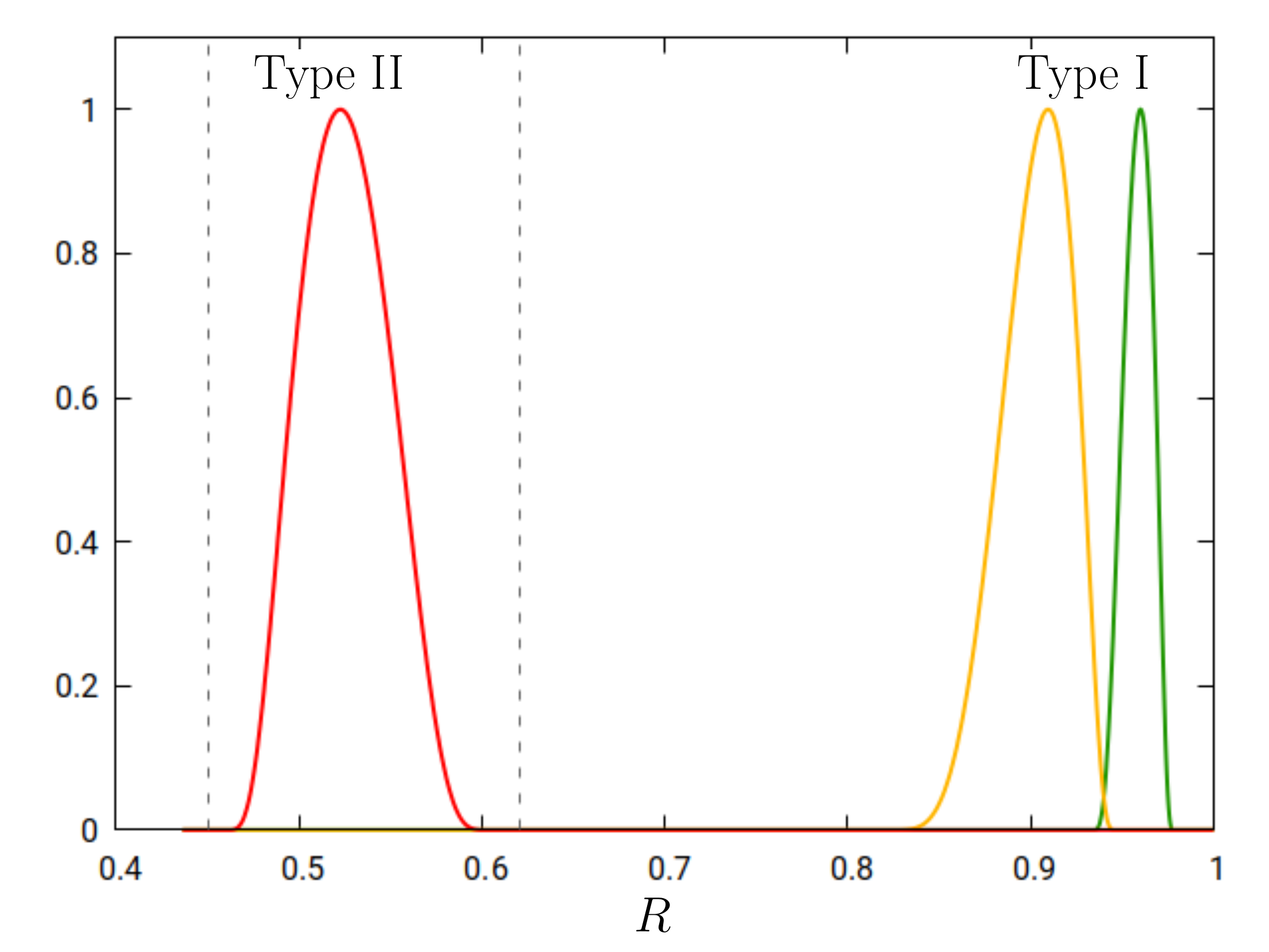}
				\caption{\footnotesize Type-I and type-II radial profiles. }
			\end{subfigure}
			\hskip.02\textwidth
			\begin{subfigure}{0.49\textwidth}
				\includegraphics[width=\textwidth]{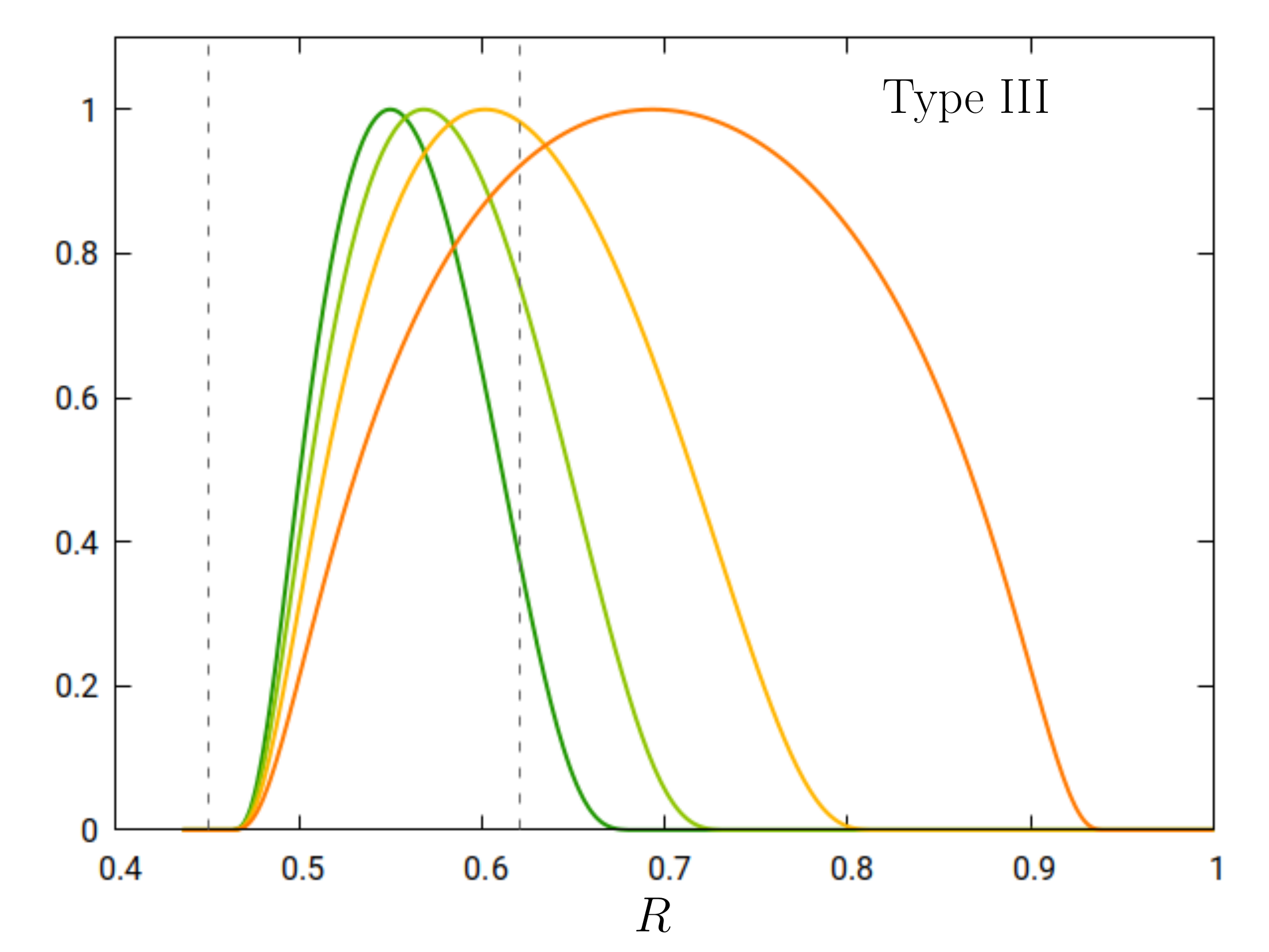}
				\caption{\footnotesize Type-III radial profiles. }
			\end{subfigure}
		}
	\end{centering}\vskip-0.2cm
	\caption{{\footnotesize  On the left the radial profiles relevant for type-I and type-II initial data are plotted, respectively. On the right the profiles pertinent for type-III data are shown. The two dashed vertical lines are to indicate the location of the horizon and the maximal radius of the ergoregion.}}
	\label{fig: supports} 
\end{figure}

As indicated in the introduction we shall apply three different types of initial data. The radial profile of each of these are of compact support. The support of type-I data is clearly separated from the ergoregion, whereas for type-II data the support is concentrated to the ergoregion. For type-III data the radial supports inter-bridge the domains of type-I and type-II data. Accordingly, they start close to the event horizon and cover, across the ergoregion, larger and larger parts of the domain of outer communication. Some of the typical $f(H_0(R))$ radial profiles are depicted on the panels of Fig.~\ref{fig: supports}.
The compact support of type-I data is visibly separated form the ergoregion while it is concentrated to the ergoregion for type-II data. 
On the left panel the two type-I radial profiles correspond to the specific values $R_1=0.8, R_0=0.9, R_2=0.95$ and $R_1=0.93, R_0=0.95, R_2=0.98$, respectively, whereas the corresponding parameters for type-II profile are $R_1=0.454, R_0=0.5, R_2=0.61$, where $R_1$ and $R_2$ denote the left and right edge of the compact support, while $R_0$ is to characterize the asymmetry of the profiles. For symmetric profiles $R_0$ also signifies the location of the maximum value, which, for these symmetric profiles, is at the mean value of $R_1$ and $R_2$, i.e.\,$R_0=(R_1+R_2)/2$. The graphs of type-III radial profiles the common left boundary is at $R_1=0.4536$, whereas the right edges are at  $R_2=0.7,0.76,0.84,0.95$, respectively. 

It is important to emphasize that analogous choices of radial profiles allow the proper frequency tuning of the solutions, and, in turn, to get the incident wave packets to be superradiant at least near to the initial data surface. We emphasize that the $\omega_0$ parameter in general differs from the actual frequency of the solution, however in the asymptotic region it proves to be a good approximation. This expectation will be verified by plotting the power spectra  
\begin{equation}\label{eq: power-spectra}
\mathcal{PS}(\omega,R)=\int_{0}^{\pi}\int_{0}^{2\pi} |\mathscr{F}\Phi^{(s)}|^2(\omega,R,\vartheta,\varphi)\,\sin\vartheta\,\mathrm{d}\vartheta
\,\mathrm{d}\varphi
\end{equation} 
for many of the inspected solutions.

\section{Numerical results}\label{sec: numerical-results}

In proceeding recall that in subsection \ref{subsec: mode-analyse} necessary conditions  guaranteeing the occurrence of superradiance were identified. Namely, it was found that superradiant scatterings cannot occur unless the frequency profile of the incident wave packet is located in the interval $0<\omega<m\Omega_{+}$, as well as, it submerges deeply into the ergoregion. We shall see that both of these conditions are of equal importance. If either of them fails to be satisfied the product $\omega(\omega- m\, \Omega_{+})\,|\mathcal{T}|^2$ on the rhs~of \eqref{eq: en-flux} may simply be small or, as seen below, in case of type-I spin $s=0$ fields, practically zero. 

\subsection{Results relevant for type-I initial data}\label{subsec: ID-separated}

The compact support of type-I initial data is always well-separated from the ergoregion as indicated on the left panel of Fig.~\ref{fig: supports}. It can be characterized by three parameters $R_1,R_2$, and $R_0$, where $R_1$ and $R_2$ denote the left and right edge of the compact support, whereas $R_0$ signifies the approximate location of the maximum value. In almost all of the simulations with type-I data these parameters are fixed as $R_1=0.93, R_0=0.95, R_2=0.98$. 
Unless otherwise stated, the ADM mass and specific angular momentum parameters of the Kerr black hole are $M=1$ and $a=0.99$, respectively. Note also that when the $T-R$- or $\omega-R$-dependencies of various quantities are plotted some of the $R=const$ dark grey dashed lines are always to indicate the equatorial radius $R_E$ of the ergoregion and the location, $R_+$, of the black hole horizon. For  $M=1$ and $a=0.99$ the specific values are $R_E=0.618$ and $R_+=0.4533$. The mean value, $\widetilde{\omega}$, of the frequency profile of the applied initial data, \eqref{eq: Initial-data-TR} and \eqref{eq: Initial-data-TR-T-der}, is expected to be below but close to the upper bound $m\,\Omega_{+}$. The specific numerical values of $\omega_0$ which is used in producing the specific spectra will always be given in each of the reported cases.

\subsubsection{Scalar field
}\label{subsubsec: type-I s=0}
	
\begin{figure}[ht!]
	\begin{centering}
		{\tiny
			\begin{subfigure}{0.49\textwidth}
				\includegraphics[width=\textwidth]{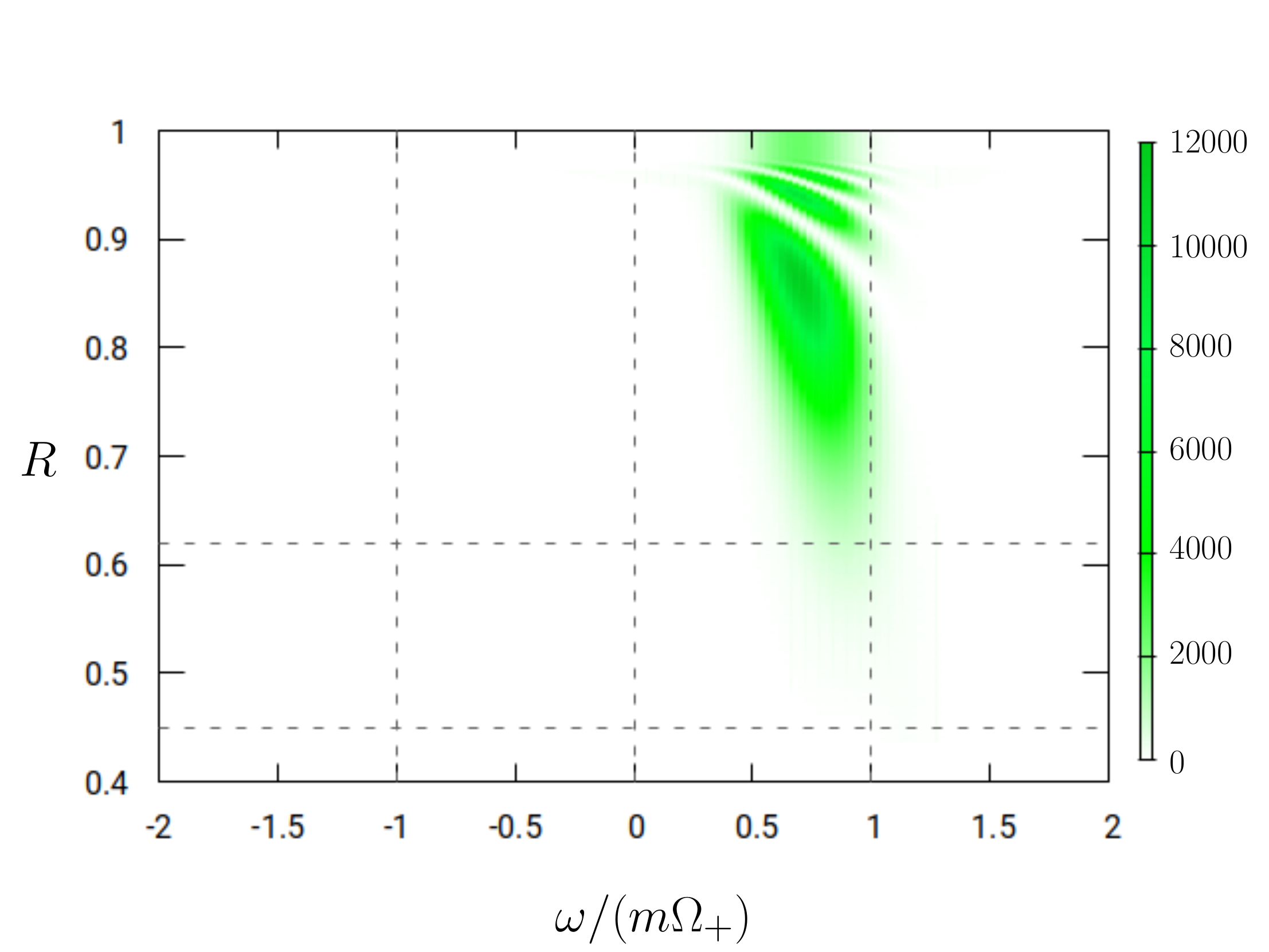}
				\caption{\footnotesize $m=+1$ }
			\end{subfigure}
			\hskip.02\textwidth
			\begin{subfigure}{0.49\textwidth}
				\includegraphics[width=\textwidth]{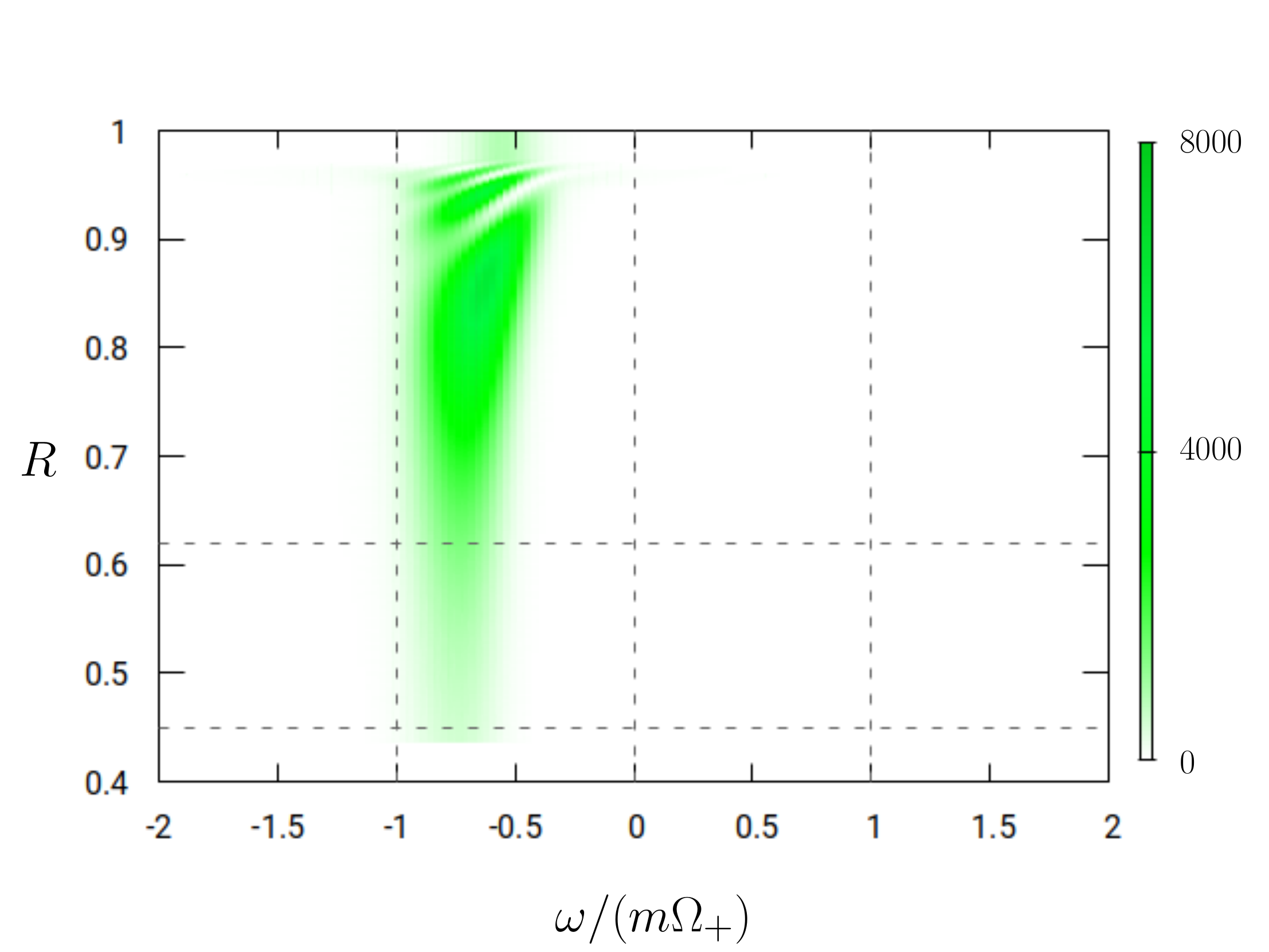}
				\caption{\footnotesize $m=-1$ }
			\end{subfigure}
		}
	\end{centering}\vskip-0.2cm
	\caption{{\footnotesize The power spectra, $\mathcal{PS}(\omega,R)$, of solutions with $m=+1$ and $ m=-1$ are shown on the left and right panel, respectively. The other characteristic parameters are  $s=0$,  $\ell=1,  \omega_0=0.3$.
			 }}\label{fig: s=0-power-m=pm1} 
\end{figure}

\begin{figure}[ht!]
	\begin{centering}
		{\tiny
			\begin{subfigure}{0.49\textwidth}
				\includegraphics[width=\textwidth]{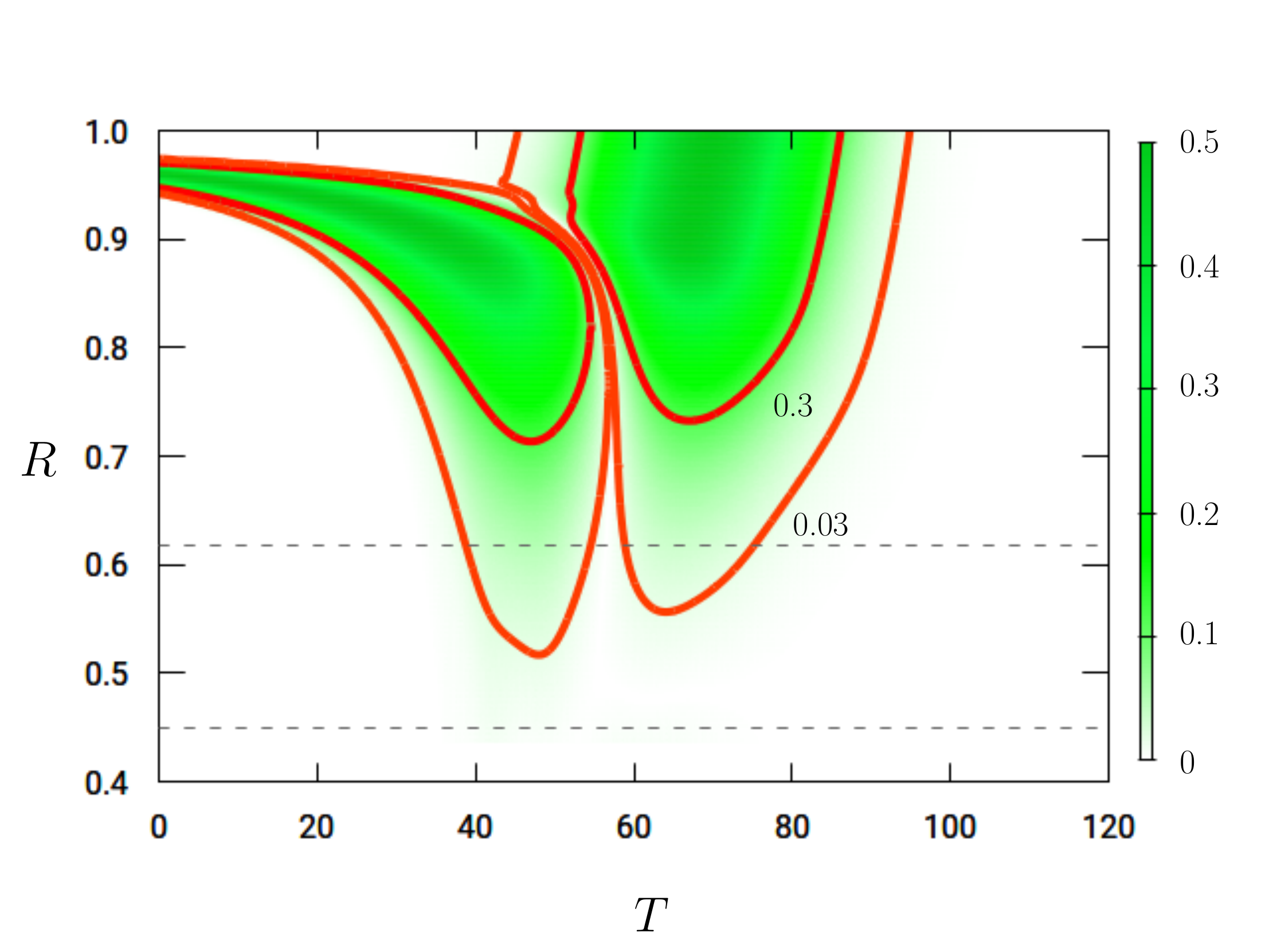}
				\vskip-0.3cm
				\caption{\footnotesize Solution with $s=0, \ell=1, m=+1$. }
			\end{subfigure}
			\hskip.02\textwidth
			\begin{subfigure}{0.49\textwidth}
				\includegraphics[width=\textwidth]{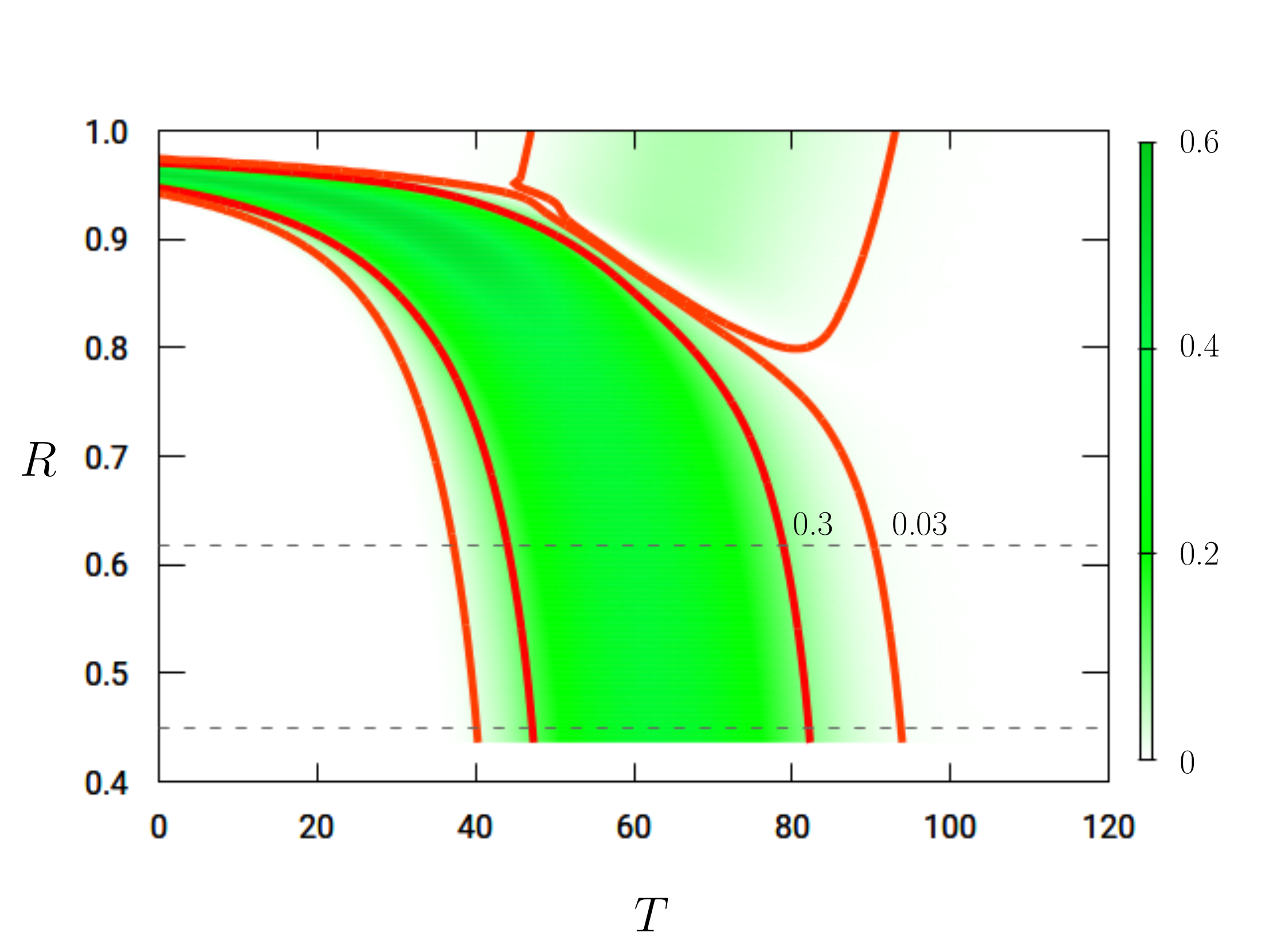}
				\vskip-0.3cm
				\caption{\footnotesize Solution with $s=0, \ell=1, m=-1$.}
			\end{subfigure}
		}
	\end{centering}\vskip-0.2cm
	\caption{{\footnotesize The volume normalized energy current densities, $\mathscr{E}=\mathscr{E}(T,R)$, are plotted for solutions with parameters $s=0, \omega_0=0.3,  \ell=1, m=\pm 1$. The corresponding values of the angular momentum  current densities, $\mathscr{J}=\mathscr{J}(T,R)$, are indicated at the colored contours. 
	}}\label{fig: s=0-curr-transport-m=pm1} 
\end{figure}

Beside $R_1,R_2$, and $R_0$ determining the radial profile the two solutions inspected in details are characterized by the physical parameters $s=0, \omega_0=0.3, \ell=1, m=\pm 1$. 
Note that the sign of the ``magnetic quantum number'' $m$ has a simple interpretation. Configurations with  $m>0$  are corotating, whereas with $m<0$ they are counterrotating with respect to the background Kerr black hole. Before inspecting the pertinent energy and angular momentum transports, it is rewarding to have a glance of the power spectra, $\mathcal{PS}=\mathcal{PS}(\omega,R)$, of the corresponding solutions depicted by the panels in Fig.\,\ref{fig: s=0-power-m=pm1}. 
\vskip-0.3cm
Notice that on the left panel with $m=+1$ the power spectra of the incident wave  packet is located---for all values of $R$ not only in the vicinity of the initial data surface where the frequency was tuned---in the desired superradiant interval, $0<\omega<m\, \Omega_{+}$. It is also remarkable that instead of going down to the black hole horizon the power spectra starts to be fading away as the ergoregion is approached. 
This behavior is confirmed by the corresponding energy and angular momentum transports plotted on the left panel of Fig.\,\ref{fig: s=0-curr-transport-m=pm1}.

The power spectra of the counterrotating solution, with $m=-1$, and with the same physical parameters, $s=0$, $\ell=1$ $\omega_0=0.3$,  is in the interval $ m\, \Omega_{+}<\omega< 0$ which is the reflection of the superradiant range to the $\omega=0$ axis. As verified by the corresponding energy and angular momentum transports plotted on the right panel of Fig.\,\ref{fig: s=0-curr-transport-m=pm1} considerably large part of the energy and angular momentum of the solution falls into the black hole directly.
On Fig.\,\ref{fig: s=0-curr-transport-m=pm1} the volume normalized energy and angular momentum current densities, $\mathscr{E}=\mathscr{E}(T,R)$ and $\mathscr{J}=\mathscr{J}(T,R)$, are depicted. As the angular momentum current densities are almost identical in their functional behavior to that of the energy current densities they will always be indicated by using colored contours and numbers signify their values at the pertinent level sets.   
The left panel of Fig.\,\ref{fig: s=0-curr-transport-m=pm1}, corresponding to the corotating solution with $m=+1$, indicates that essentially all the energy and angular momentum content of the incident wave  packet bounces back before entering the ergoregion. As opposed to this on the right panels, corresponding to the counterrotating solution with $m=-1$, apparently almost all the energy and angular momentum content of the incident wave  packet falls directly into the black hole. 

\medskip	

\begin{figure}[ht!] 
	\begin{centering}
		{\tiny
			\begin{subfigure}{0.49\textwidth}
				\includegraphics[width=\textwidth]{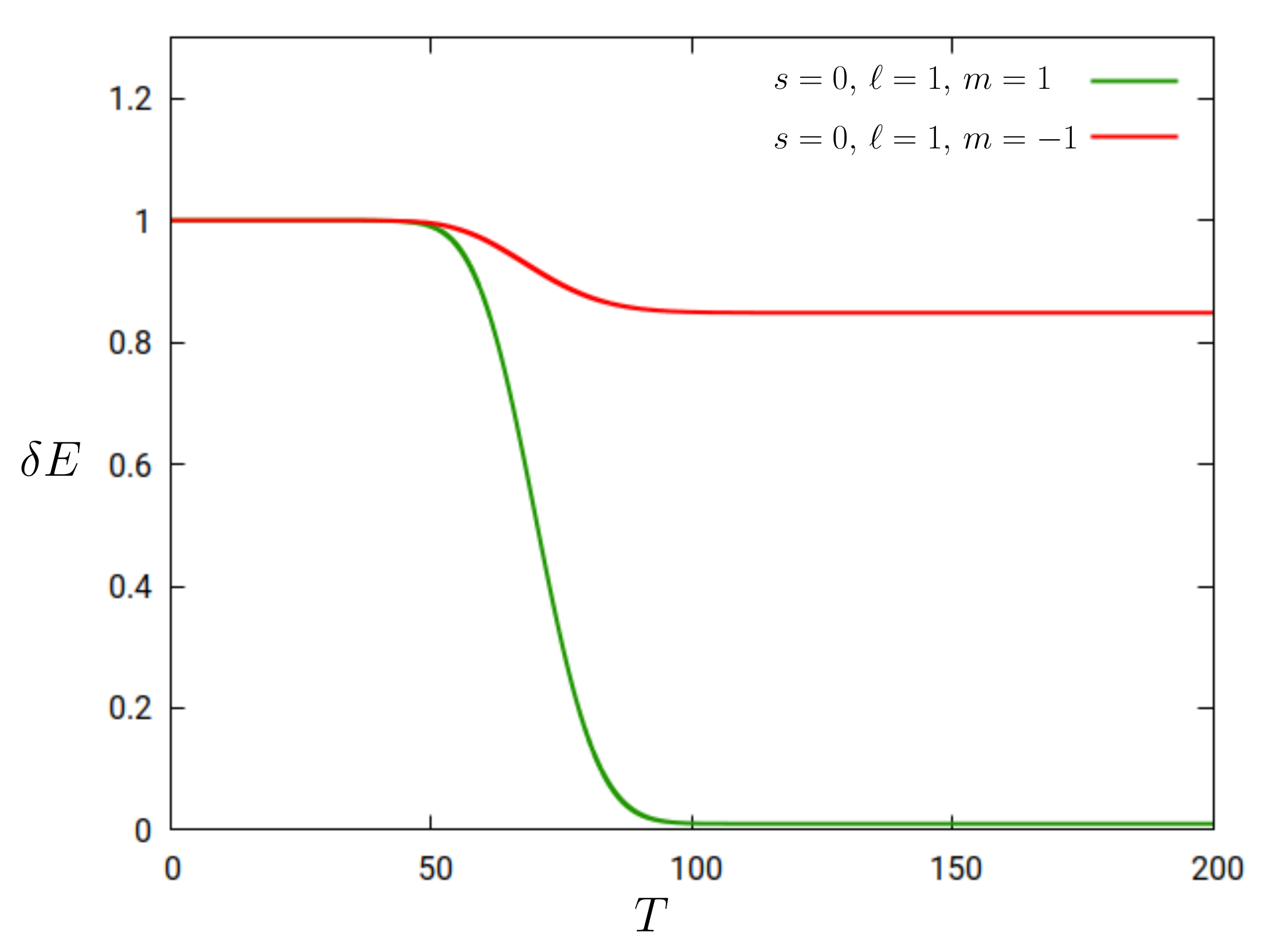}
				\caption{\footnotesize The energy gain. }
			\end{subfigure}
			\hskip.02\textwidth
			\begin{subfigure}{0.49\textwidth}
				\includegraphics[width=\textwidth]{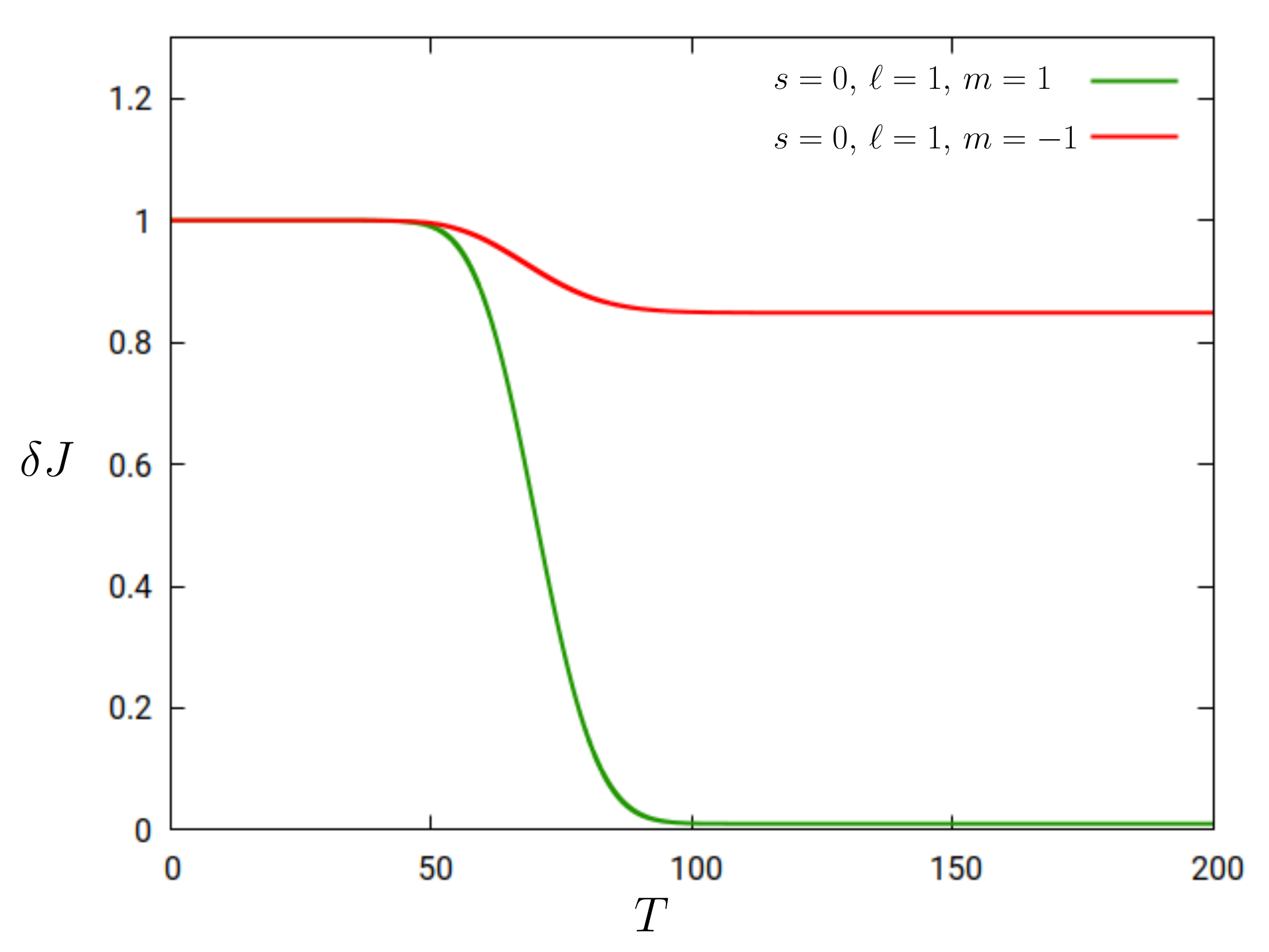}
				\caption{\footnotesize The angular momentum gain. }
			\end{subfigure}
		}
	\end{centering}\vskip-0.2cm
	\caption{{\footnotesize The time dependence of the energy and angular momentum gains, $\delta E=(E_0-E_{out})/E_0$ and $\delta J=(J_0-J_{out})/J_0$, are plotted for the investigated solutions with $m=+1$ and $m=-1$. 
	}}\label{fig: s=0-super-m=pm1} 
\end{figure}
All the foregoing observations are also confirmed by the panels of Fig.\,\ref{fig: s=0-super-m=pm1} depicting the respective time dependence of the energy and angular momentum gains, $\delta E=(E_0-E_{out})/E_0$ and $\delta J=(J_0-J_{out})/J_0$. 
In virtue of the graphs on the panels of Fig.\,\ref{fig: s=0-super-m=pm1} for the solution with  $m=+1$---in accordance with the left panel of Fig.\,\ref{fig: s=0-curr-transport-m=pm1}---the incident wave  packet go through (in accordance with the claims in \cite{Csizmadia:2012kq,Andras:2014kq}) an almost perfect reflection. Accordingly, no energy or angular momentum fall into the black hole,  i.e.~the energy and angular momentum content leave the domain through null infinity. As opposed to this, the counterrotating solution with $m=-1$ about $16\%$ of the energy and about $17\%$ of the angular momentum contained by the incident wave  packet leaves the  domain through $\scrip$ while the rest is falling directly into the black hole as it is visible on the right panel of Fig.\,\ref{fig: s=0-curr-transport-m=pm1}. 	
	
\medskip	

It is of interest to know what happens if for a type-I  counterrotating solution $\omega_0$ is replaced by $-\omega_0$. One may suspect that this will shift the power spectra of the pertinent solution into the superradiant range. To see that indeed this happens follows from the choice we made for the initial data $(\phi^{(s)},\phi_T^{(s)})$, in \eqref{eq: Initial-data-TR} and \eqref{eq: Initial-data-TR-T-der}, and by applying the relation $\overline{{}_{s}Y{}_{\ell}{}^{m}}= (-1)^{s+m}\cdot {}_{-s}Y{}_{\ell}{}^{-m}$. It is straightforward to see then that the initial data of a counterrotating scalar configuration with  $-\omega_0$  is $(-1)^{m}$ times of the complex conjugate of the initial data of a corotating solution with $\omega_0$. For this reason the insight provided by the inspection of time evolution of scalar initial data yielded by the simultaneous replacements $m \rightarrow -m$, $\omega_0 \rightarrow -\omega_0$ is very limited. Note also that for type-I counterrotating solutions with positive $\omega_0$ the power spectra is always out of the superradiant range. Accordingly, their behavior does not differ significantly from those of the corresponding counter-rotating solutions. For this reason, in the rest of subsection \ref{subsec: ID-separated}, we shall not investigate these type of configurations either. As seen later counterrotating solutions do not play significant role in case of type-II configurations either, nevertheless, they will be investigated in some extent in subsection \ref{subsec: ID-overlapping}.

\subsubsection{Outgoing electromagnetic and gravitational perturbations
}

\begin{figure}[ht!]
	\vskip-0.4cm
	\begin{centering}
		{\tiny
			\begin{subfigure}{0.49\textwidth}
				\includegraphics[width=\textwidth]{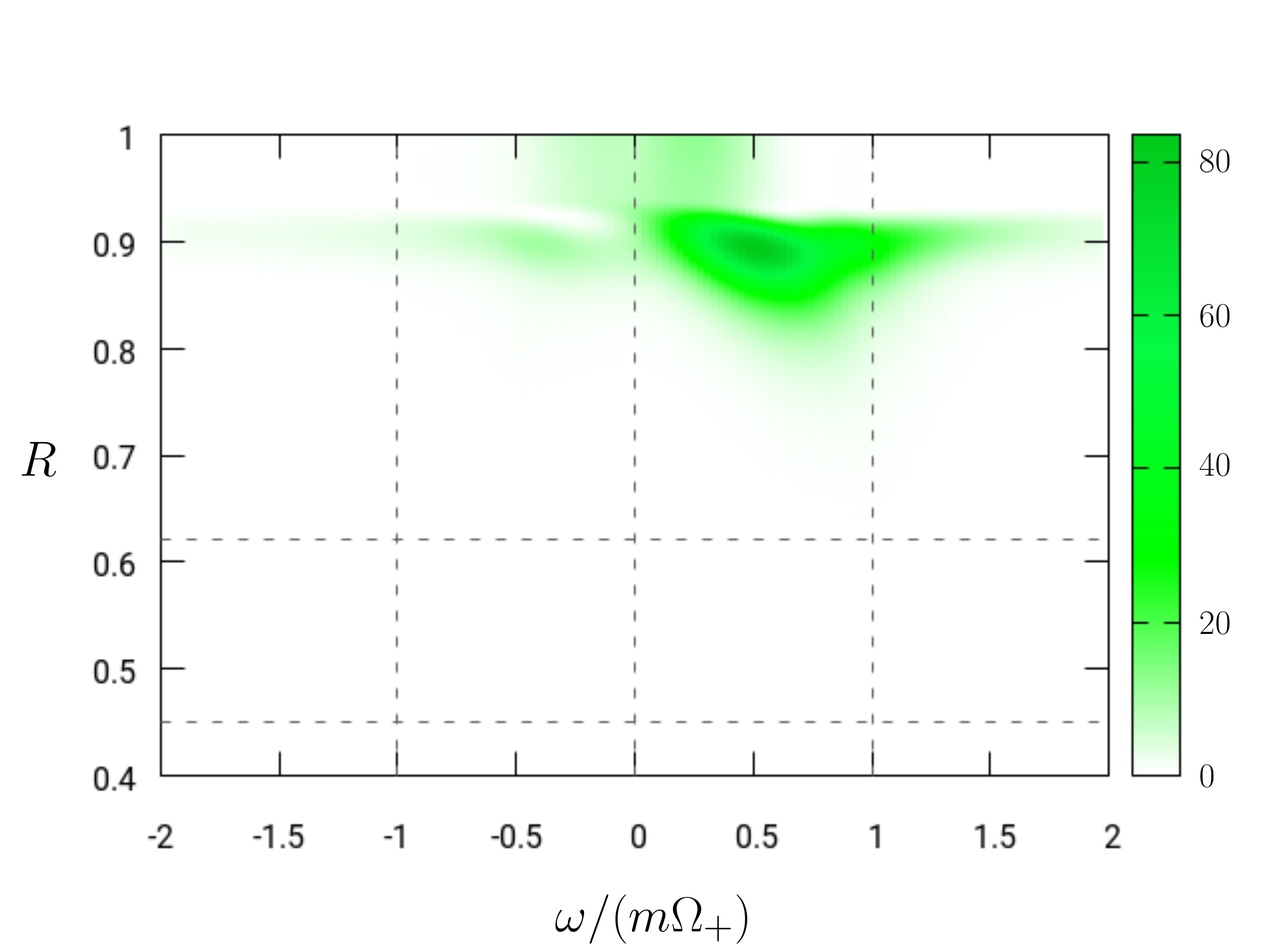}
				\caption{\footnotesize $\mathcal{PS}$
				 for solution with $s=+1, m=+1$. }
			\end{subfigure}
			\hskip.02\textwidth
			\begin{subfigure}{0.49\textwidth}
				\includegraphics[width=\textwidth]{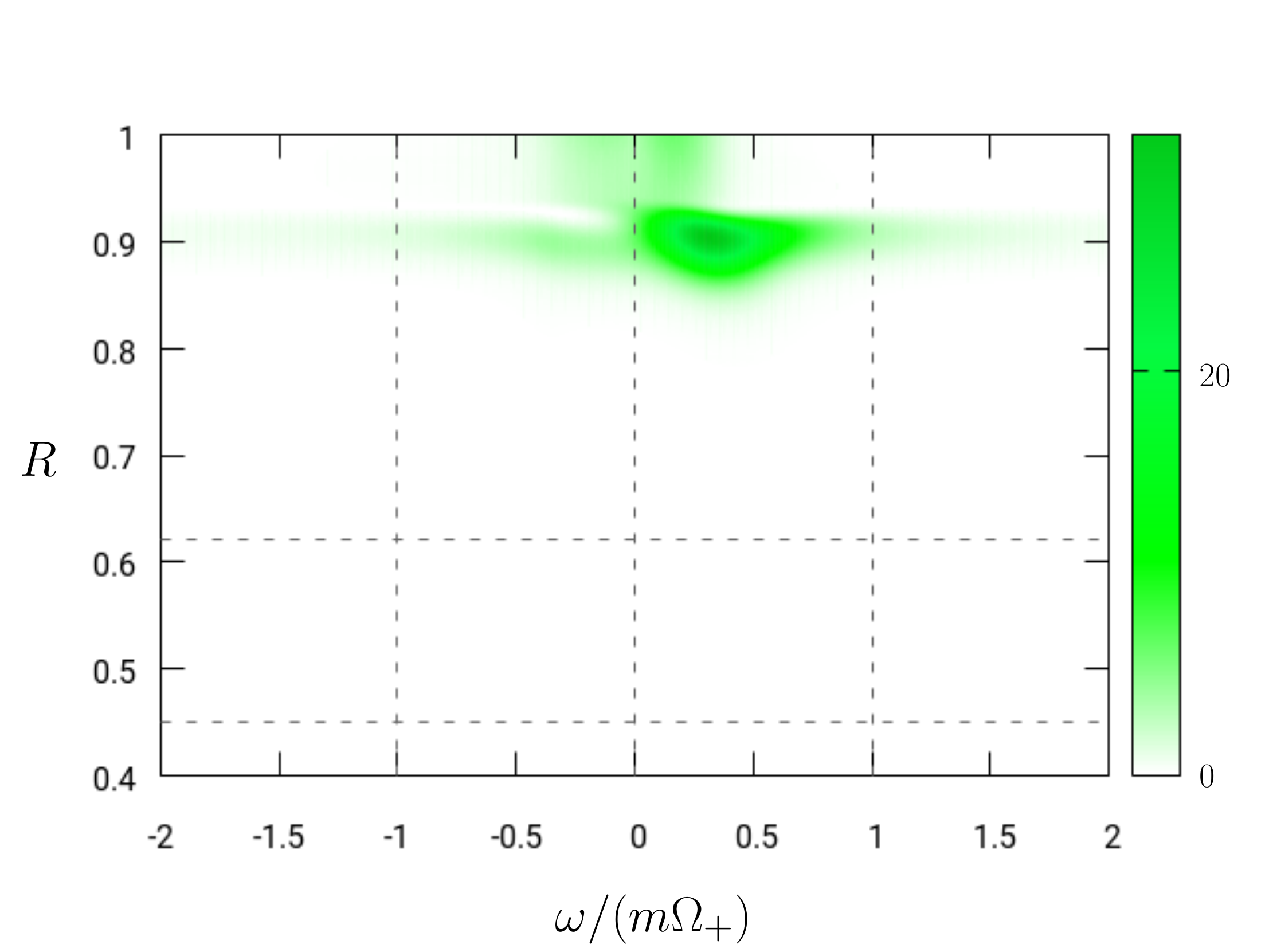}
				\caption{\footnotesize $\mathcal{PS}$
					for solution with $s=+2, m=+2$ }
			\end{subfigure}
		}
	\end{centering}\vskip-0.2cm
	\caption{\footnotesize The power spectra of the solutions with parameters $s=+1, \omega_0=0.21339,  \ell=1, m=+1$ and $s=+2, \omega_0=0.21339,  \ell=2, m=+2$ are shown on the left and right panel, respectively. 
	}\label{fig: s=1-power-m=pm1pm2} 
\end{figure}

The spin $s=+1$ electromagnetic and spin $s=+2$ gravitational perturbations are usually referred to as outgoing modes. In this subsection, for both cases, the applied radial profile is fixed by the  values $R_1=0.8, R_0=0.9, R_2=0.95$, where $R_1$ and $R_2$ denote the left and right edge of the compact support (see Fig.\,\ref{fig: supports}).
Since for the positive spin value there is no significant difference between the functional behavior of the electromagnetic and gravitational perturbations they are treated together in this subsection. The wave  packets were frequency tuned as it is depicted by the power spectras, $\mathcal PS$, plotted on Fig.\,\ref{fig: s=1-power-m=pm1pm2}. It is clearly visible on both panels of Fig.\,\ref{fig: s=1-power-m=pm1pm2} that the incident wave  packets are well-tuned to be superradiant.

\begin{figure}[ht!]
	\begin{centering}
		{\tiny
			\begin{subfigure}{0.49\textwidth}
				\includegraphics[width=\textwidth]{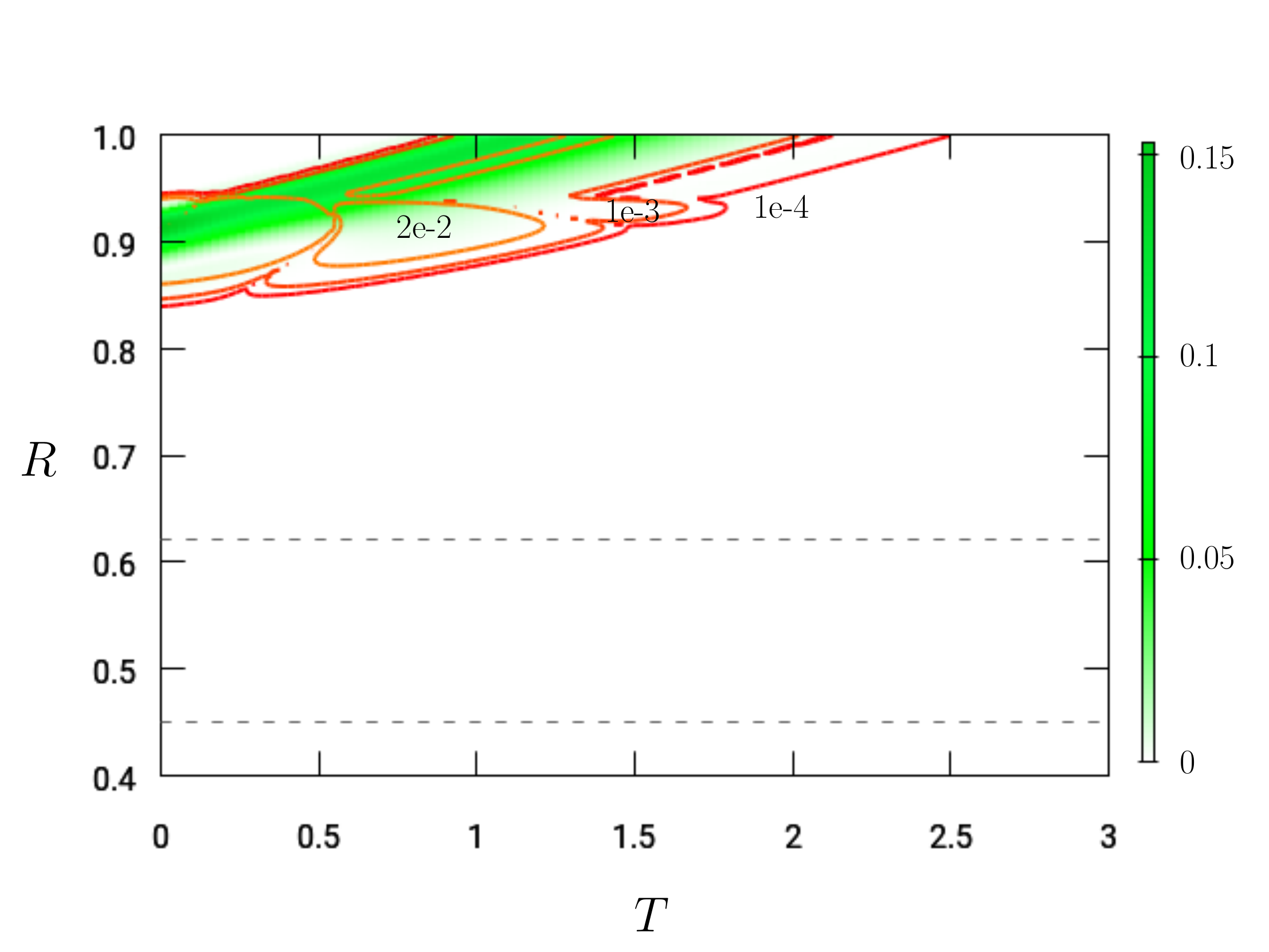}
				\vskip-0.3cm
				\caption{\footnotesize $\mathscr{E}$ and $\mathscr{J}$ with $s=+1$ and $m=+1$. }
			\end{subfigure}
			\hskip.02\textwidth
			\begin{subfigure}{0.49\textwidth}
				\includegraphics[width=\textwidth]{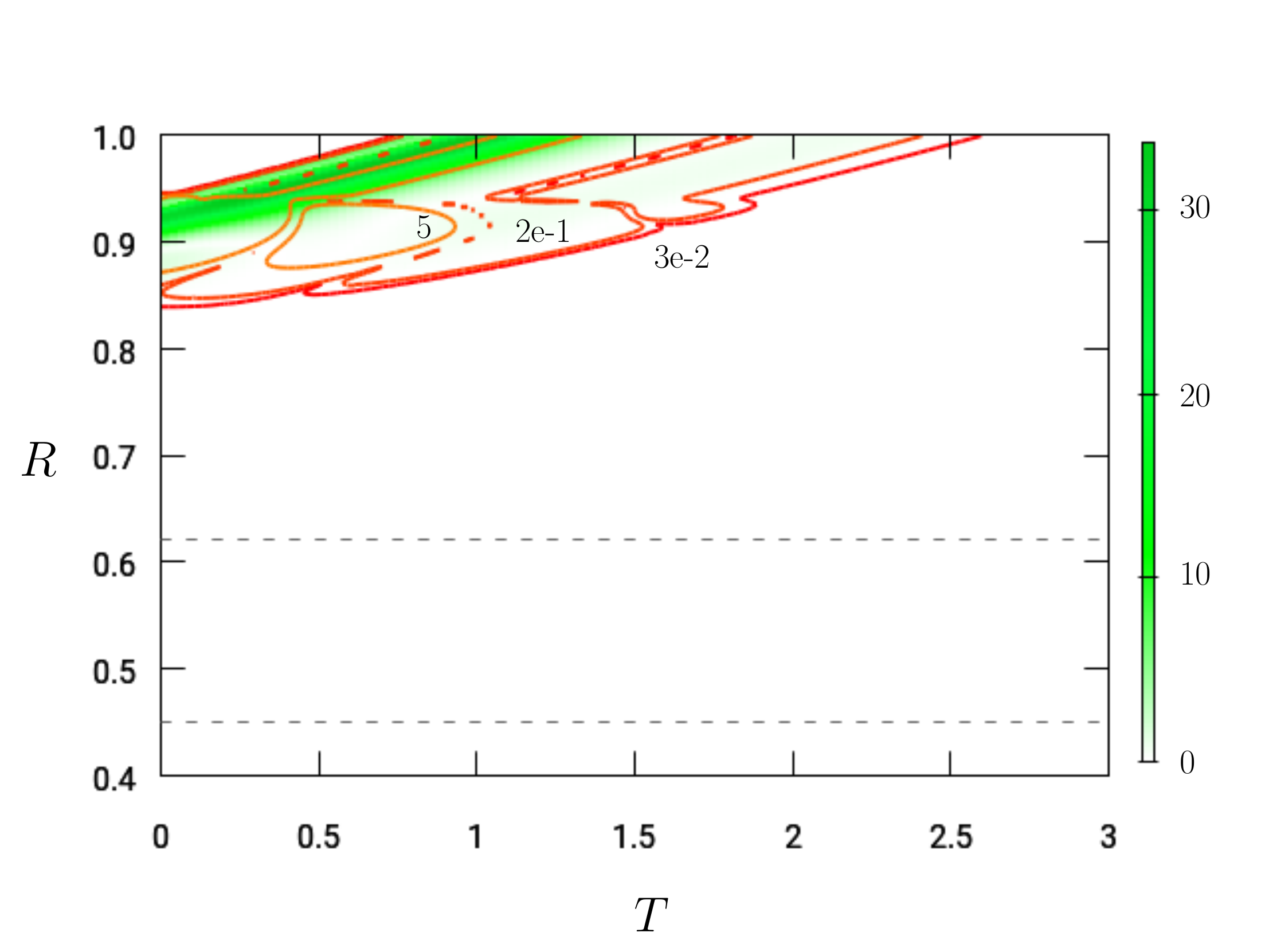}
				\vskip-0.3cm
				\caption{\footnotesize $\mathscr{E}$ and $\mathscr{J}$ with $s=+2$ and $m=+2$. }
			\end{subfigure}
		}
	\end{centering}\vskip-0.2cm
	\caption{{\footnotesize The normalized energy and angular momentum current densities, $\mathscr{E}(T,R)$ and $\mathscr{J}(T,R)$, are plotted for outgoing electromagnetic and gravitational perturbations with physical parameters $s=+1,\ell=1,m=+1,\omega_0= 0.21339$  and $s=+2,\ell=2,m=+2,\omega_0=0.21339$, respectively. 
	}}\label{fig: s=1-curr-transport-m=pm1pm2} 
\end{figure}

The corresponding energy and angular momentum transports (see the panels of Fig.\,\ref{fig: s=1-curr-transport-m=pm1pm2}) 
do also indicate that the $s=+1$electromagnetic and $s=+2$ gravitational perturbations behave analogously. In accordance with their reputation, for these type-I configurations the wave  packets start to move directly toward null infinity and they quickly leave the domain of outer communication. 
Note also that this does not allow us to have energy and angular-momentum gains in either case.  It is also worth emphasizing that all the foregoing observations are unaffected by the concrete value of the spin as only the outgoing character of considered electromagnetic and gravitational perturbations what matters. The above conclusions are insensitive to the sign of $m$, i.e.~to the co- or counterrotating character of these perturbations. 	

\medskip

\begin{figure}[ht!]
	\begin{center}
		{\tiny
			\begin{subfigure}{0.48\textwidth}
				\includegraphics[width=\textwidth]{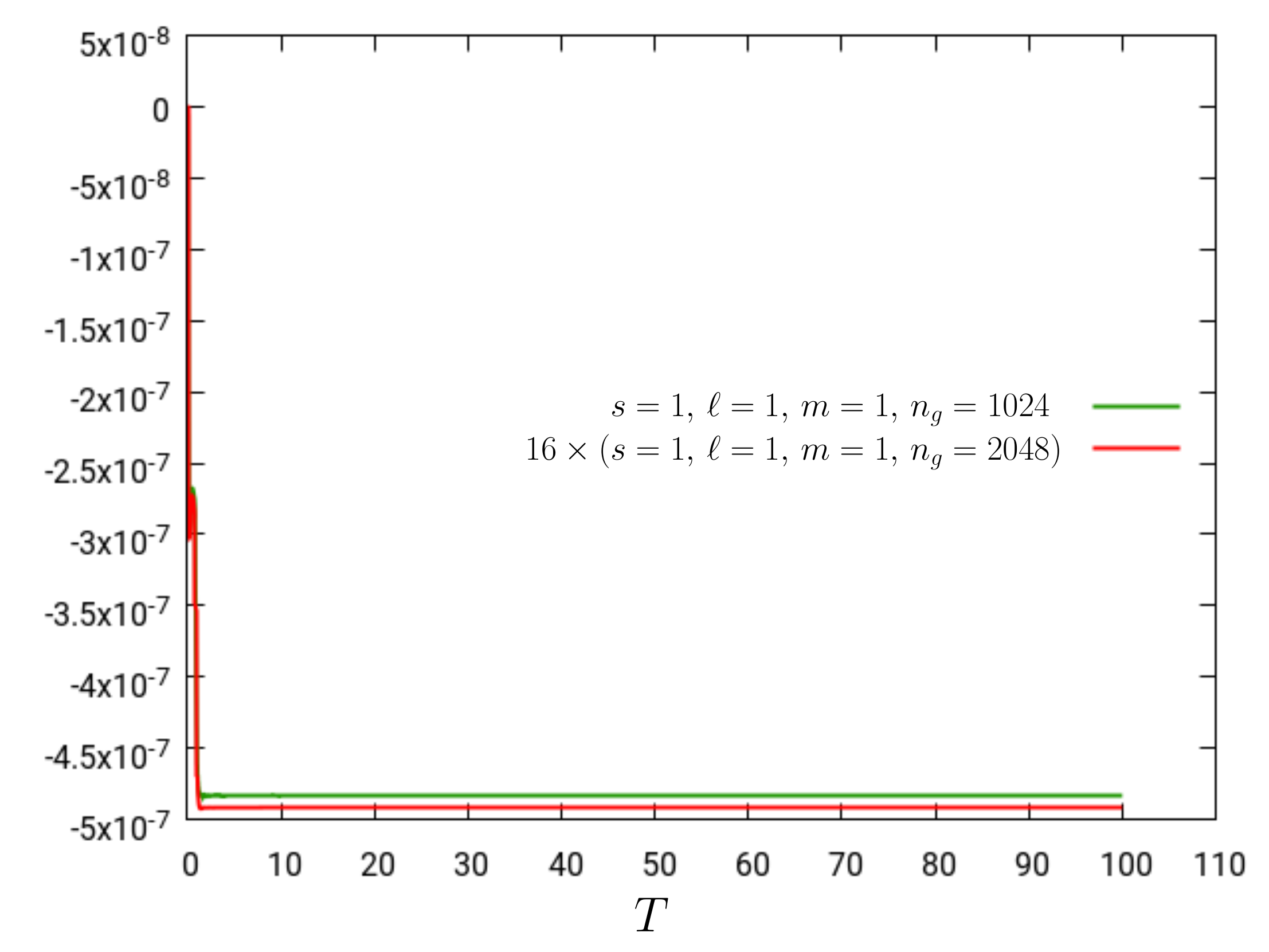}
				\vskip-0.3cm
				\caption{\footnotesize The convergence rate in the $T-R$ section is verified by plotting the real part of the energy balance for the indicated two radial resolutions.  }
			\end{subfigure}
			\hskip.02\textwidth
			\begin{subfigure}{0.48\textwidth}
				\includegraphics[width=\textwidth]{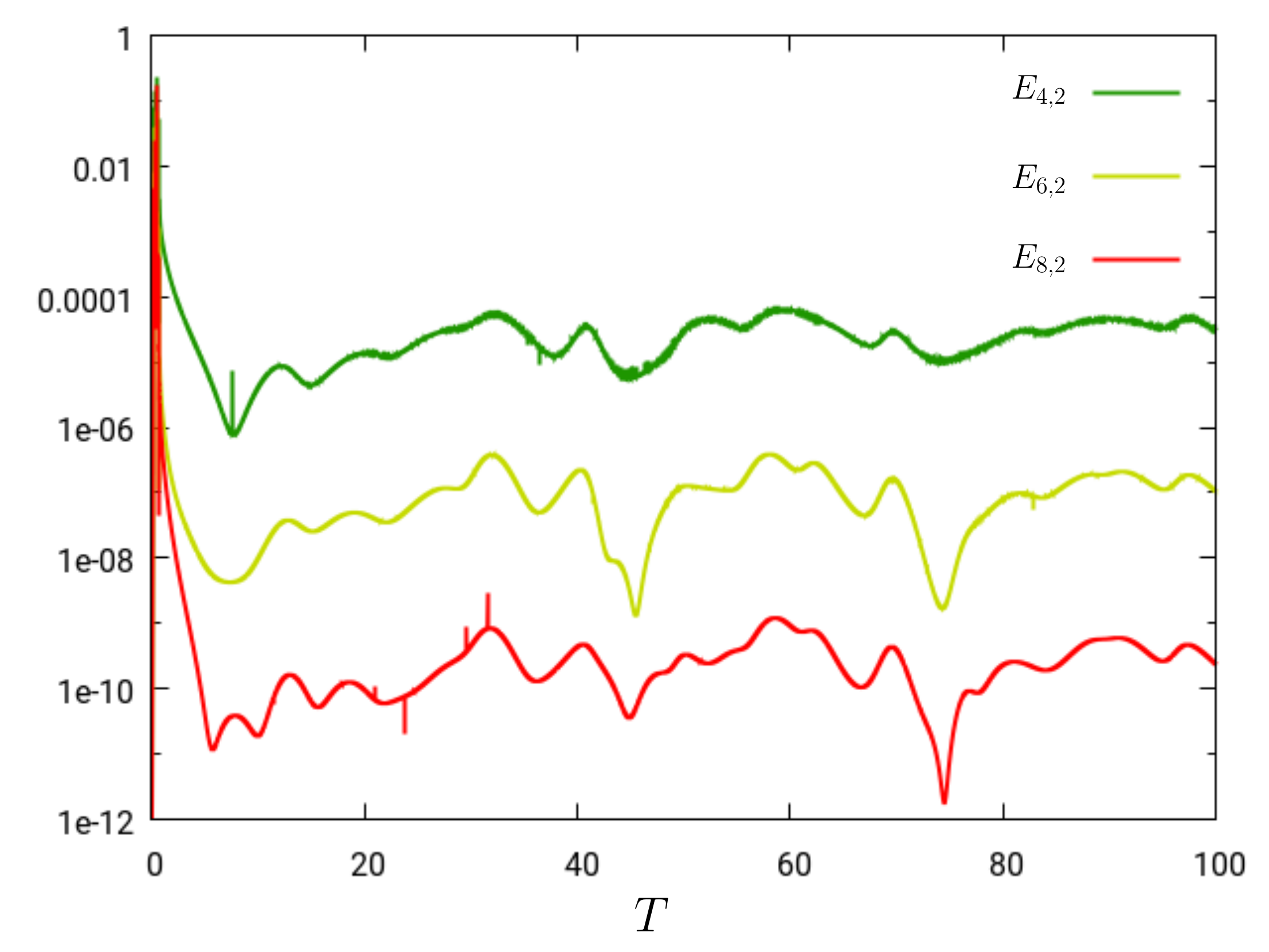}
				\vskip-0.3cm
				\caption{\footnotesize The relative error $E_{\ell_{max}, \Delta \ell_{max}}(\Phi^{(-1)})$  is plotted, for the indicated $\ell_{max}$ values, at the location $R=0.8$ and on the interval $0 \leq T \leq 100$. }
			\end{subfigure}
		}
	\end{center}
\vskip-0.5cm
	\caption{{\footnotesize The convergence rates in the $T-R$ and in the angular sections are verified for the co-rotating solution---with parameters $s=+1, \omega_0=0.21339,  \ell=1$ and $m=+1$, respectively. The convergence rate in the $T-R$ section is of $4^{th}$ order whereas in the angular section it is exponential. 
	}}\label{fig: s=+1-ang-conv} 
\end{figure}

In closing this section it is informative to have a glance of the energy and angular momentum balances. 
The right panel of Fig.\,\ref{fig: s=+1-ang-conv} is to verify the fourth order convergence, in the $T-R$ section,  of the real part of the energy balance  (see also the analogous figures, e.g. Fig.\,4, in \cite{cskritgzs-2019}). 
The convergence rate in the angular  section is also of obvious interest. This---in applying the multipole expansions, as described in section \ref{sec: initialization}---can be monitored by varying the value of $\ell_{max}$, i.e.,~the value of $\ell$ where the infinite series is truncated in \eqref{eq: mult-exp}. To characterize the  $\ell_{max}$ dependence the relative error of the basic variable $\Phi^{(s)}$, defined via the ratio,
\begin{equation}
E_{\ell_{max}, \Delta \ell_{max}}(\Phi^{(s)})=\frac{ || \Phi^{(s)}_{ \ell_{max}} - \Phi^{(s)}_{ \ell_{max}+\Delta
		\ell_{max}} || }{|| \Phi^{(s)}_{ \ell_{max}+\Delta  \ell_{max}} ||}
\end{equation}
is monitored, where $||\cdot ||$ denotes the $C^0$ norm. On the right panel of Fig.\,\ref{fig: s=+1-ang-conv} the relative error, relevant for the sequence of values $\ell_{max}=4, 6, 8, 10$, is plotted for an outgoing electromagnetic perturbation with  physical parameters  $s=+1, \omega_0=0.21339,  \ell=1, m=+1$ for the time interval $0 \leq T \leq 100$. The apparent linear shift in the succeeding graphs indicates that the rate of convergence in the angular section is, as expected, exponential.

\subsubsection{Ingoing electromagnetic and gravitational perturbations
}

\begin{figure}[ht!]
	\vskip-0.4cm
	\begin{centering}
		{\tiny
			\begin{subfigure}{0.49\textwidth}
				\includegraphics[width=\textwidth]{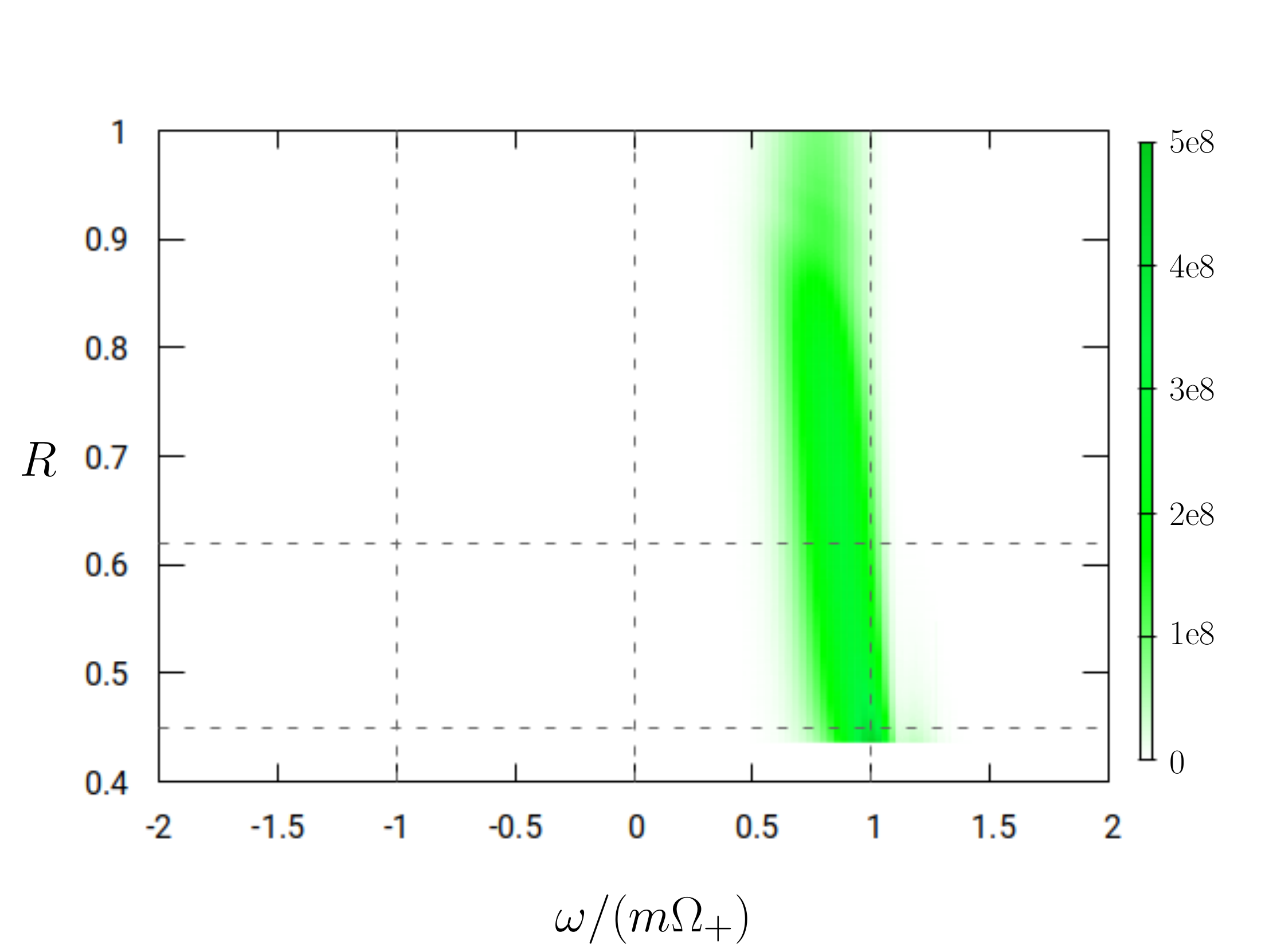}
				\caption{\footnotesize $\mathcal{PS}$ for solution with $s=-1, m=+1$. }
			\end{subfigure}
			\hskip.02\textwidth
			\begin{subfigure}{0.49\textwidth}
				\includegraphics[width=\textwidth]{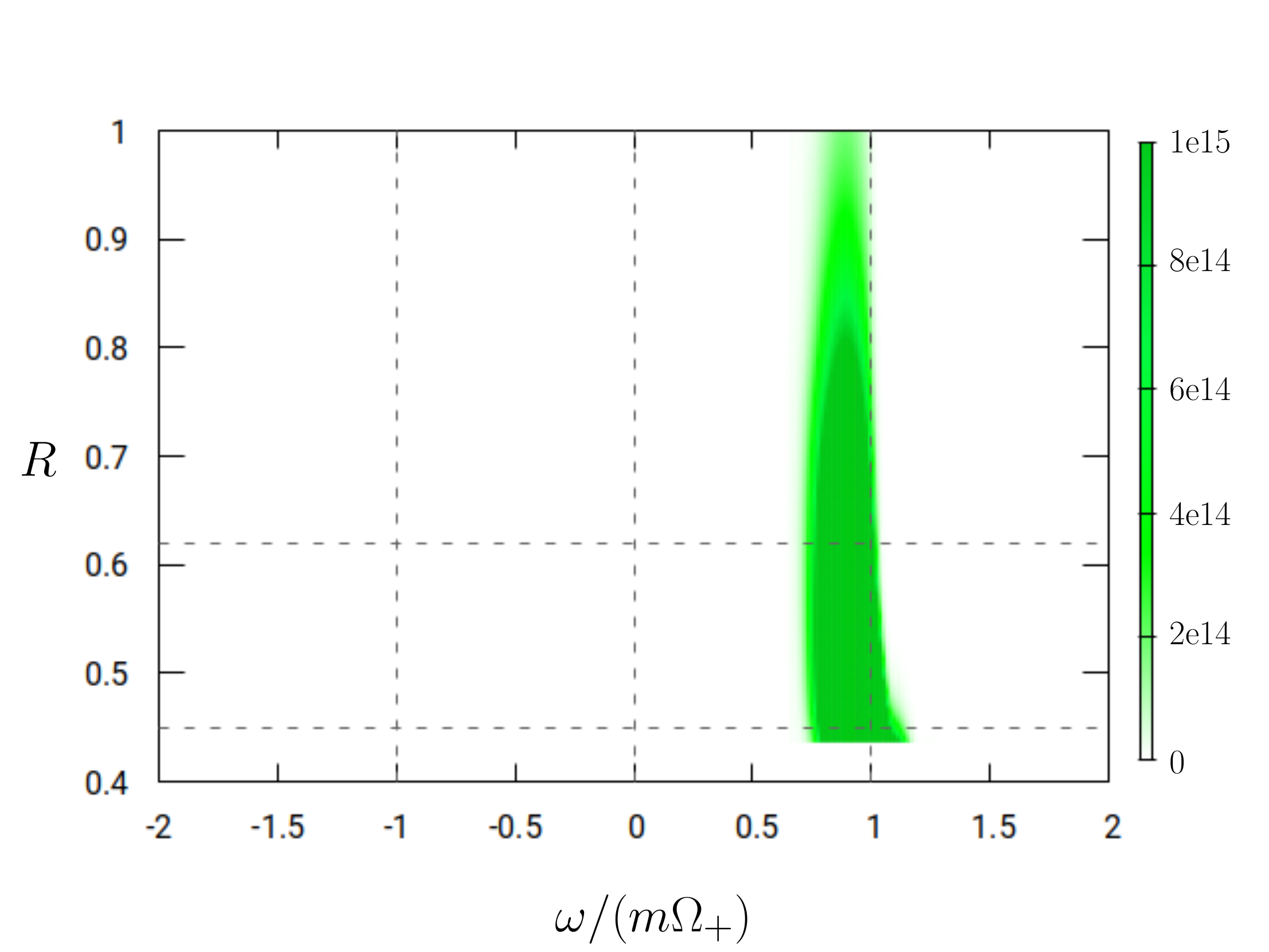}
				\caption{\footnotesize $\mathcal{PS}$ for solution with $s=-2, m=+2$. }
				\label{fig:s-2in}
			\end{subfigure}
		}
	\end{centering}\vskip-0.2cm
	\caption{{\footnotesize The power spectra of ingoing and corotating electromagnetic and gravitational perturbations---with physical parameters $s=-1, \omega_0=0.28,  \ell=1, m=+1$ and  $s=-2, \omega_0=0.65,  \ell=2, m=+2$, are plotted, respectively. Notice that considerably large fraction of both of the incident wave packets, are in the superradiant regime and remain there for the entire evolution. 
	}}\label{fig: s=-1-power-m=pm1} 
\end{figure}

The applied radial profile is determined here again by the parameters $R_1=0.93, R_0=0.95, R_2=0.98$ (see Fig.\,\ref{fig: supports}). The other physical parameters are $s=-1, \omega_0=0.28,  \ell=1, m=+1$ and $s=-2, \omega_0=0.65,  \ell=2, m=+2$, respectively. The story is very similar to the one reported in case of ingoing corotating scalar fields. The power spectra  $\mathcal{PS}=\mathcal{PS}(\omega,R)$ of the ingoing electromagnetic and gravitational perturbations are shown on the two panels of Fig.\,\ref{fig: s=-1-power-m=pm1}.
The power spectra for the corotating electromagnetic and gravitational perturbations are located entirely in the superradiant interval $0< \omega < m\,\Omega_{+}$ not only in the vicinity of the compact support of the initial data surface. 
\begin{figure}[ht!]
	\begin{centering}
		{\tiny
			\begin{subfigure}{0.49\textwidth}
				\includegraphics[width=\textwidth]{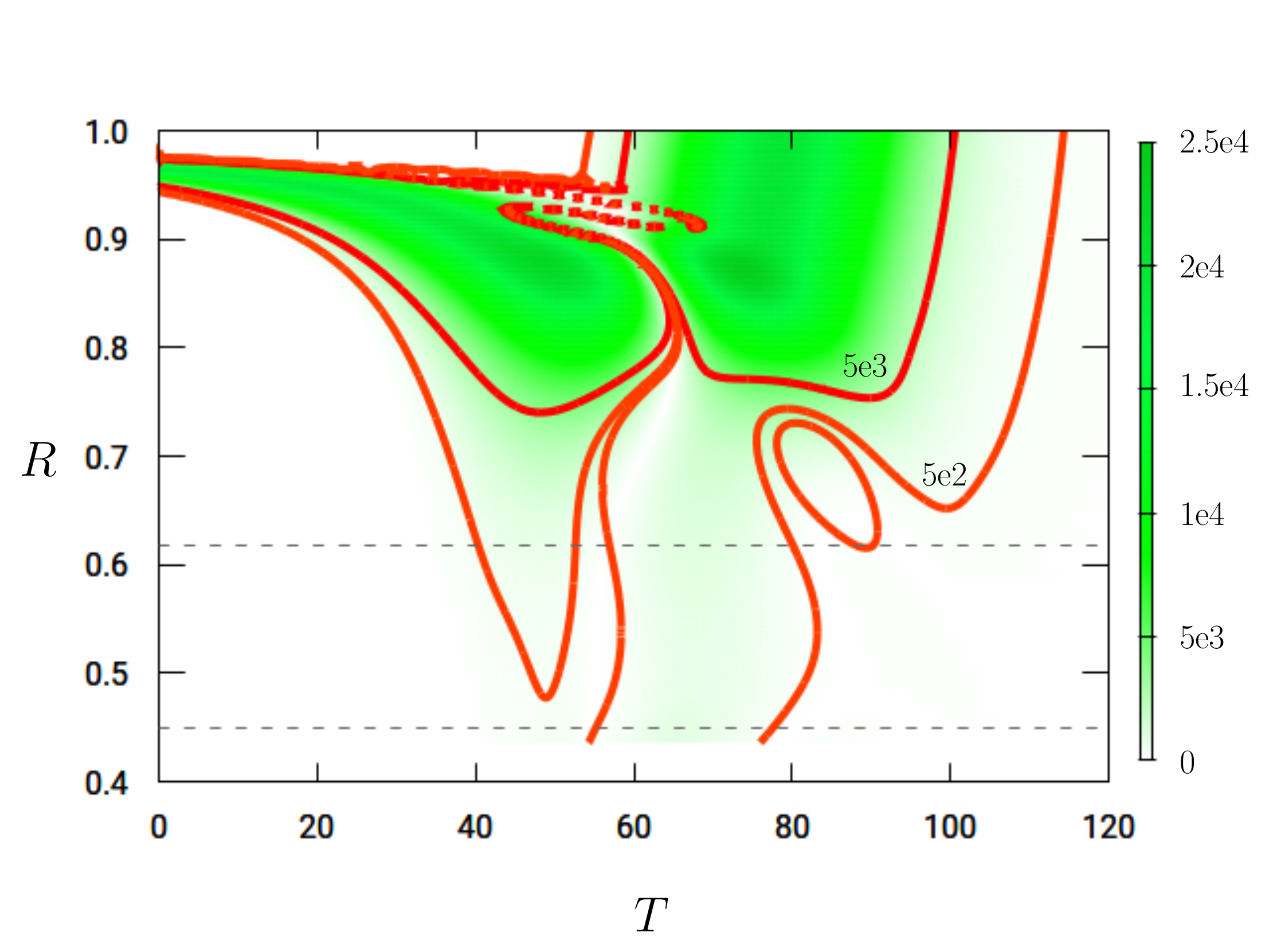}
				\vskip-0.3cm
				\caption{\footnotesize $\mathscr{E}$ and $\mathscr{J}$ for $s=-1$ and $m=+1$. }
			\end{subfigure}
			\hskip.02\textwidth
			\begin{subfigure}{0.49\textwidth}
				\includegraphics[width=\textwidth]{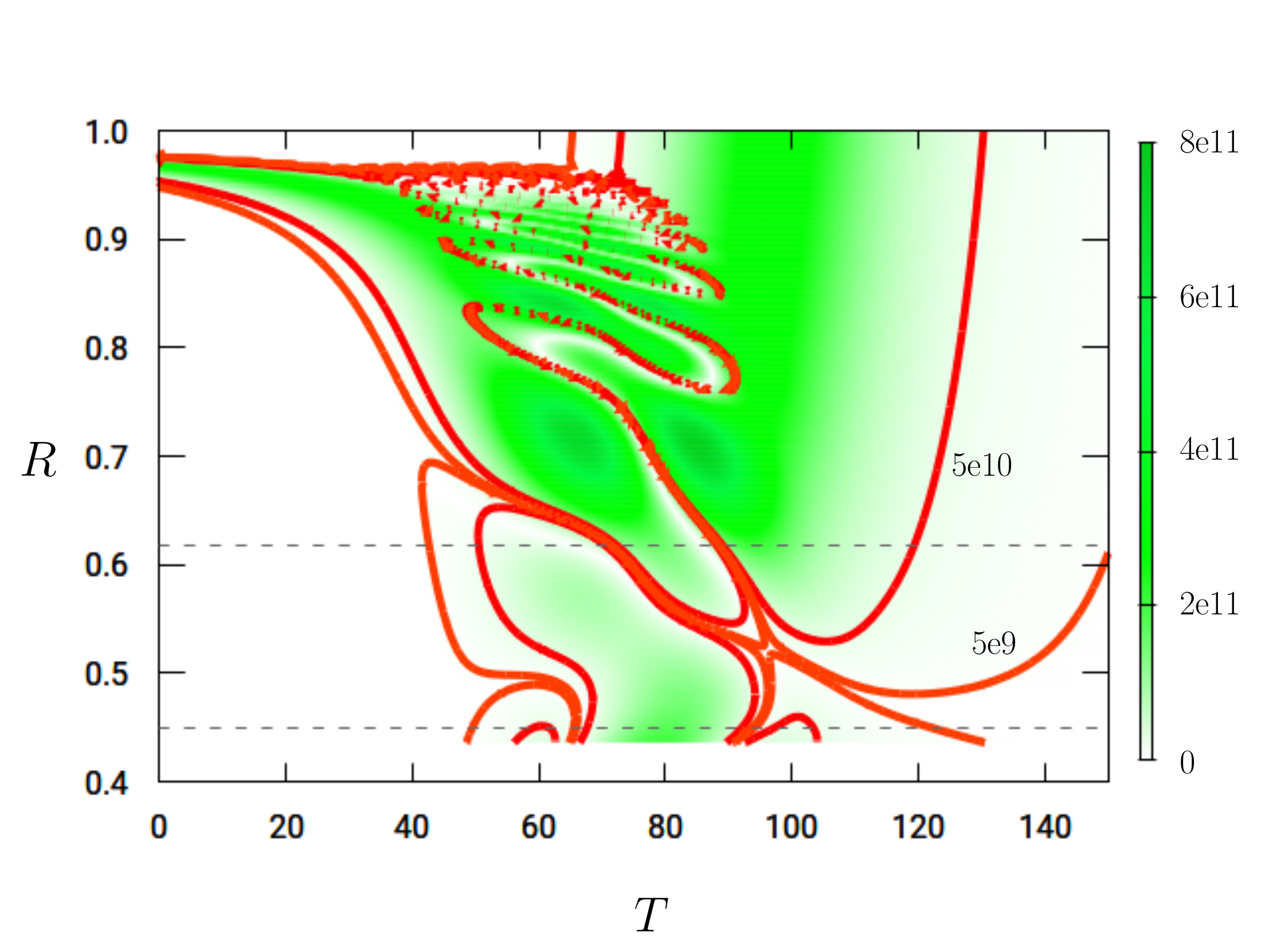}
				\vskip-0.3cm
				\caption{\footnotesize $\mathscr{E}$ and $\mathscr{J}$ for $s=-2$ and $m=+2$. }
			\end{subfigure}
		}
	\end{centering}\vskip-0.2cm
	\caption{{\footnotesize The volume normalized energy and angular momentum current densities, $\mathscr{E}(T,R)$ and  $\mathscr{J}(T,R)$ are plotted for ingoing and corotating  electromagnetic and gravitational perturbations, respectively. The physical parameters are exactly the same as those applied in producing  Fig.\,\ref{fig: s=-1-power-m=pm1}.
	}}\label{fig: s=-1-curr-transport-m=pm1} 
\end{figure}
Accordingly, only a small part of the energy and angular momentum of the incident wave falls into the black hole. There are also small fractions of the power spectra with slightly larger frequency than $m\,\Omega_{+}$. These, by determining the power spectra relevant only for the late time period, can be seen to correspond to the quasinormal oscillations of the electromagnetic and gravitational perturbations, respectively. 

\medskip

The energy and angular momentum transports are in accordance with the comments have just made by inspecting the power spectras.  
On the left and right panels of Fig.\,\ref{fig: s=-1-curr-transport-m=pm1} the volume normalized energy and angular momentum current densities, $\mathscr{E}=\mathscr{E}(T,R)$ and  $\mathscr{J}=\mathscr{J}(T,R)$ are shown for the inspected ingoing and corotating electromagnetic and gravitational perturbations, respectively.

As before, the angular momentum current densities are indicated only by colored contours such that the pertinent values are marked by the numbers next to these contours. 
\begin{figure}[ht!] 
	\begin{centering}
		{\tiny
			\begin{subfigure}{0.49\textwidth}
				\includegraphics[width=\textwidth]{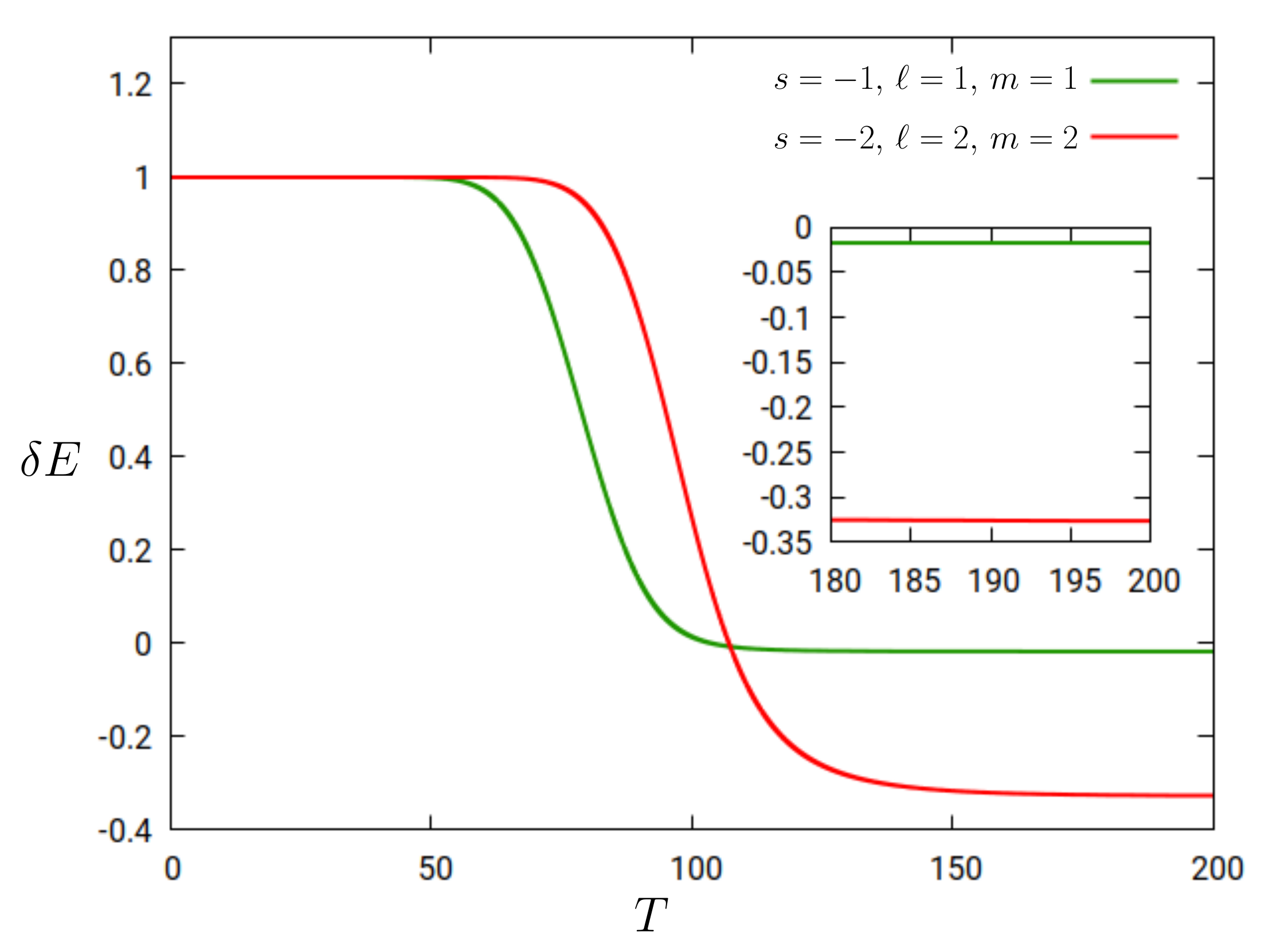}
				\caption{\footnotesize The energy gains. }
			\end{subfigure}
			\hskip.02\textwidth
			\begin{subfigure}{0.49\textwidth}
				\includegraphics[width=\textwidth]{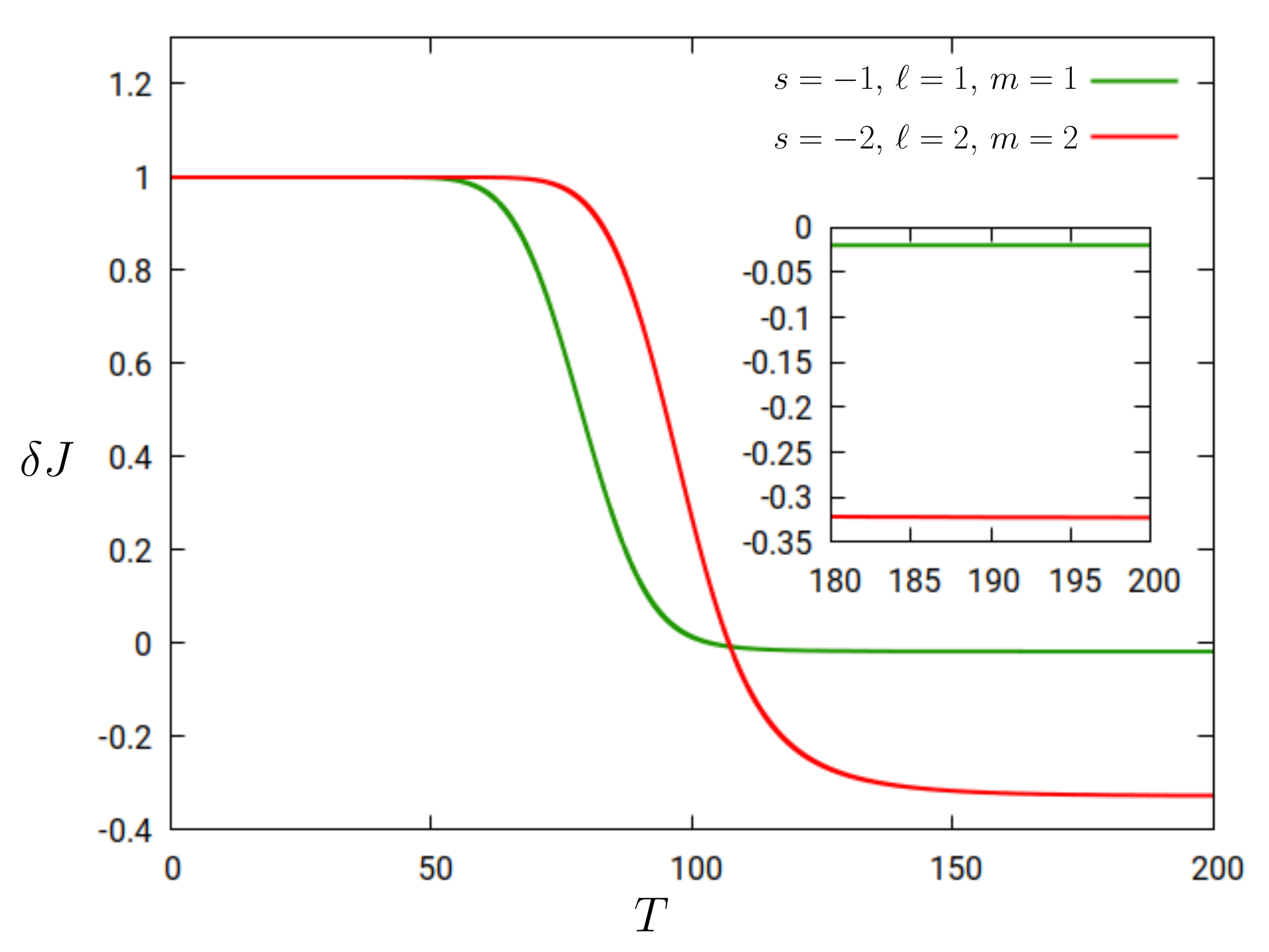}
				\caption{\footnotesize The angular momentum gains. }
			\end{subfigure}
		}
	\end{centering}\vskip-0.2cm
	\caption{{\footnotesize The time dependence of the energy and angular momentum gains, $\delta E=(E_0-E_{out})/E_0$ and $\delta J=(J_0-J_{out})/J_0$, are plotted for the investigated ingoing and corotating  electromagnetic and gravitational perturbations, respectively. As it is indicated by these panels the gains are still modest as they are about $2\%$ and $32.5\%$ for electromagnetic and gravitational perturbations, respectively.
	}}\label{fig: s=-1-super-m=pm1} 
\end{figure}
Both of the panels verify that considerably large fraction of the energy and angular momentum of the ingoing, corotating configurations bounced back before entering to the ergoregion and leave the domain of outer communication via null infinity $\mathscr{I}^+$, that is represented by the $R=1$ surface. Nevertheless, it is also clearly visible that dynamical processes occur also in the ergoregion which indicates that superradiance might also occur in these cases.


\medskip

Recall that to quantify superradiance the energy and angular momentum gains,  $\delta E=(E_0-E_{out})/E_0$ and $\delta J=(J_0-J_{out})/J_0$, are used. On the two panels of  Fig.\,\ref{fig: s=-1-super-m=pm1}  
the time dependence of these quantities are plotted for the investigated ingoing and co-rotating  electromagnetic and gravitational perturbations, respectively. 
As noted earlier, if either $\delta E$ or $\delta J$ gets negative it means that superradiance does indeed occur. Although the   energy and angular momentum gains, $\delta E$ and $\delta J$, are modest $\sim2\%$ for electromagnetic perturbations, whereas the scale of superradiance is about $~32.5\%$ for gravitational perturbations.

\subsection{Results relevant for type-II initial data}\label{subsec: ID-overlapping}

In this section type-II configurations will be considered. This means that the compact support of the initial data and the ergoregion overlap significantly. Actually, in all the reported cases we shall choose the outer radius of the compact support to be slightly smaller than the pertinent equatorial radius of the ergoregion. In practice, the characteristic parameters of the radial profile of the compact support will be set as $R_1=0.4536, R_2=0.615$, and $R_0=0.5$, where $R_1$ and $R_2$ denote the left and right edges of the compact support (see Fig.\,\ref{fig: supports}). 
\begin{figure}[ht!] 	
	\begin{centering}	
		{\tiny
			\begin{subfigure}{0.49\textwidth}
				\includegraphics[width=\textwidth]{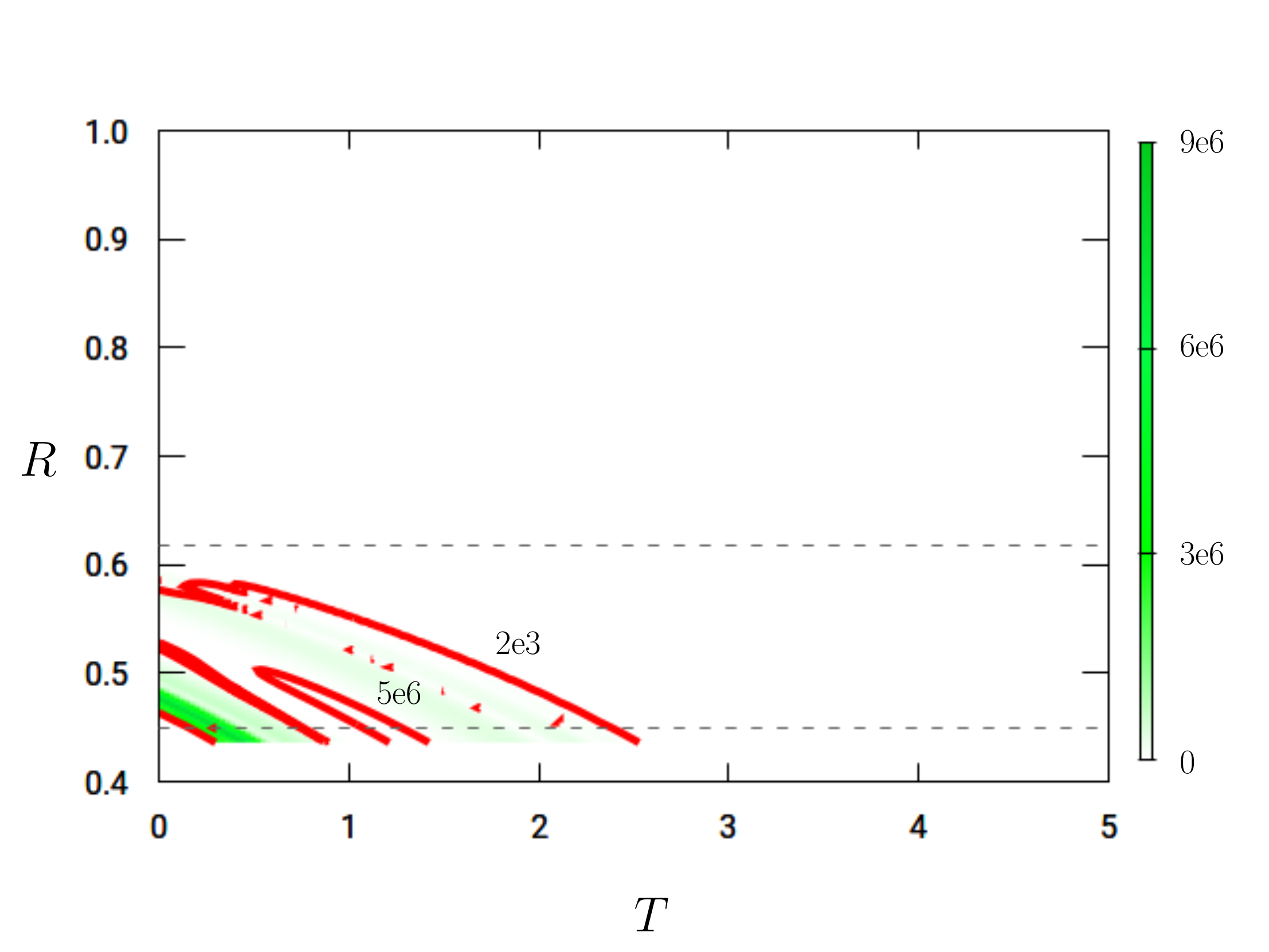}
				\vskip-0.2cm
				\caption{\footnotesize  $\mathscr{E}$ and $\mathscr{J}$ with $s=-2$ and $m=+2$.}
			\end{subfigure}
			\hskip.02\textwidth
			\begin{subfigure}{0.49\textwidth}
				\includegraphics[width=\textwidth]{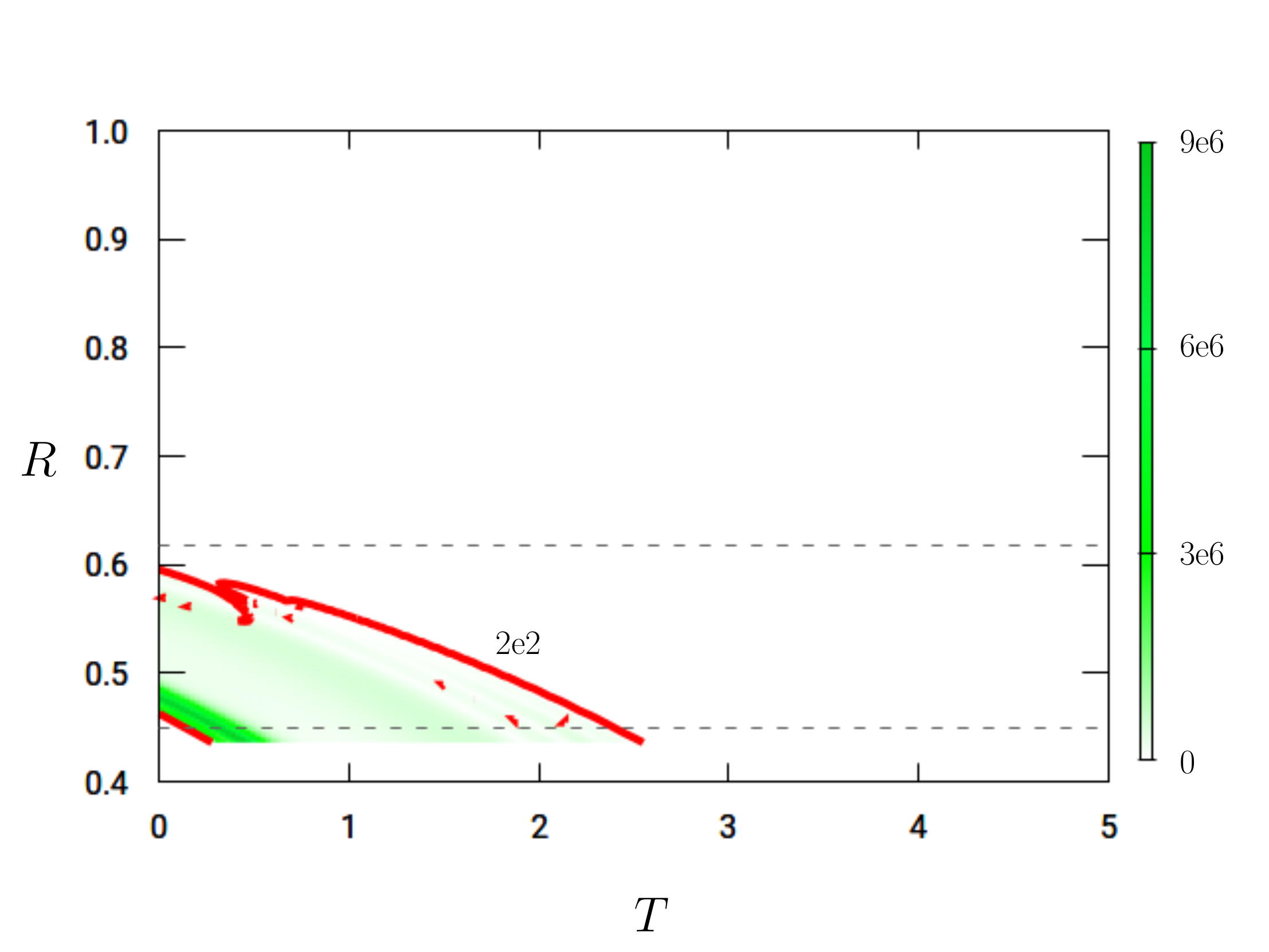}
				\vskip-0.2cm
				\caption{\footnotesize $\mathscr{E}$ and $\mathscr{J}$ with $s=-2$ and $m=-2$.}
			\end{subfigure}
		}
	\end{centering}\vskip-0.2cm
	\caption{\footnotesize The volume normalized energy and angular momentum current densities, $\mathscr{E}(T,R)$ and  $\mathscr{J}(T,R)$ are plotted for ingoing  electromagnetic and gravitational perturbations, respectively. The physical parameters are $s=-2, \omega_0=0.725,  \ell=2, m=\pm2$.
	}\label{fig: s=-2-ing-ebr-m=pm2} 
\end{figure}
Note also that, as in the previous section, the mass and specific angular momentum parameters of the Kerr black hole are $M=1$ and $a=0.99$, respectively. These physical parameters set the horizon radius as $R_+=0.4533$ while the equatorial radius of the ergoregion is $R_E=0.618$.

\subsubsection{Ingoing scalar, electromagnetic and gravitational perturbations
}\label{subsubsec: type-II s=0,-1,2}
\begin{figure}[ht!] 
	\vskip-.3cm
	\begin{centering}
		{\tiny
			\begin{subfigure}{0.49\textwidth}
				\includegraphics[width=\textwidth]{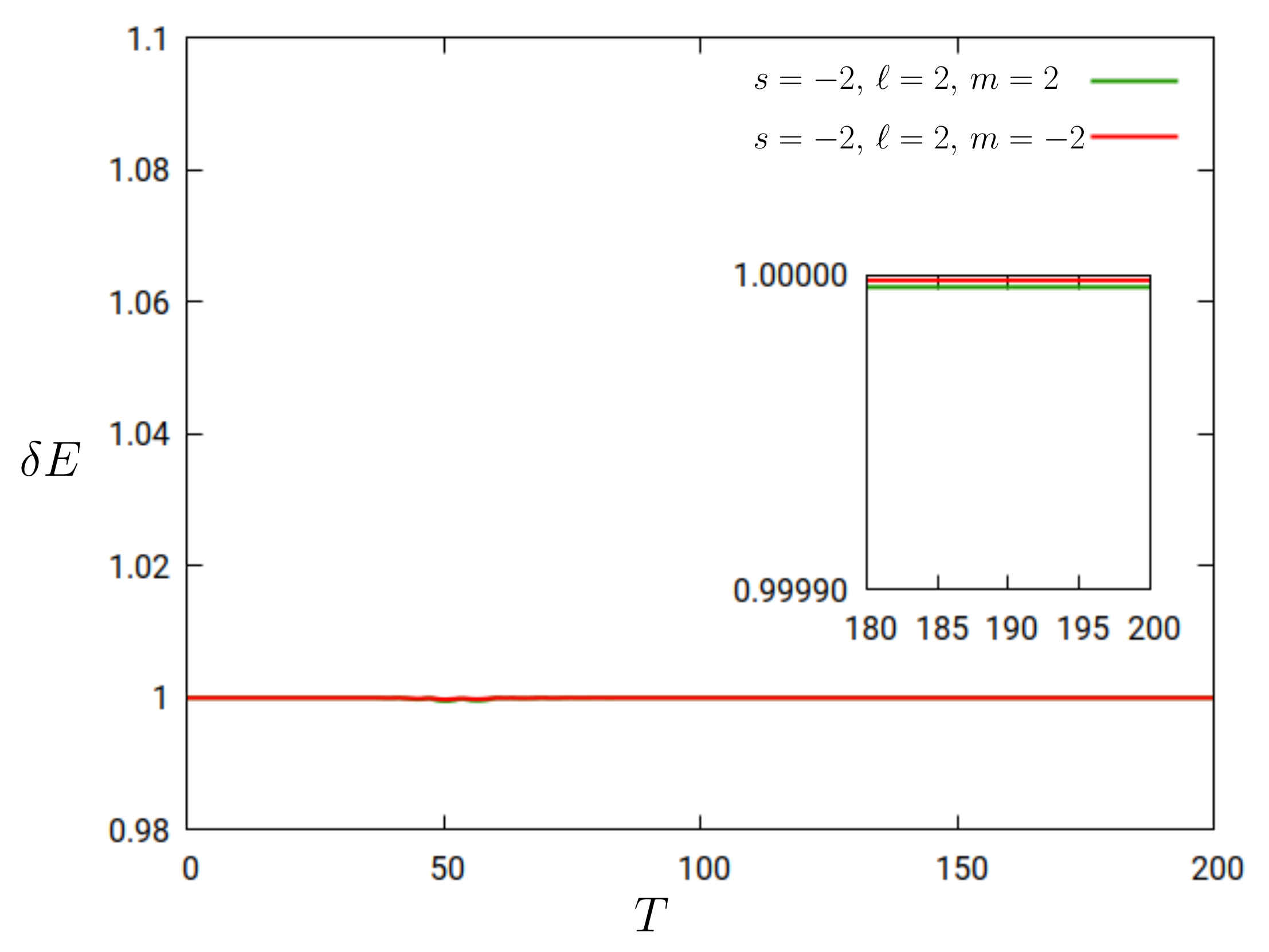}
				\caption{\footnotesize Energy gains: $s=-2, m=\pm 2$. }
			\end{subfigure}
			\hskip.02\textwidth
			\begin{subfigure}{0.49\textwidth}
				\includegraphics[width=\textwidth]{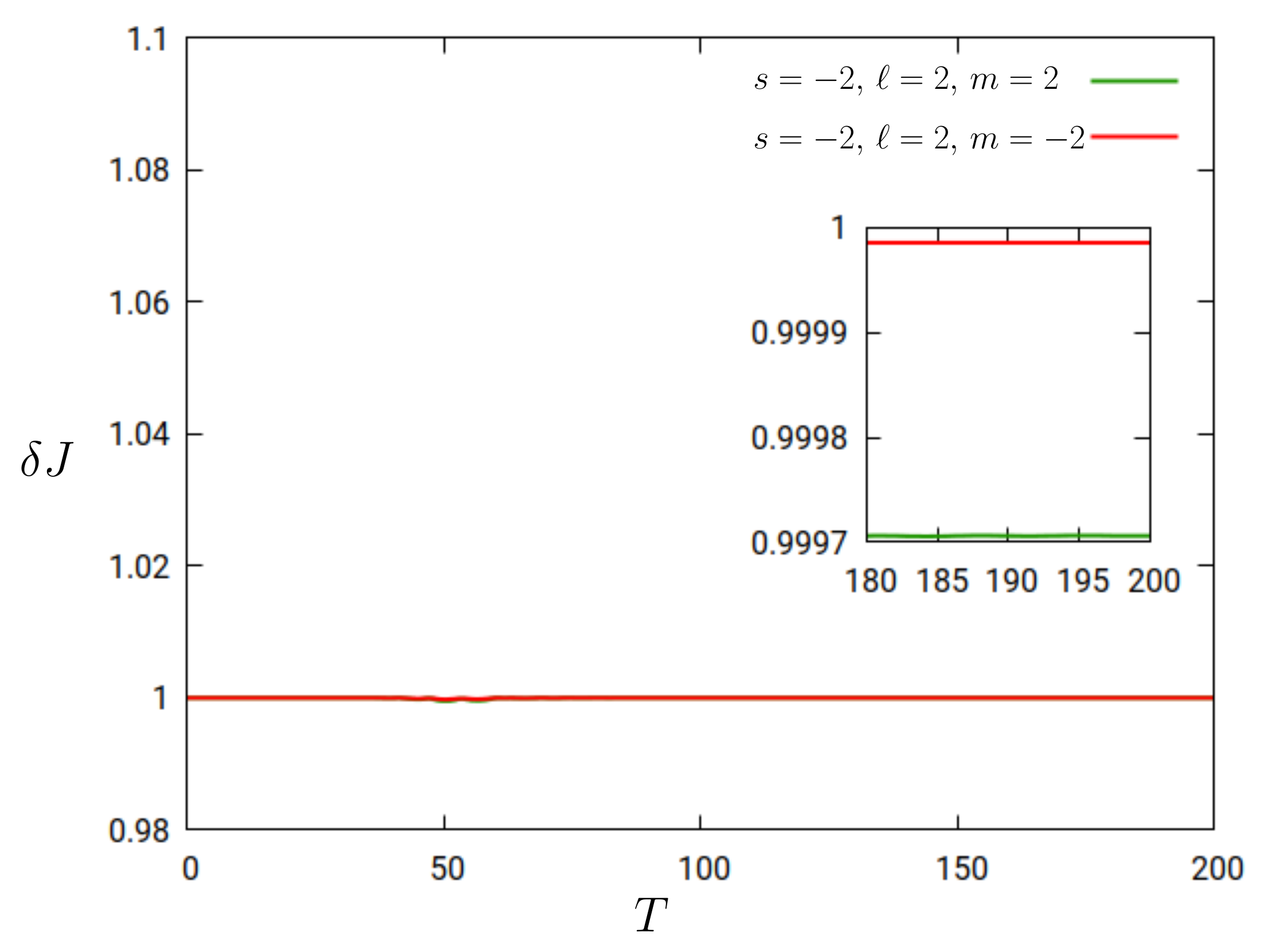}
				\caption{\footnotesize Angular momentum gains: $s=-2, m=\pm 2$. }
			\end{subfigure}
		}
	\end{centering}\vskip-0.2cm
	\caption{{\footnotesize The time dependence of the energy and angular momentum gains, $\delta E=(E_0-E_{out})/E_0$ and $\delta J=(J_0-J_{out})/J_0$, are plotted for the investigated ingoing electromagnetic and gravitational perturbations, respectively. These panels verify that no gains thereby no superradiance happens. 
	}}\label{fig: s=-2-super-m=pm2} 
\end{figure}
The most important difference between the type-I and type-II configurations is that while the 
ingoing type-I electromagnetic or gravitational perturbations were superradiant they do not show superradiance at all when type-II initial data is used. Instead, the incident wave  packets fall directly toward the black hole region and leave the computational domain through the black hole horizon.

Accordingly, neither of the power spectra of these type-II ingoing spin $s=0$, $s=-1$ and $s=-2$ configurations is informative because in each of these cases the truly dynamical part of the evolution is rather short. 

This interpretation is verified by the panels on Fig.\,\ref{fig: s=-2-ing-ebr-m=pm2}
where the energy and angular momentum transports of a gravitational perturbation with physical parameters $s=-2, \omega_0=0.725,  \ell=2, m=\pm2$ are plotted. 

These figures are typical in the sense that completely analogous processes occur in case of ingoing type-II spin $s=0$ and $s=-1$ configurations. 
\begin{figure}[ht!] 
	\vskip-0.4cm
	\begin{centering}
		{\tiny
			\begin{subfigure}{0.49\textwidth}
				\includegraphics[width=\textwidth]{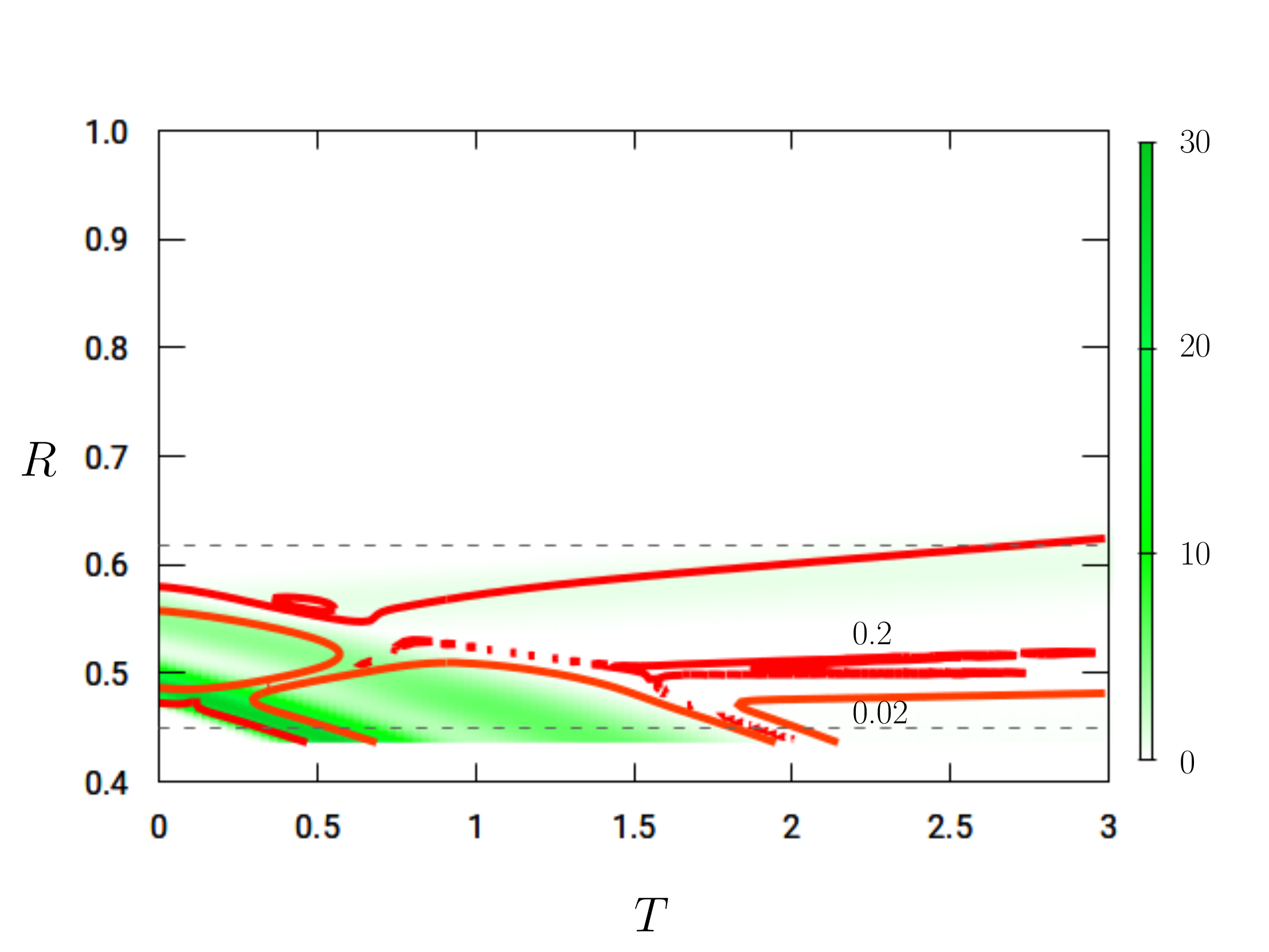}
				\caption{\footnotesize $\mathscr{E}$ and $\mathscr{J}$ with $s=0, m=+1$. }
			\end{subfigure}
			\hskip.02\textwidth
			\begin{subfigure}{0.49\textwidth}
				\includegraphics[width=\textwidth]{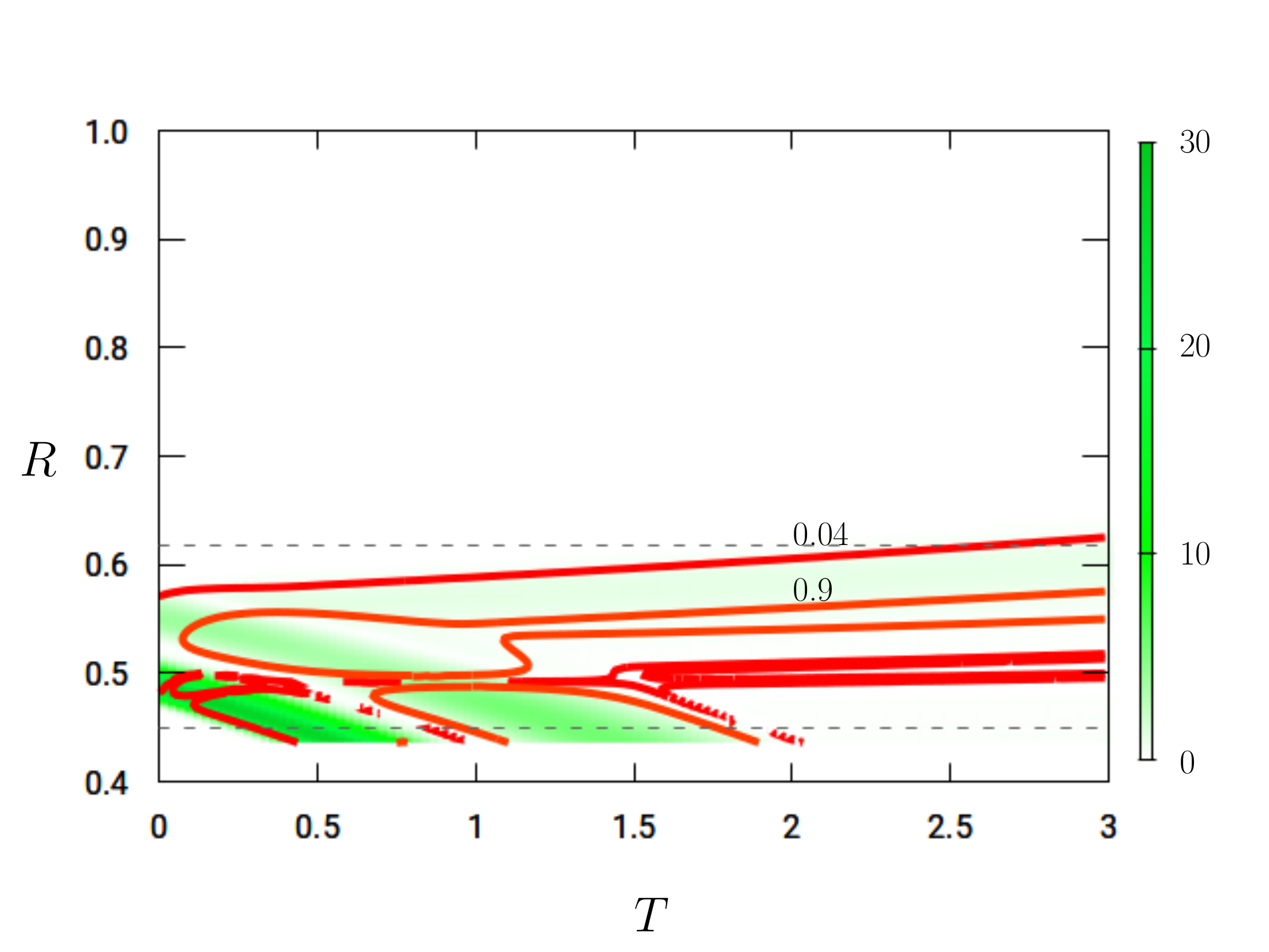}
				\caption{\footnotesize $\mathscr{E}$ and $\mathscr{J}$ with $s=0, m=-1$. }
			\end{subfigure}
		}
	\end{centering}\vskip-0.2cm
	\caption{{\footnotesize The volume normalized energy and angular momentum current densities, $\mathscr{E}(T,R)$ and  $\mathscr{J}(T,R)$ are plotted for ingoing scalar perturbations with reversed time derivative and with physical parameters  $s=0, \omega_0=0.3,  \ell=1, m=\pm1$. }}
	\label{fig: s=0-energy-m=pm1} 
\end{figure}
Exactly the same conclusion can be drawn by inspecting the energy and angular momentum gains, $\delta E=(E_0-E_{out})/E_0$ and $\delta J=(J_0-J_{out})/J_0$, plotted on the panels of Fig.\,\ref{fig: s=-2-super-m=pm2}. 
It is clearly visible that essentially the entire energy and angular momentum stored by the initial data leave, in a very short period, the domain of outer communication via the black hole event horizon.
\begin{figure}[ht!] 
	\vskip-0.4cm
	\begin{centering}
		{\tiny
			\begin{subfigure}{0.49\textwidth}
				\includegraphics[width=\textwidth]{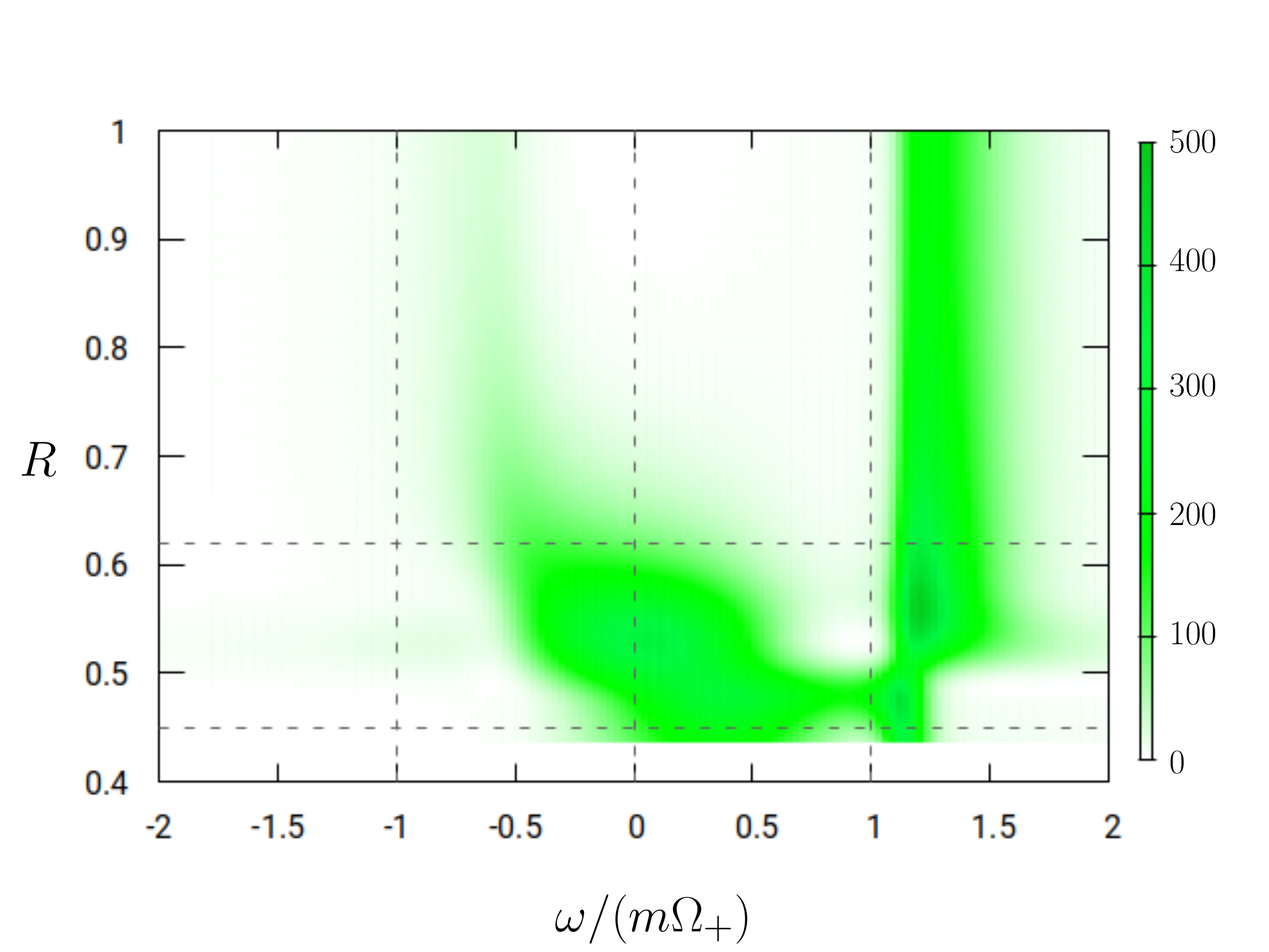}\vskip-0.3cm
				\caption{\footnotesize $\mathcal{PS}$ for solution with  $s=0, m=+1$.}
			\end{subfigure}
			\hskip.02\textwidth
			\begin{subfigure}{0.49\textwidth}
				\includegraphics[width=\textwidth]{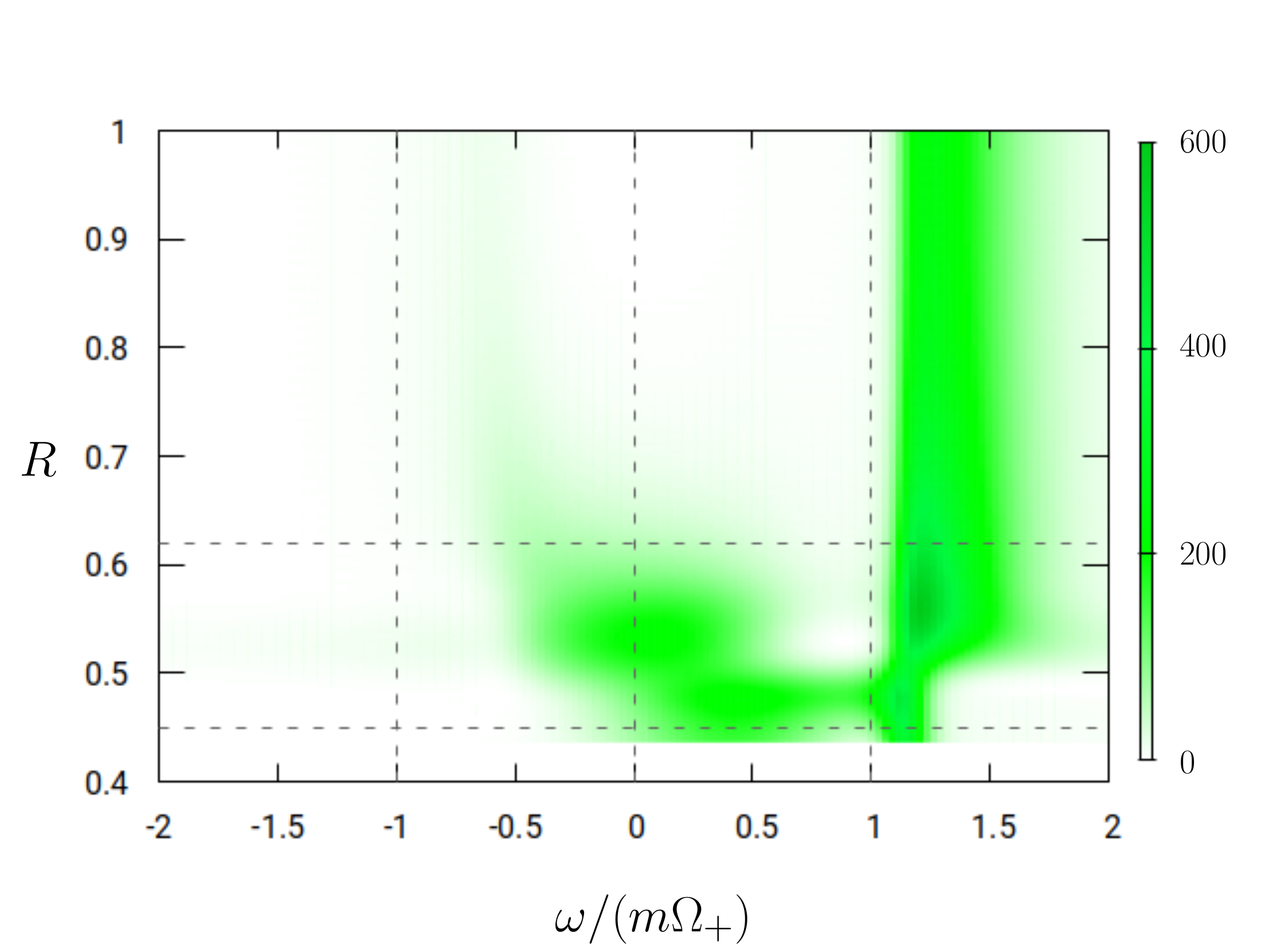}\vskip-0.3cm
				\caption{\footnotesize $\mathcal{PS}$ for solution with  $s=0, m=-1$.}
			\end{subfigure}
		}
	\end{centering}\vskip-0.2cm
	\caption{{\footnotesize The power spectra of the ingoing scalar perturbations---with reversed time derivative and with physical parameters  $s=0, \omega_0=0.3,  \ell=1, m=\pm1$---are plotted, respectively. 
	}}\label{fig: s=0-ft-m=pm1} 
\end{figure}
\begin{figure}[ht!] 
	\vskip-.5cm
	\begin{centering}
		{\tiny
			\begin{subfigure}{0.49\textwidth}
				\includegraphics[width=\textwidth]{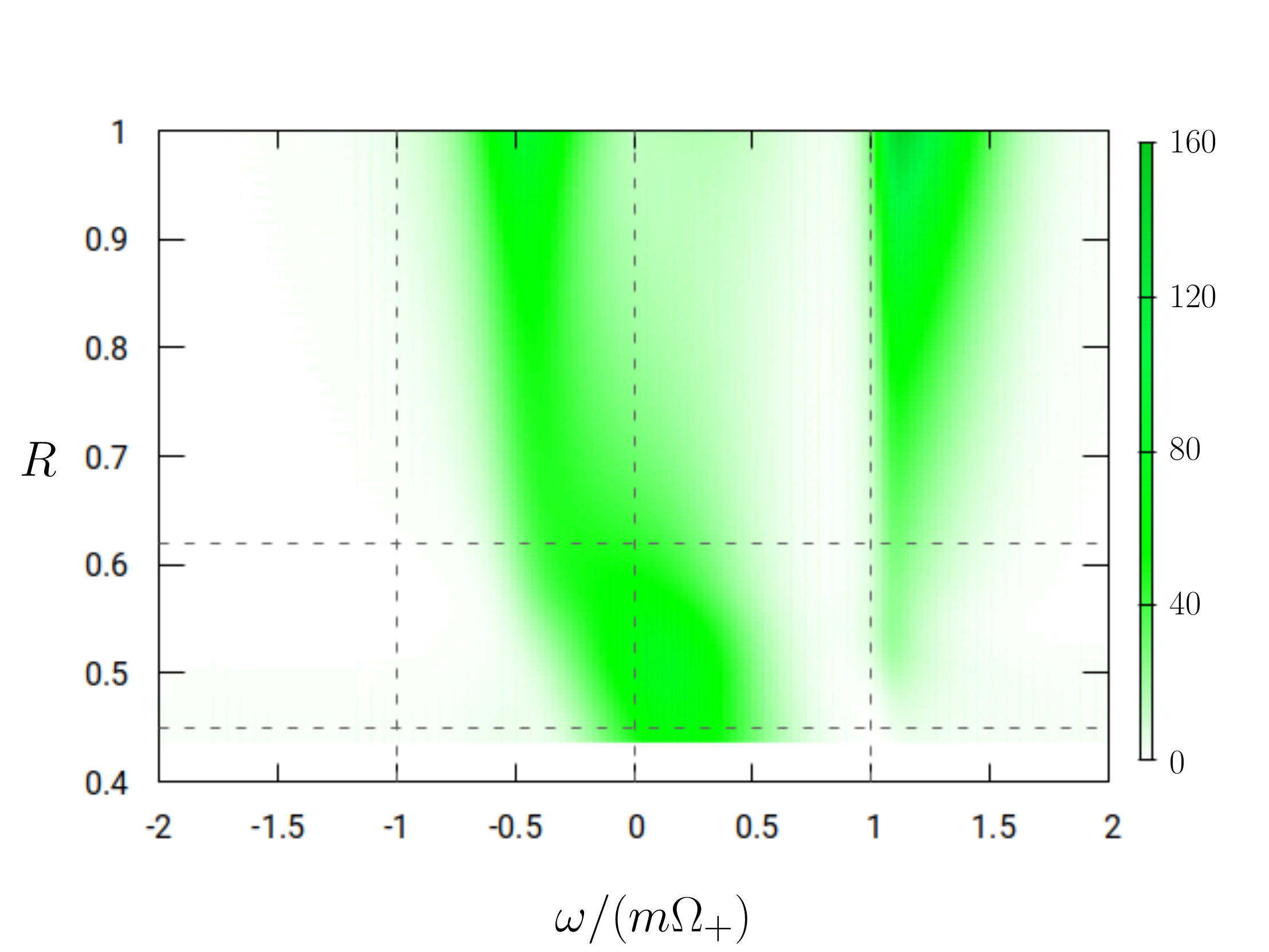}
				\vskip-0.2cm\caption{\footnotesize $\mathcal{PS}$ for solution with  $s=+1, m=+1$.}
			\end{subfigure}
			\hskip.02\textwidth
			\begin{subfigure}{0.49\textwidth}
				\includegraphics[width=\textwidth]{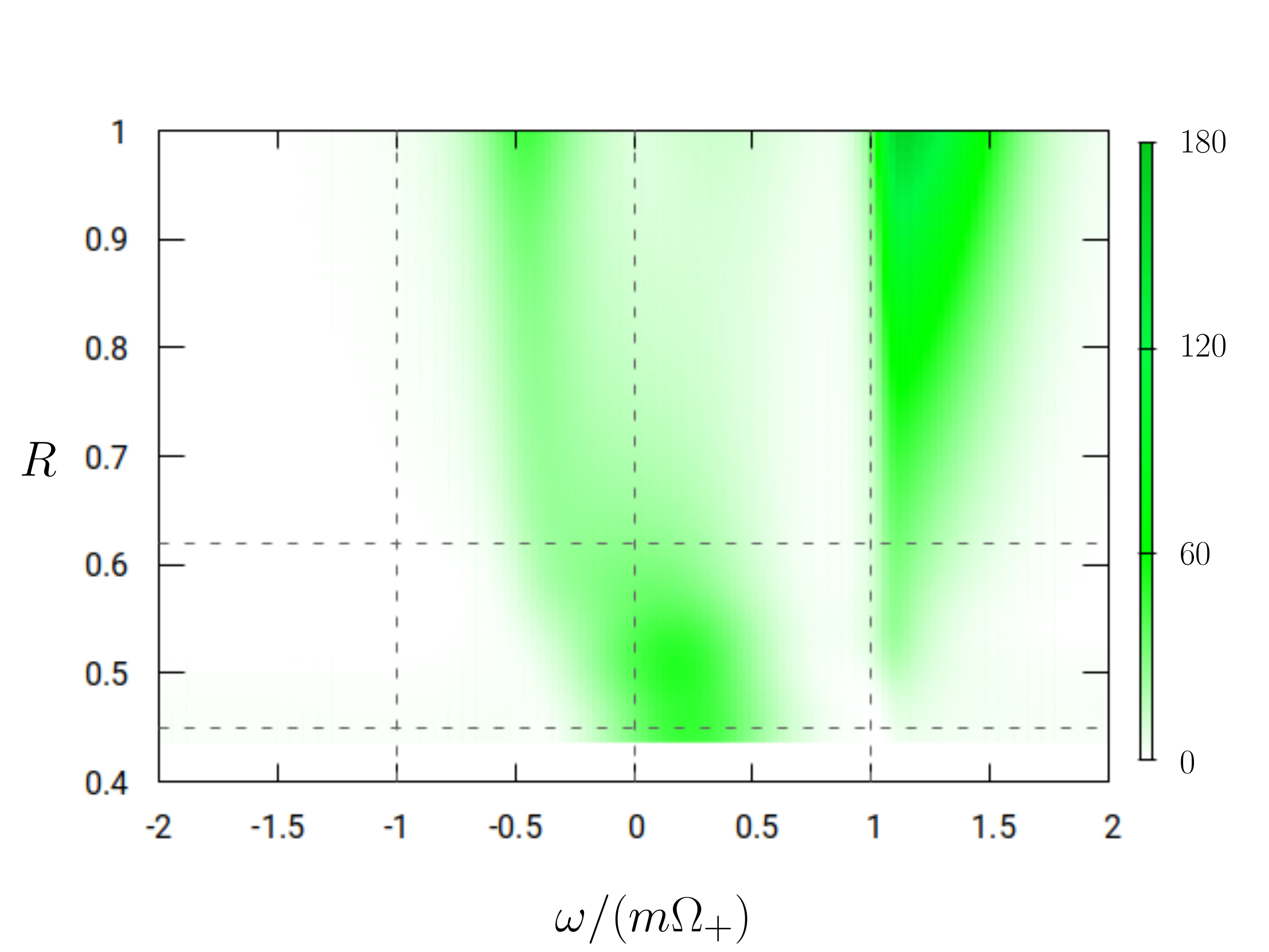}
				\vskip-0.2cm\caption{\footnotesize $\mathcal{PS}$ for solution with  $s=+1, m=-1$.}
			\end{subfigure}
			\begin{subfigure}{0.49\textwidth}
				\includegraphics[width=\textwidth]{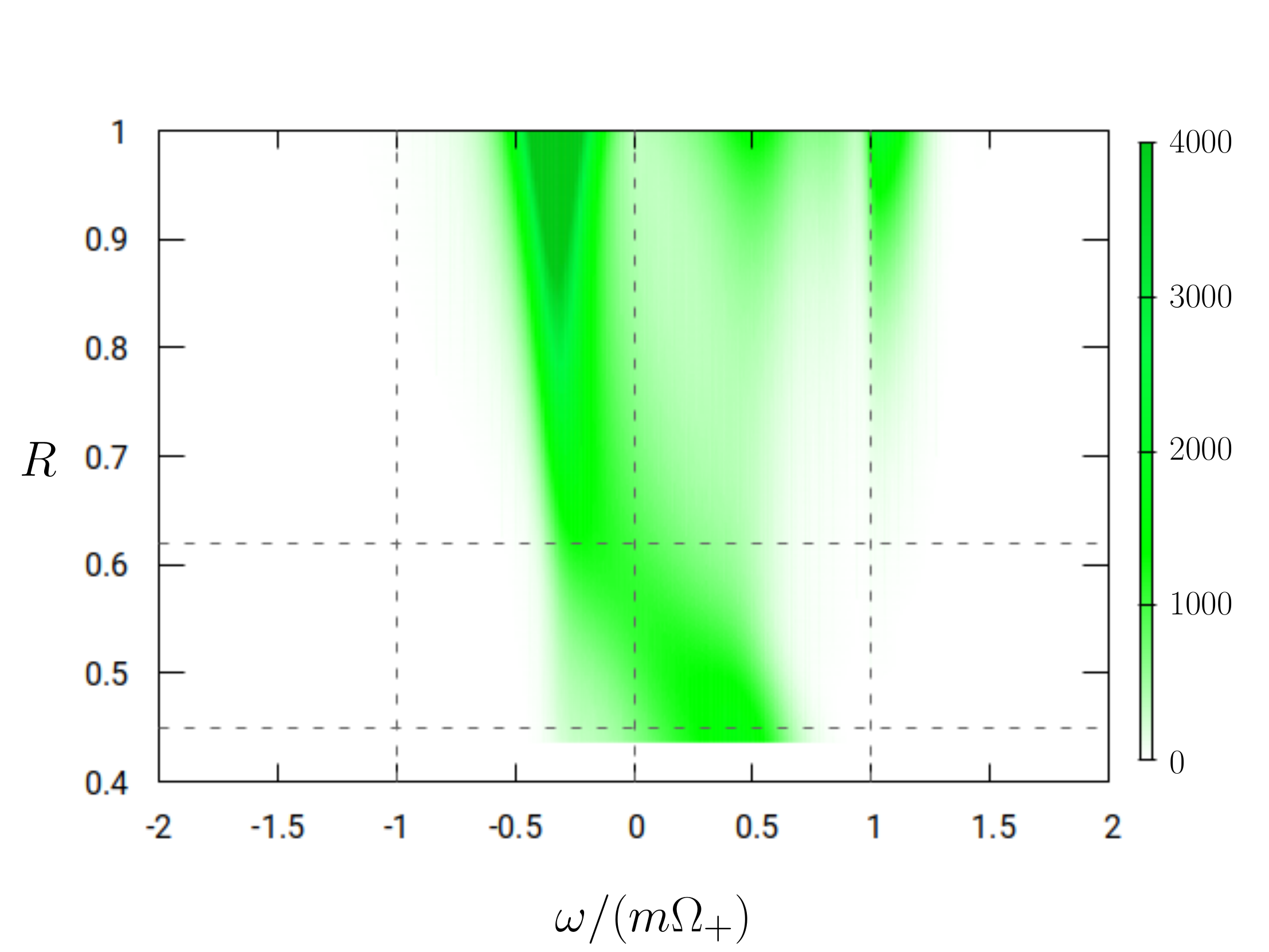}
				\vskip-0.2cm\caption{\footnotesize $\mathcal{PS}$ for solution with  $s=+2, m=+2$.}
			\end{subfigure}
			\hskip.02\textwidth
			\begin{subfigure}{0.49\textwidth}
				\includegraphics[width=\textwidth]{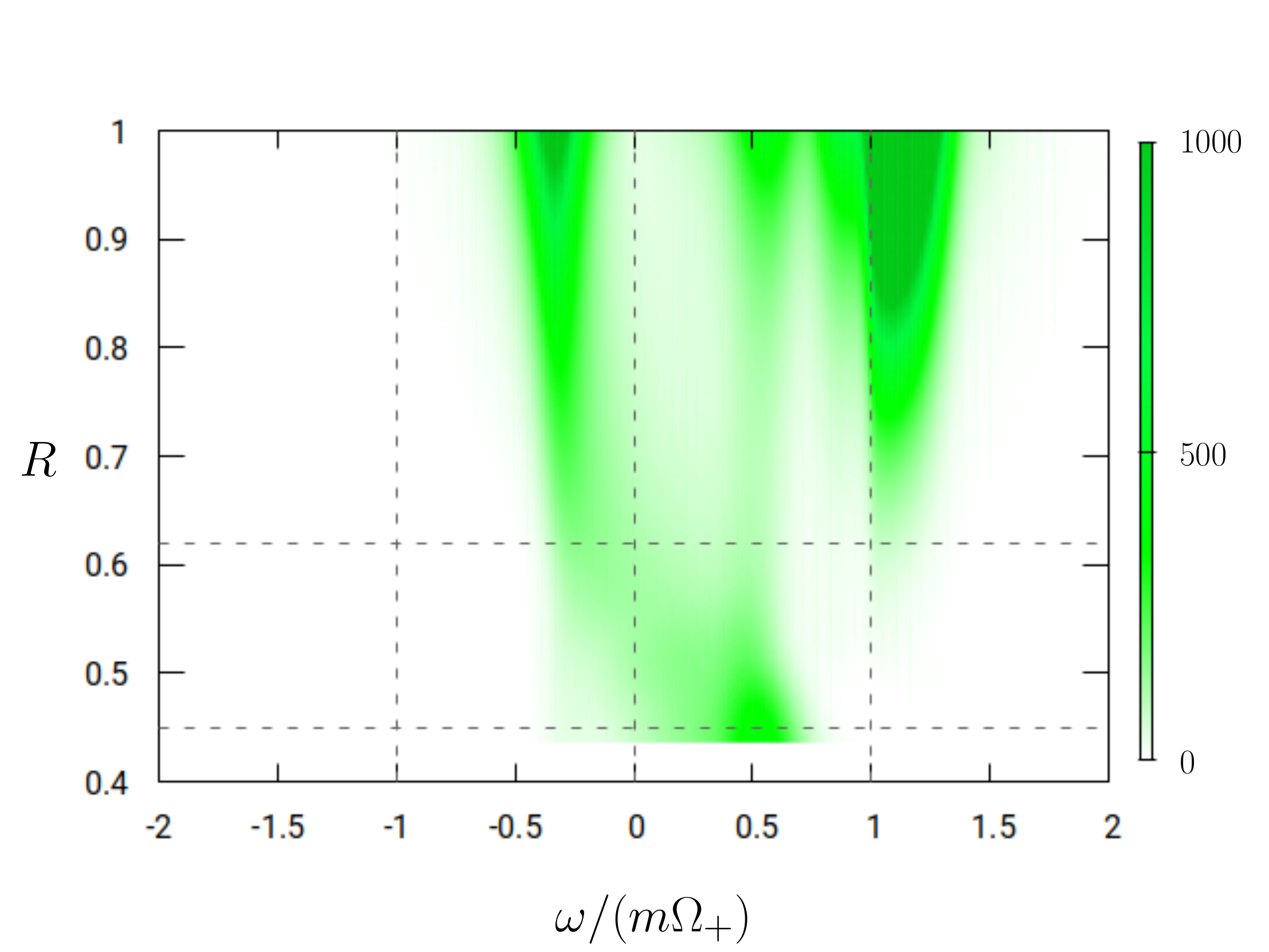}
				\vskip-0.2cm\caption{\footnotesize $\mathcal{PS}$ for solution with  $s=+2, m=-2$.}
			\end{subfigure}
		}
	\end{centering}\vskip-0.3cm
	\caption{{\footnotesize The power spectra of the outgoing electromagnetic and gravitational perturbations---with physical parameters $s=-1, \omega_0=0.25,  \ell=1, m=\pm1$ and  $s=-2, \omega_0=0.65,  \ell=2, m=\pm2$---are plotted, respectively. 
	}}\label{fig: s=+2-ft-m=pm2} 
\end{figure}
As nothing interesting happened in these ingoing type-II configurations it was tempting to ask what happens if in the initial data $(\phi^{(s)},\phi_T^{(s)})$ the time derivative is reversed, i.e.~the initial data $(\phi^{(s)},-\phi_T^{(s)})$ is used. 
One might naively expect that this also reverses the direction of energy and angular momentum transports of the pertinent configurations.
On contrary to this expectation---indicated by the panels on Fig.\,\ref{fig: s=0-energy-m=pm1}---only tiny fragments of the wave  packets start to move outwards, in spite of the fact that the power spectra shown on the panels of Fig.\,\ref{fig: s=0-ft-m=pm1} could allow more interesting dynamics. It is also somewhat counterintuitive that no energy gain occurs while the angular momentum gains are significant for both co- and counterrotating configurations.
It is also counterintuitive that, in the considered ingoing time derivative reversed cases, the angular momentum gain is the strongest for scalar configurations it is much modest for electromagnetic perturbations and even this type of gain is missing for linear $s=-2$ metric perturbations.
The angular momentum gain for scalar field---with physical parameters  $s=0, \omega_0=0.3,  \ell=1, m=\pm1$ and with the replacement $\phi_T^{(s)}\rightarrow-\phi_T^{(s)}$---is about $40\%$ for the corotating while it is about $315\%$ for counterrotating configuration. 

\subsubsection{Outgoing electromagnetic and gravitational perturbations
}
The type-II outgoing scalar configurations were already considered in the previous subsection where time evolution with reversed time derivative was discussed. Whence, here only the electromagnetic and gravitational perturbations will be considered. As noted before the main difference between the type-I and type-II configurations is that the role of the ingoing and outgoing configurations gets reversed.
\begin{figure}[ht!] 
	\vskip-0.5cm\begin{centering}
		{\tiny
			\begin{subfigure}{0.49\textwidth}
				\includegraphics[width=\textwidth]{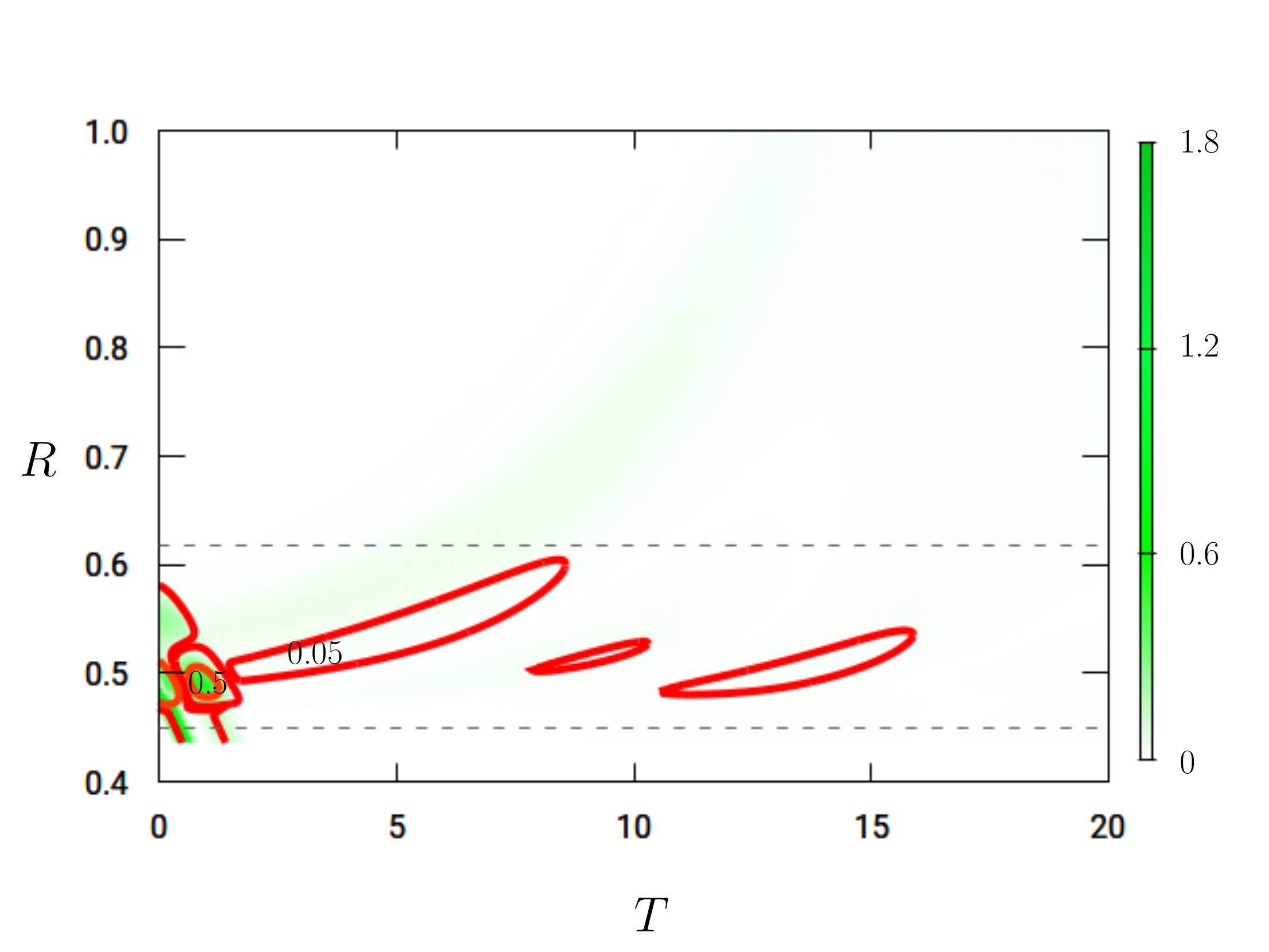}
				\vskip-0.2cm
				\caption{\footnotesize  $\mathscr{E}$ and $\mathscr{J}$ with $s=+1$ and $m=+1$.}
			\end{subfigure}
			\hskip.02\textwidth
			\begin{subfigure}{0.49\textwidth}
				\includegraphics[width=\textwidth]{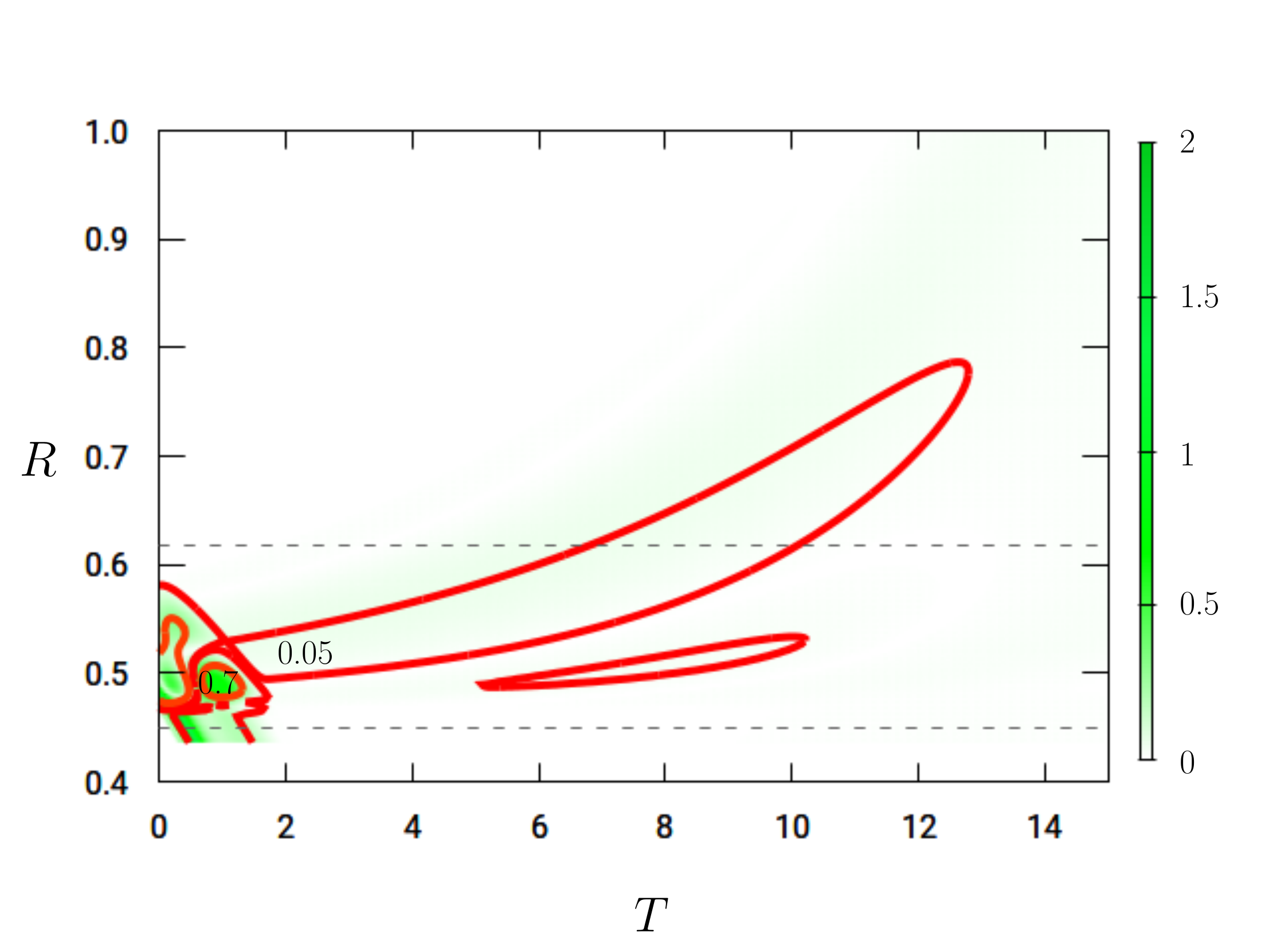}
				\vskip-0.2cm
				\caption{\footnotesize  $\mathscr{E}$ and $\mathscr{J}$ with $s=+1$ and $m=-1$.}
			\end{subfigure}
			\begin{subfigure}{0.49\textwidth}
				\includegraphics[width=\textwidth]{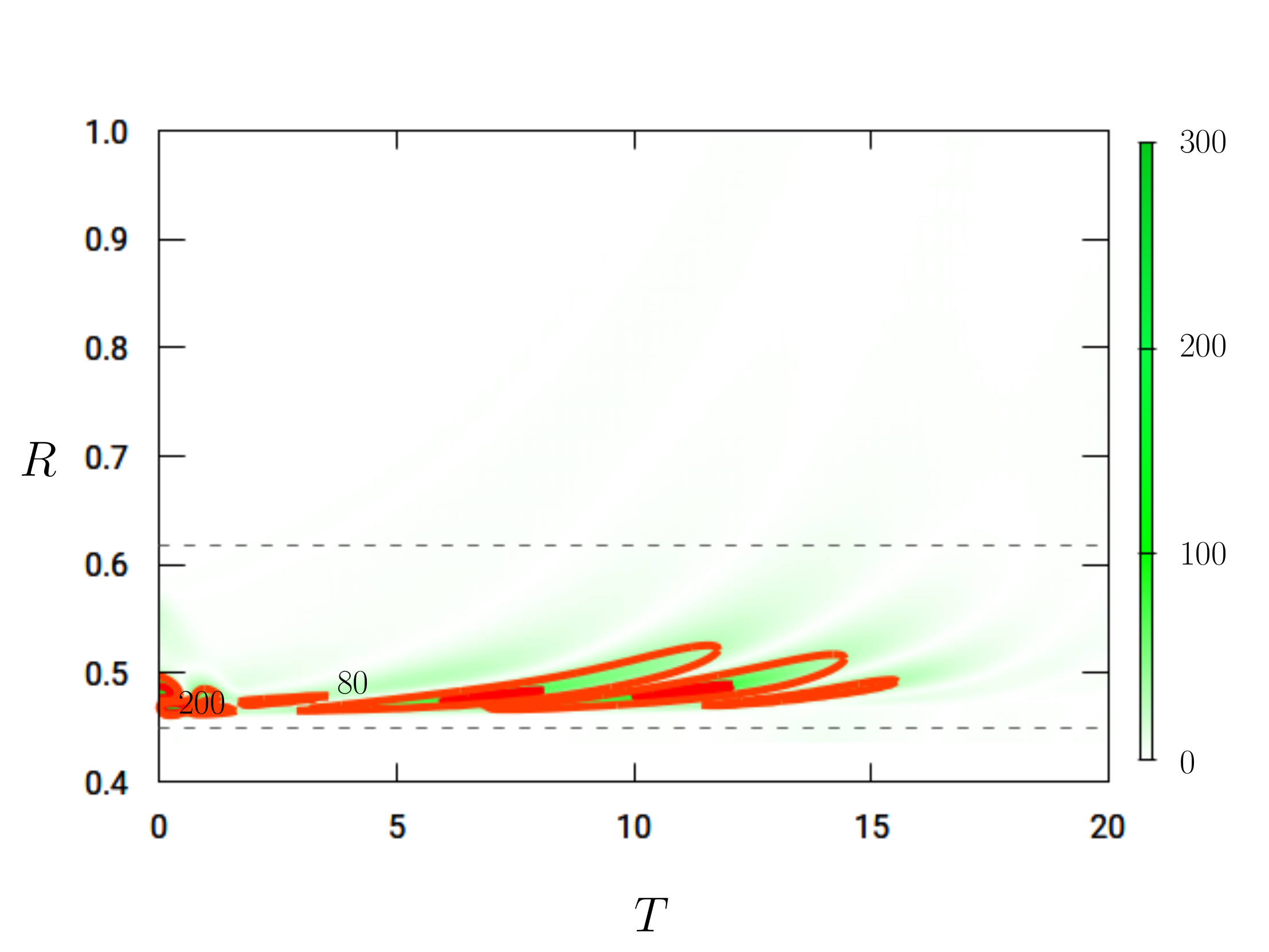}
				\vskip-0.2cm
				\caption{\footnotesize $\mathscr{E}$ and $\mathscr{J}$ with $s=+2$ and $m=+2$.}
			\end{subfigure}
			\hskip.02\textwidth
			\begin{subfigure}{0.49\textwidth}
				\includegraphics[width=\textwidth]{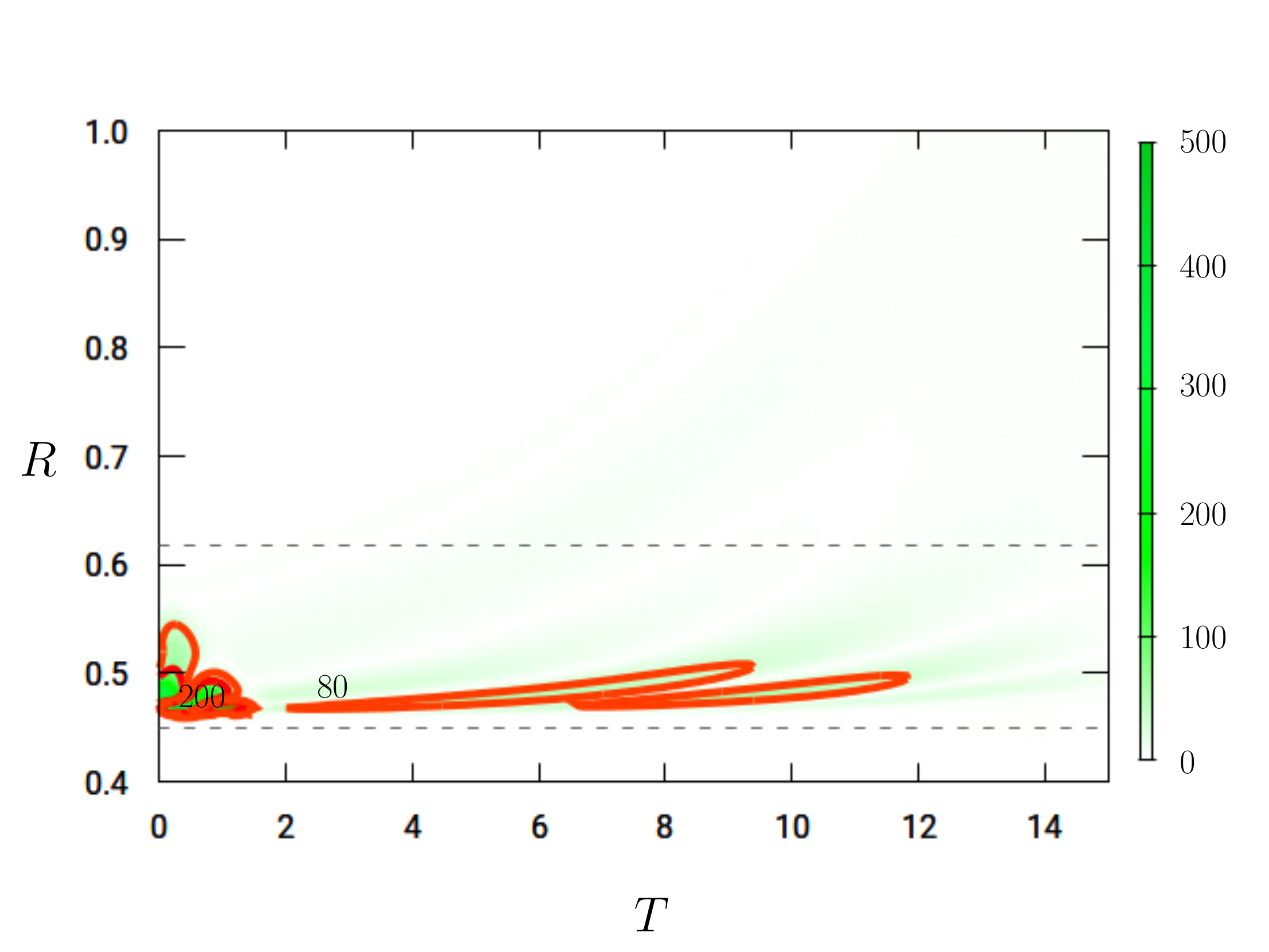}
				\vskip-0.2cm
				\caption{\footnotesize  $\mathscr{E}$ and $\mathscr{J}$ with $s=+2$ and $m=-2$.}
			\end{subfigure}
		}
	\end{centering}\vskip-0.2cm
	\caption{{\footnotesize The volume normalized energy and angular momentum current densities, $\mathscr{E}(T,R)$ and  $\mathscr{J}(T,R)$ are plotted for type-II outgoing  electromagnetic and gravitational perturbations, respectively. The physical parameters are exactly the same as those applied in producing Fig.\,\ref{fig: s=+2-ft-m=pm2} .
	}}\label{fig: s=+2-ebr-m=pm2} 
\end{figure}
More specifically, while the outgoing configurations did not show any sort of superradiance for type-I electromagnetic or gravitational perturbations they step forward to be the main actors in superradiant processes of type-II configurations.  It is not surprising that the strength of superradiance of outgoing type-II configurations does also depend on spin. It is rather modest for $s=1$ electromagnetic fields, whereas it is getting considerable for $s=2$ gravitational perturbations.  


The power spectra of the outgoing electromagnetic and gravitational perturbations---with physical parameters $s=1, \omega_0=0.28,  \ell=1, m=\pm1$ and  $s=2, \omega_0=0.725,  \ell=2, m=\pm2$---, depicted on Fig.\,\ref{fig: s=+2-ft-m=pm2}, are not extremely informative. There are some not too intensive spots in the critical $0< \omega< m\,\Omega_{+}$ interval, while the intense ones outside of this interval are associated with lasting quasinormal ringings.   

\medskip

The energy and angular momentum transfers, depicted on Fig. \ref{fig: s=+2-ebr-m=pm2}, do indicate a long lasting breathing structure that takes place mainly in the ergoregion and that is supposedly a manifestation of superradiant processes. 

\begin{figure}[ht!] 
	\begin{centering}
		{\tiny
			\begin{subfigure}{0.49\textwidth}
				\includegraphics[width=\textwidth]{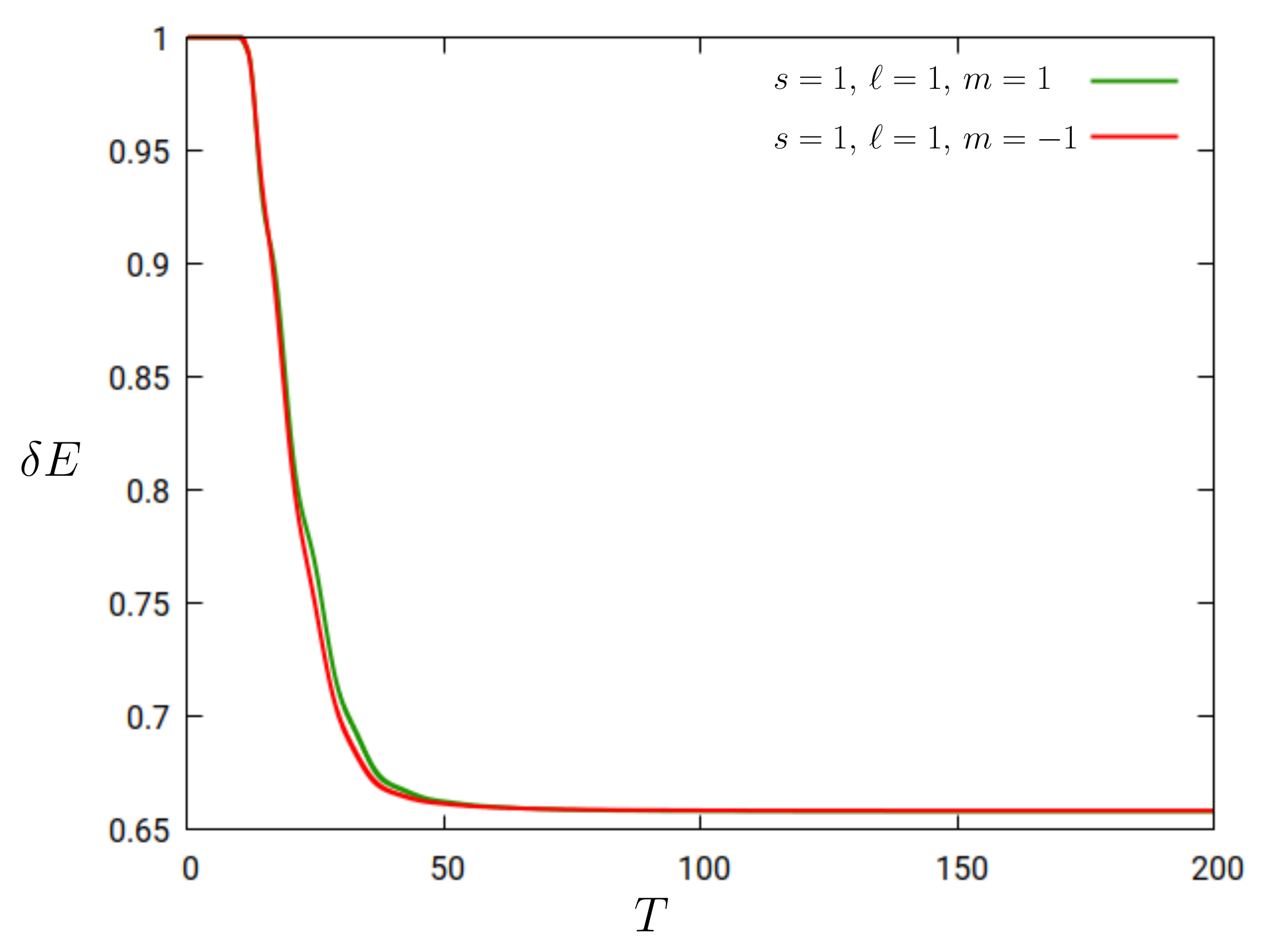}
				\caption{\footnotesize Energy gains: $s=+1, m=\pm 1$. }
			\end{subfigure}
			\hskip.02\textwidth
			\begin{subfigure}{0.49\textwidth}
				\includegraphics[width=\textwidth]{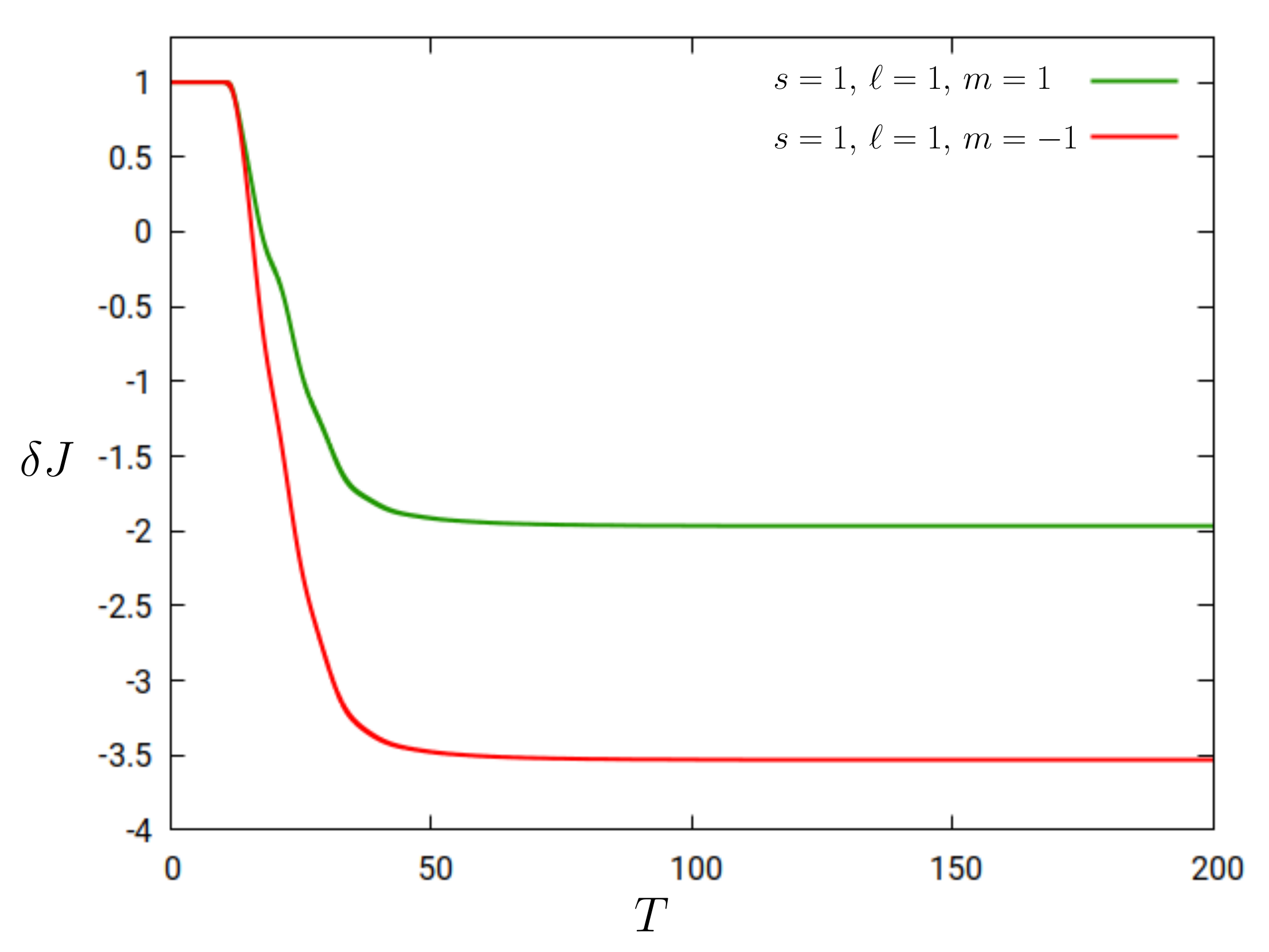}
				\caption{\footnotesize Angular momentum gains: $s=+1, m=\pm 1$. }
			\end{subfigure}
			\begin{subfigure}{0.49\textwidth}
				\includegraphics[width=\textwidth]{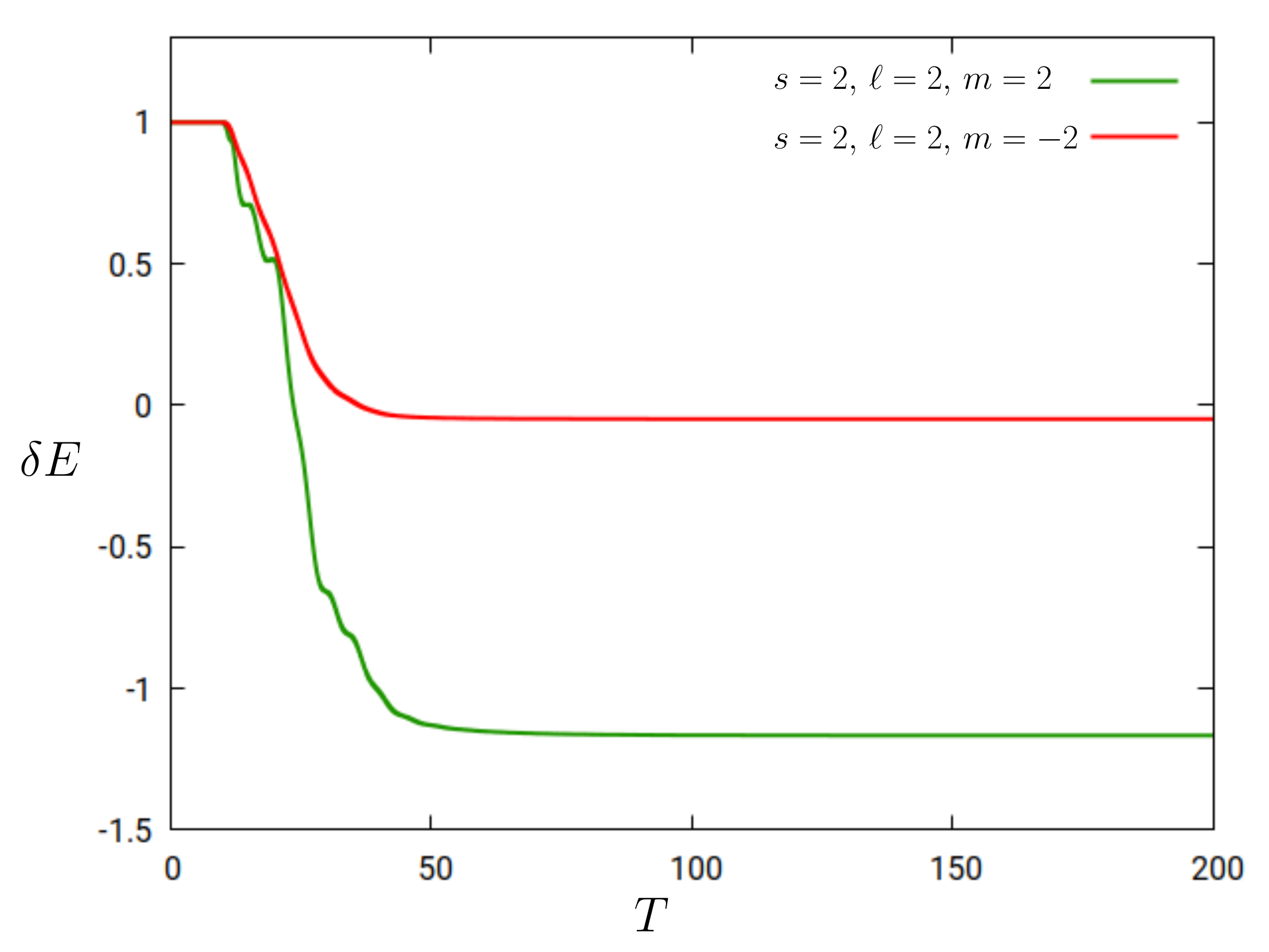}
				\caption{\footnotesize Energy gains: $s=+2, m=\pm 2$.}
			\end{subfigure}
			\hskip.02\textwidth
			\begin{subfigure}{0.49\textwidth}
				\includegraphics[width=\textwidth]{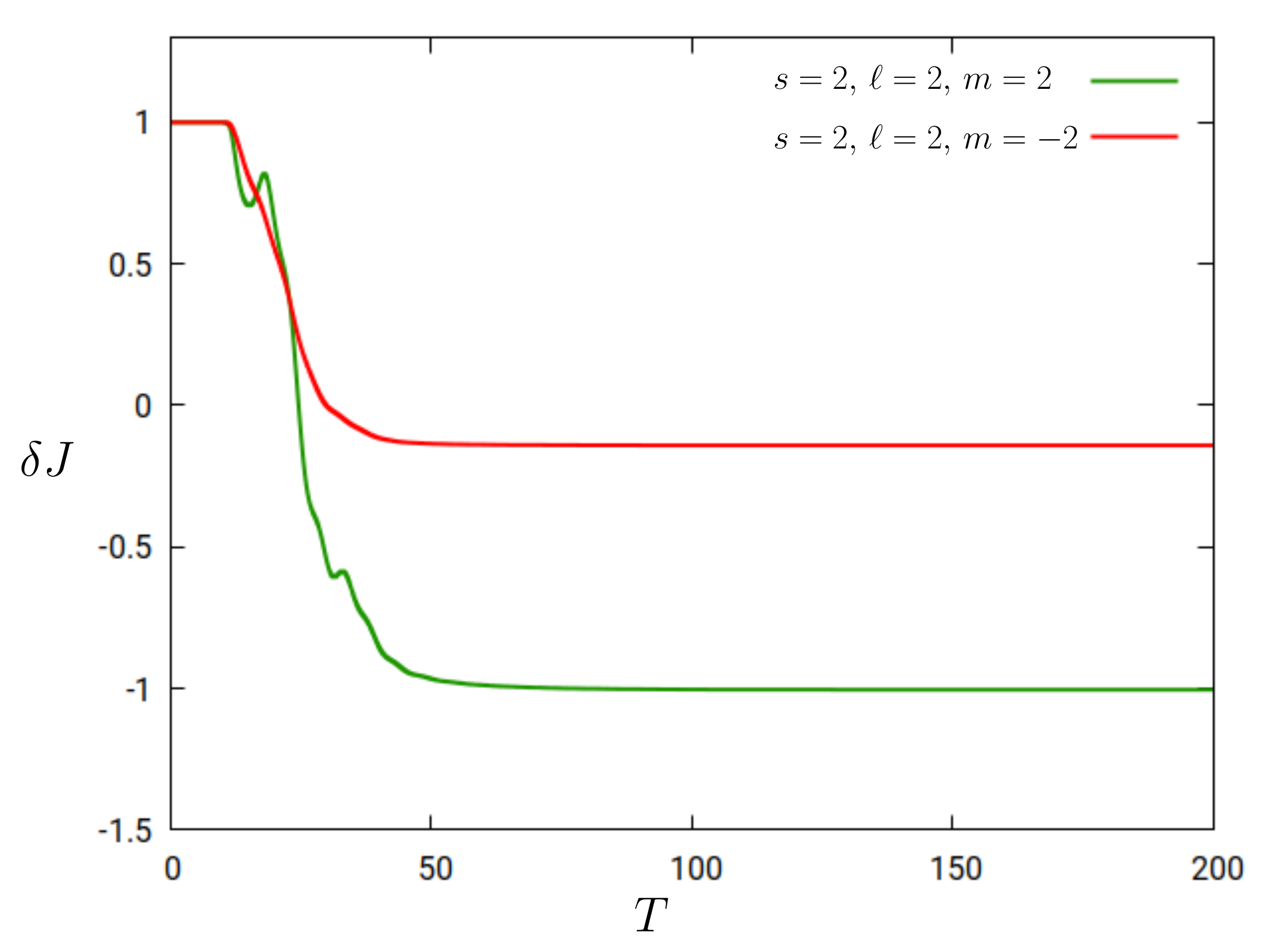}
				\caption{\footnotesize Angular momentum gains: $s=+2, m=\pm 2$. }
			\end{subfigure}
		}
	\end{centering}\vskip-0.2cm
	\caption{{\footnotesize 	The time dependence of the energy and angular momentum gains, $\delta E=(E_0-E_{out})/E_0$ and $\delta J=(J_0-J_{out})/J_0$, are plotted for the investigated outgoing and corotating  electromagnetic and gravitational perturbations, respectively. As it is indicated by these panels there is only angular momentum gain for  electromagnetic  perturbations whereas there are both energy and angular momentum gains   for gravitational perturbations.
	}}\label{fig: s=+2-super-m=pm2} 
\end{figure}
\begin{figure}[ht!] 
	\vskip-0.4cm
	\begin{centering}
		{\tiny
			\begin{subfigure}{0.49\textwidth}
				\includegraphics[width=\textwidth]{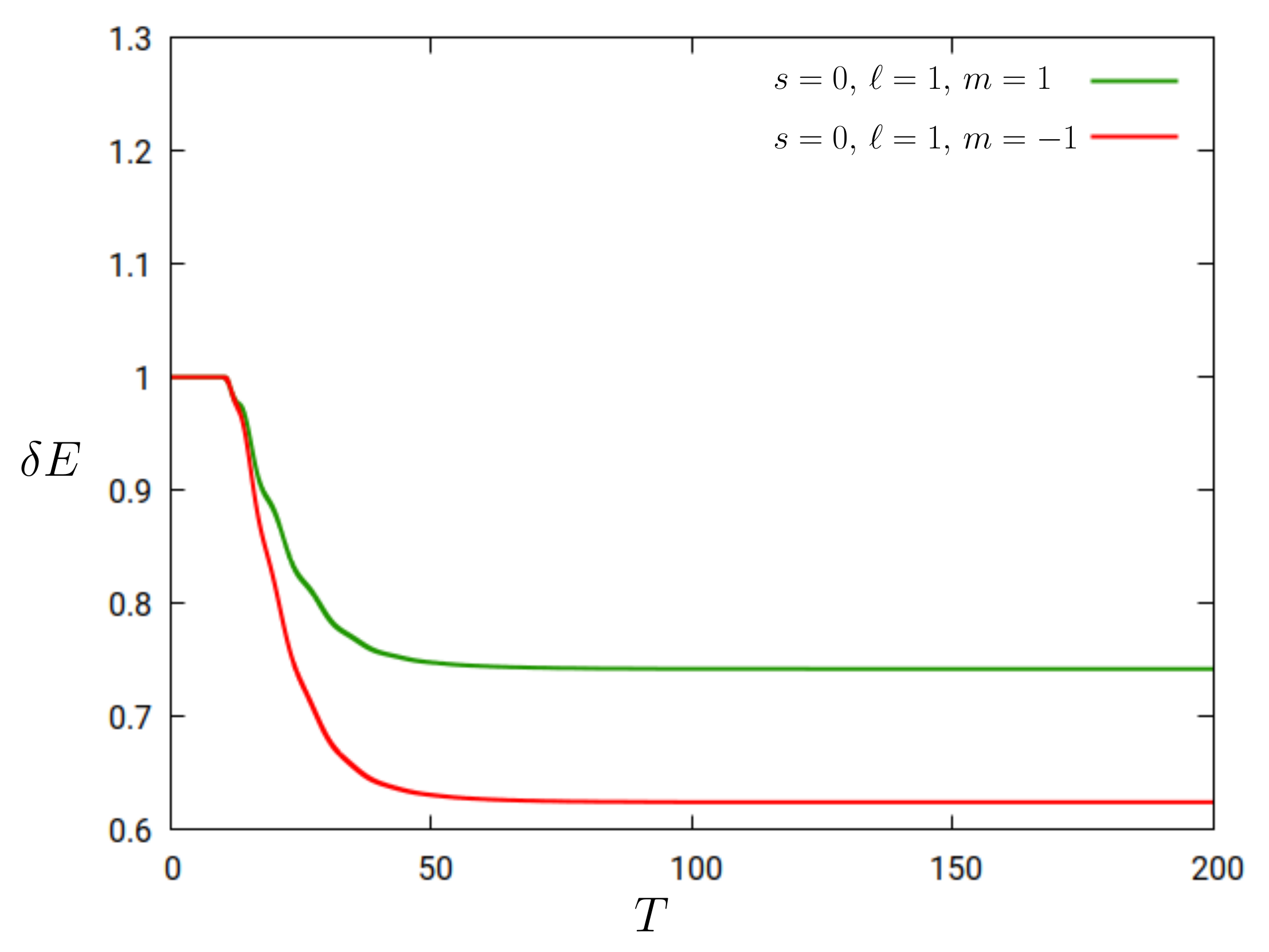}
				\caption{\footnotesize Energy gains: $s=0, m=\pm 1$. }
			\end{subfigure}
			\hskip.02\textwidth
			\begin{subfigure}{0.49\textwidth}
				\includegraphics[width=\textwidth]{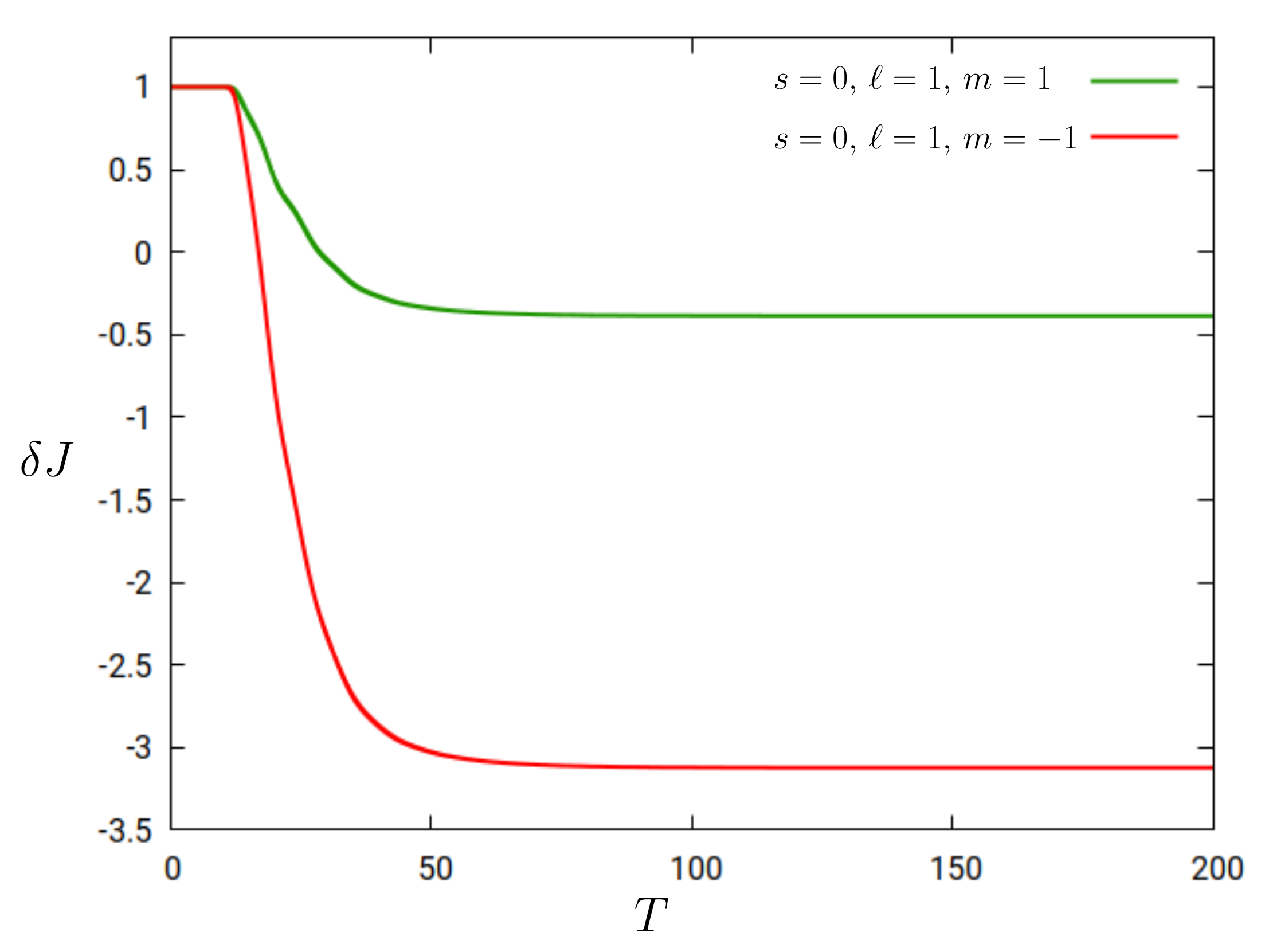}
				\caption{\footnotesize Angular momentum gains: $s=0, m=\pm 1$. }
			\end{subfigure}
		}
	\end{centering}\vskip-0.2cm
	\caption{{\footnotesize The time dependence of the energy and angular momentum gains, $\delta E=(E_0-E_{out})/E_0$ and $\delta J=(J_0-J_{out})/J_0$, are plotted for ingoing scalar perturbations (with reversed time derivative), respectively. The physical parameters are as those are on Figs.\,\ref{fig: s=0-ft-m=pm1} and \ref{fig: s=0-energy-m=pm1}. 
	}}\label{fig: s=0o-super-m=pm1} 
\end{figure}
Recall that the ratios $\delta E=(E_0-E_{out})/E_0$ and $\delta J=(J_0-J_{out})/J_0$ were designed to scale superradiance. If energy superradiance and angular momentum gain occurs in scattering processes then the energy and angular momentum, $E_{out}$ and  $J_{out}$, received at null infinity, $\scrip$, have to be larger than the energy and angular momentum, $E_0$ and $J_0$, of the initial wave packet, which means that $\delta E$ and $\delta J$ get to be negative. Notably, on the top panels of Fig.\,\ref{fig: s=+2-super-m=pm2} depicting the electromagnetic case,
(likewise on the analogous panels of Fig.\,\ref{fig: s=0o-super-m=pm1}) only $\delta J$  gets negative. The angular momentum gain is $\sim 200\%$ for corotating, while it is  $\sim 350\%$ for counterrotating electromagnetic perturbations. Both energy superradiance and angular momentum gain occur in case of gravitational perturbations. The energy gain is $\sim 116.7\%$ for corotating and $\sim6\%$ for counterrotating configurations, whereas the corresponding angular momentum gains are $\sim 100\%$ and $\sim 5\%$, respectively. Remarkably, the angular momentum gain (again likewise in case of type-II outgoing scalar configurations with reversed time derivative) is larger for counterrotating configurations while in case of gravitational perturbations, the situation is quite the opposite, the corotating configurations are found to be more superradiant. 
	
\subsection{Results relevant for type-III initial data}\label{subsec: ID-III}

So far in using type-I and type-II initial data only the $\omega_{0}$ characteristic frequency was fine tuned to amplify the effect of superradiance. The type-III initial data, considered in this section, is to probe the strength of superradiance with respect to the variation of the width of the radial profile. In doing so we shall modify the type-II radial profile by keeping the left boundary at $R_1=0.4536$ fixed but moving the right edge outwards such that $R_2$ takes the values  $R_2=0.7,0.76,0.84,0.95$. The maximum of the radial profile is located approximately at the mean value of $R_1$ and  $R_2$, i.e.\,$R_0=(R_1+R_2)/2$ for each of the considered configurations as shown on the right panel of Fig.\,\ref{fig: supports}. 	
	
\subsubsection{Ingoing and outgoing scalar wave packets
}	

It is somewhat unexpected but very little change occurs when type-III initial data is replacing the type-II ones in the scalar field case. More specifically, no superradiance occurred when either of the four type-III co- or counterrotating incident wave  packets were applied in the ``ingoing'' case with initial data as specified in \eqref{eq: Initial-data-TR} and \eqref{eq: Initial-data-TR-T-der}. 

\medskip

As seen earlier somewhat more interesting dynamics may occur when time derivative reversed type-II scalar wave  packets were investigated.  
\begin{figure}[ht!] 
	\vskip-0.4cm
	\begin{centering}
		{\tiny
			\begin{subfigure}{0.49\textwidth}
				\includegraphics[width=\textwidth]{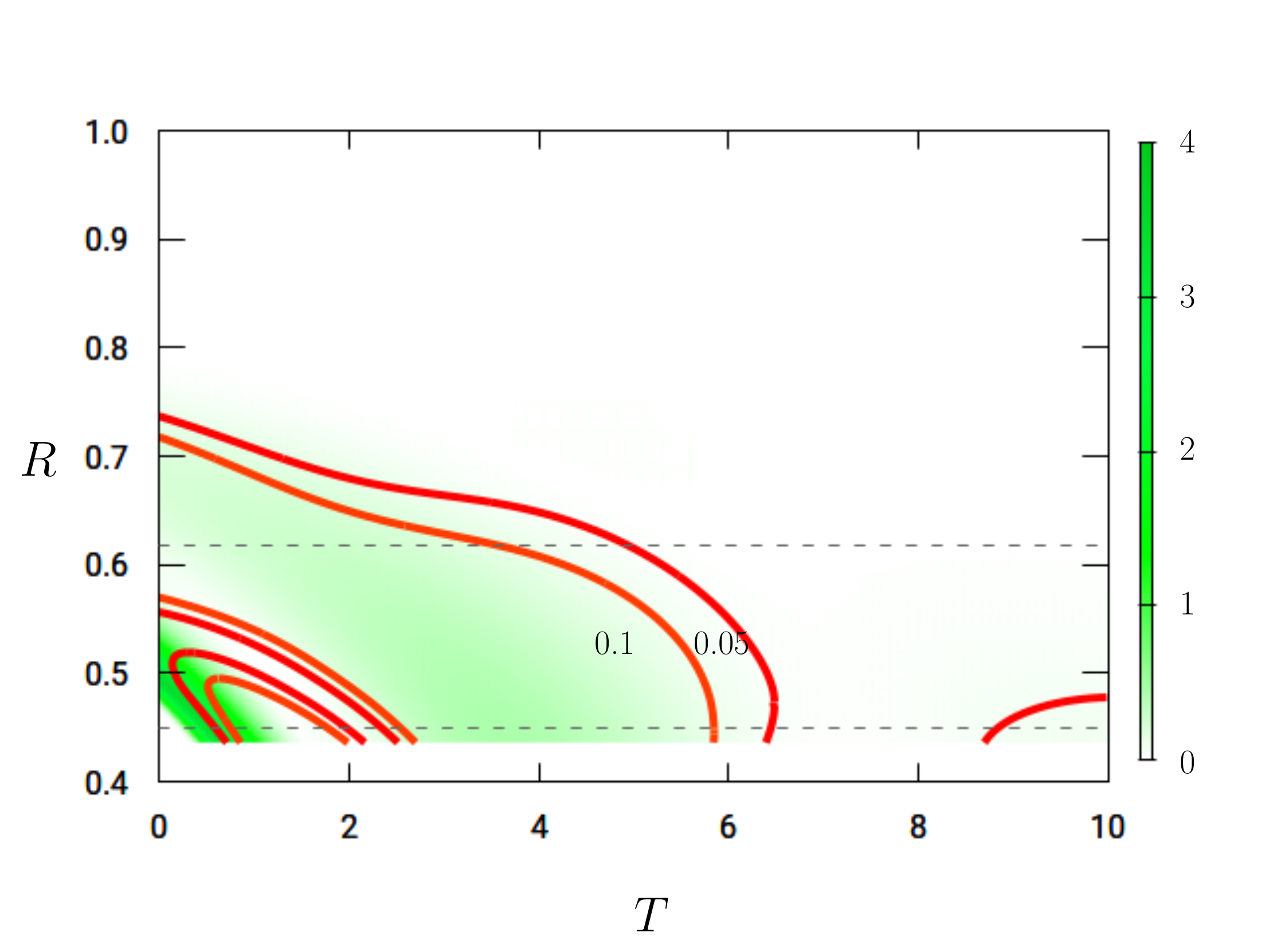}
				\vskip-0.2cm
			\end{subfigure}
			\hskip.02\textwidth
			\begin{subfigure}{0.49\textwidth}
				\includegraphics[width=\textwidth]{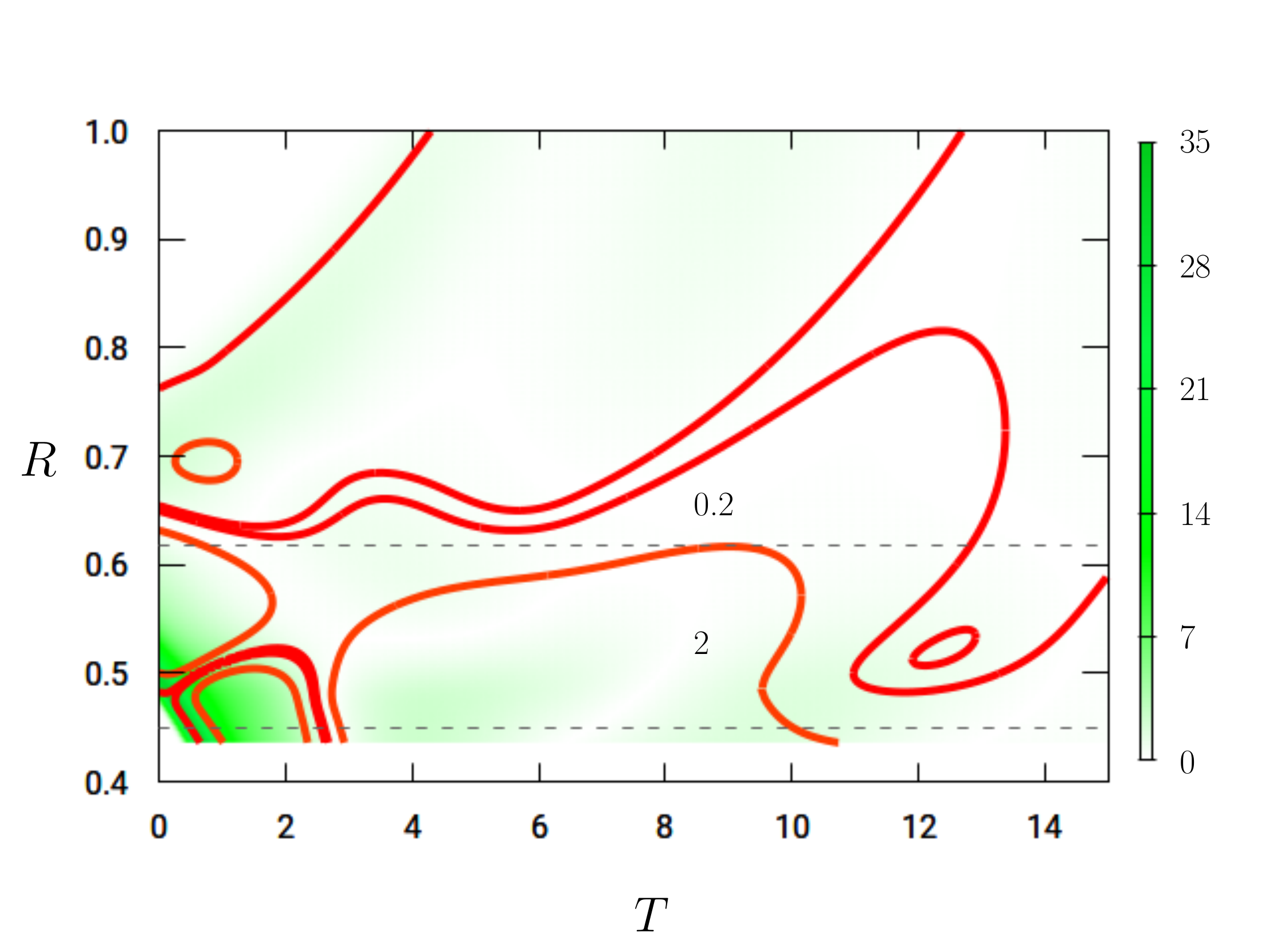}
				\vskip-0.2cm
			\end{subfigure}
		}
	\end{centering}\vskip-0.4cm
	\caption{{\footnotesize The volume normalized energy and angular momentum current densities, $\mathscr{E}(T,R)$ and  $\mathscr{J}(T,R)$ are plotted for in- and outgoing type-III scalar wave packets, respectively.
	}}\label{fig: III-s=0-ebr-m=pm1} 
\end{figure}
These time derivative reversed scalar field configurations may be considered as being ``outgoing'' ones.     
\begin{figure}[ht!] 
	\begin{centering}
		{\tiny
			\begin{subfigure}{0.49\textwidth}
				\includegraphics[width=\textwidth]{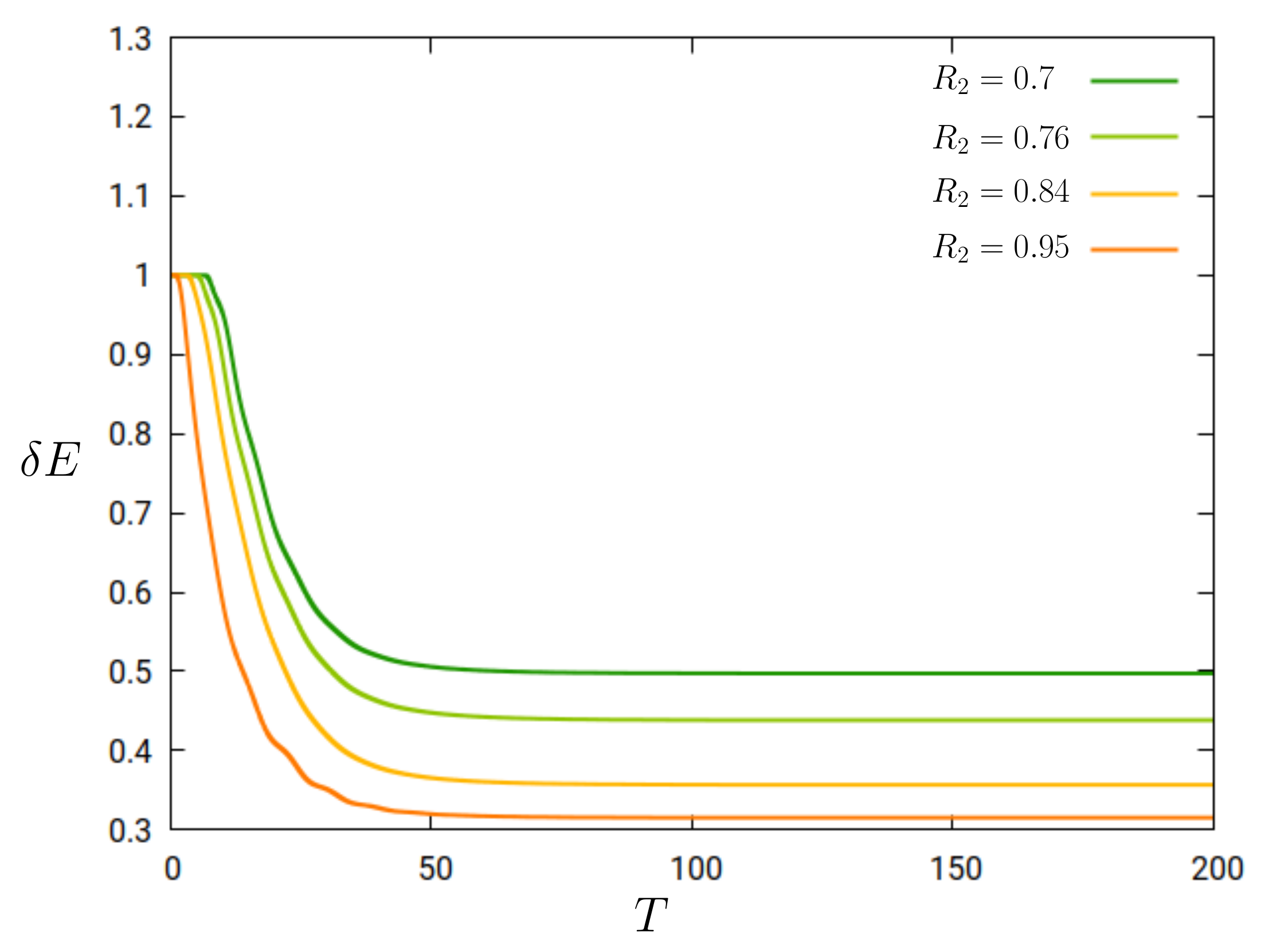}
				\vskip-0.2cm\caption{\footnotesize Energy gains: $s=0, m=-1$. }
			\end{subfigure}
			\hskip.02\textwidth
			\begin{subfigure}{0.49\textwidth}
				\includegraphics[width=\textwidth]{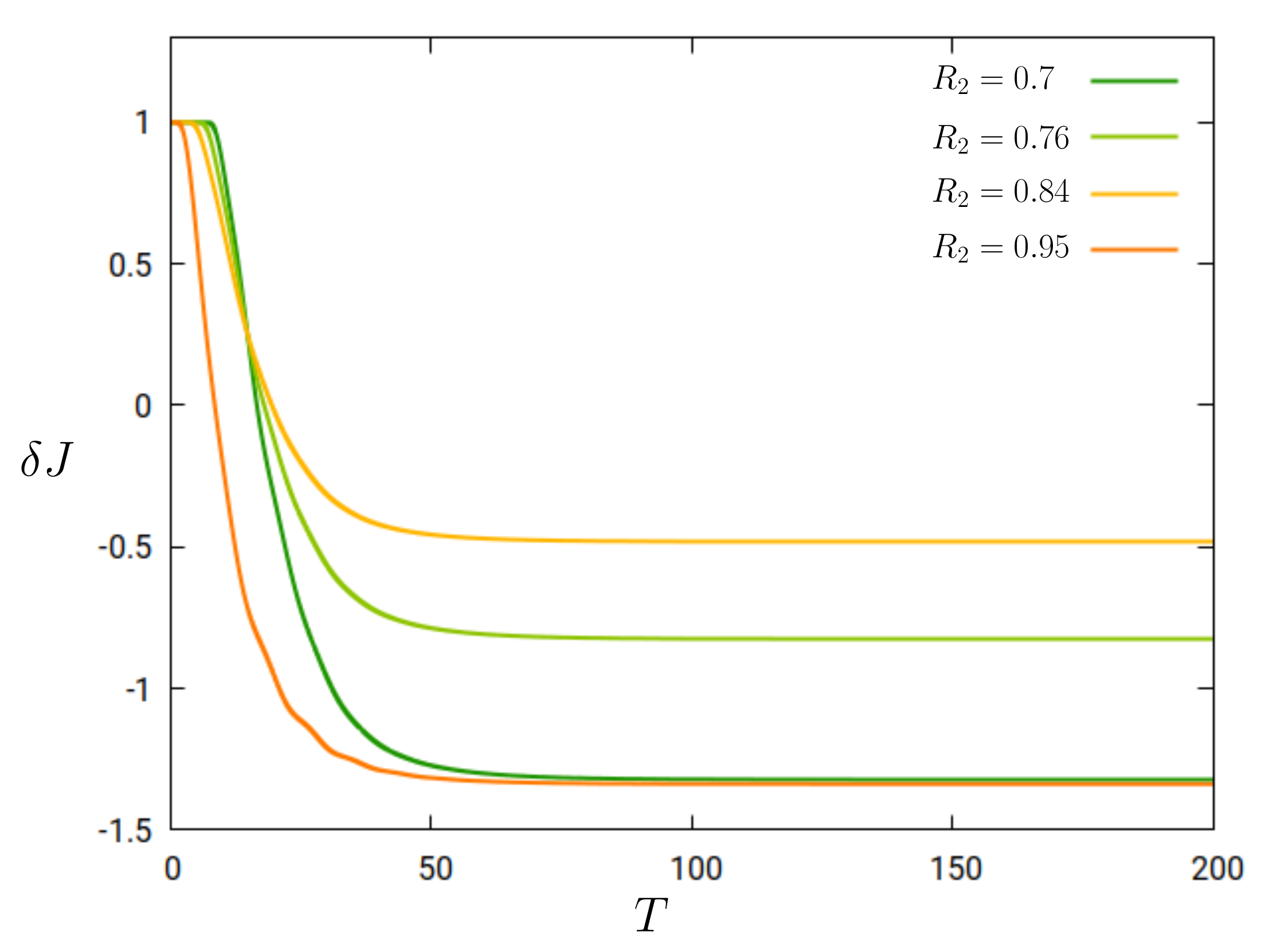}
				\vskip-0.2cm\caption{\footnotesize Angular momentum gains: $s=0, m=-1$. }
			\end{subfigure}
		}
	\end{centering}\vskip-0.2cm
	\caption{{\footnotesize No energy but notable angular momentum gains occur for counterrotating type-III outgoing scalar perturbations. 
	}}\label{fig: III-s=0-super-m=-1} 
\end{figure}
To verify the rightfulness of the use of the terminology ingoing and outgoing one it is rewarding to inspect the energy transports of scalar wave  packets shearing the compact support and the value of $\omega_{0}$, with $R_2=0.84$ and $\omega_{0}=0.3$, shown by the two panels of Fig.\,\ref{fig: III-s=0-ebr-m=pm1}. While on the left panel the energy and angular momentum transfer, with initial data given as in \eqref{eq: Initial-data-TR} and \eqref{eq: Initial-data-TR-T-der}, is visibly ingoing for the time derivative reversed case, shown on the right panel, the energy and angular momentum move to the outer direction. 	
  
Notably, what we have experienced in case of counterrotating outgoing type-II scalar field configurations manifest itself for the type-III wave  packets too. More specifically, for time derivative reversed scalar fields---with physical parameters $s=0, \omega_0=0.3, \ell=1, m=-1$ ---while no energy superradiance occurs the angular momentum gain ranges between $49\%$  and $135\%$ as shown on the right panel of Fig.\,\ref{fig: III-s=0-super-m=-1}.
	
\subsubsection{Ingoing electromagnetic and gravitational perturbations
}	
It was a generic feature of type-II ingoing, $s=-1$ and $s=-2$, electromagnetic and gravitational perturbations that neither of them shows superradiance. Instead considerable large part of the initial pulse was leaving the domain of outer communication in a short period via the event horizon. Exactly the same phenomenon shows up in case of type-III ingoing  configurations. The only notable difference is that the evolution gets somewhat longer. Neither energy nor angular momentum gains show up regardless whether co- or counterrotating ingoing $s=-1$ electromagnetic or $s=-2$ gravitational perturbations were investigated. 

\subsubsection{Outgoing electromagnetic and gravitational perturbations
}
The corotating outgoing electromagnetic and gravitational perturbations are not too much exciting either. There are no energy or angular momentum gains for $s=+1, \ell=1, m=+1$ electromagnetic  perturbations. The evolution relevant for the four radial profiles differ only by the ratio of the energy or angular momentum that leave the domain of outer communication either through the event horizon or via null infinity.  

\medskip 

As opposed to this the evolution of counterrotating type-III outgoing $s=+1$  electromagnetic perturbations show somewhat more exciting behavior. 
\begin{figure}[ht!] 
	\vskip-0.4cm
	\begin{centering}
		{\tiny
			\begin{subfigure}{0.49\textwidth}
				\includegraphics[width=\textwidth]{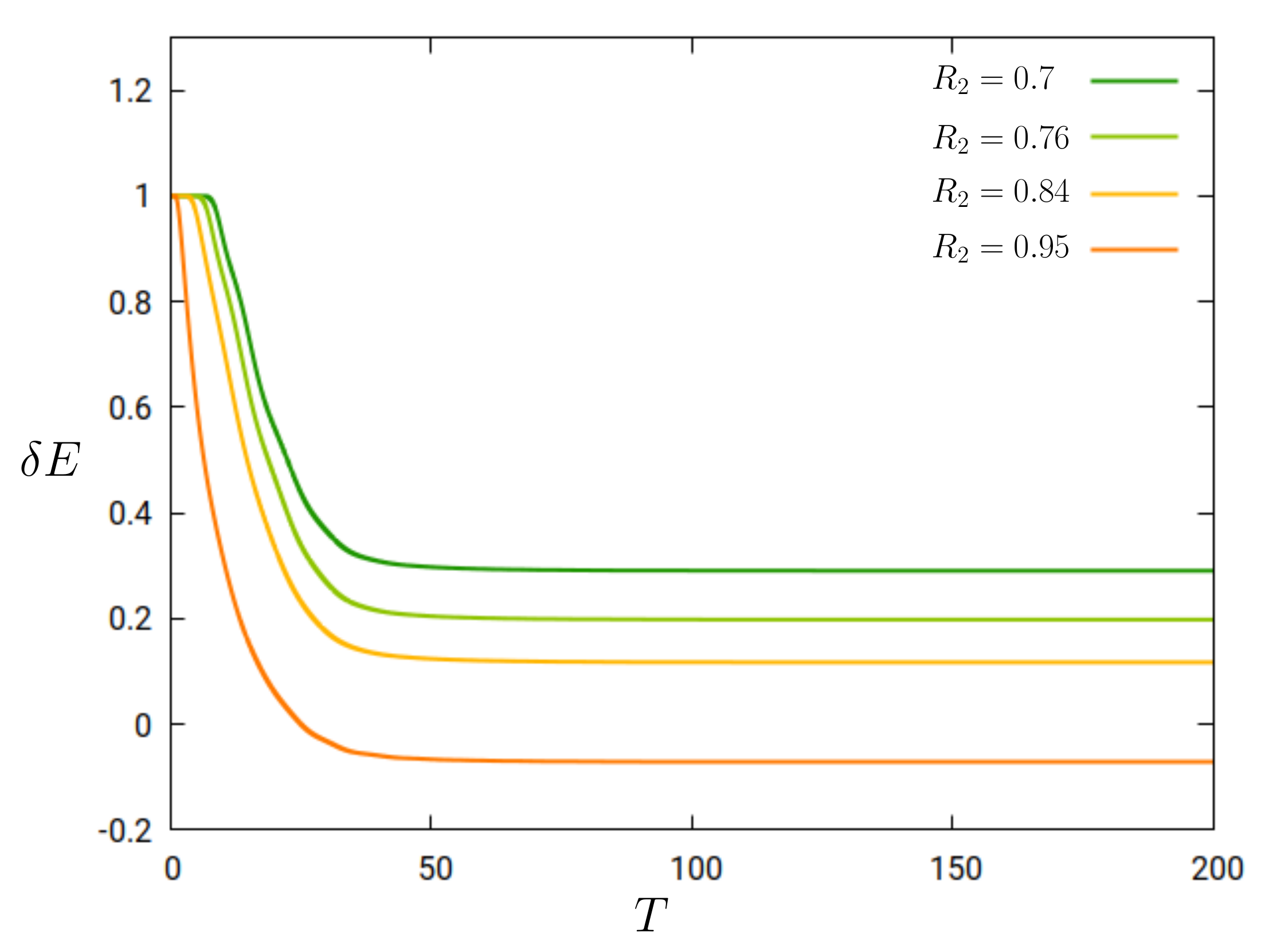}
				\vskip-0.2cm\caption{\footnotesize Energy gains: $s=+1, m=-1$.}
			\end{subfigure}
			\hskip.02\textwidth
			\begin{subfigure}{0.49\textwidth}
				\includegraphics[width=\textwidth]{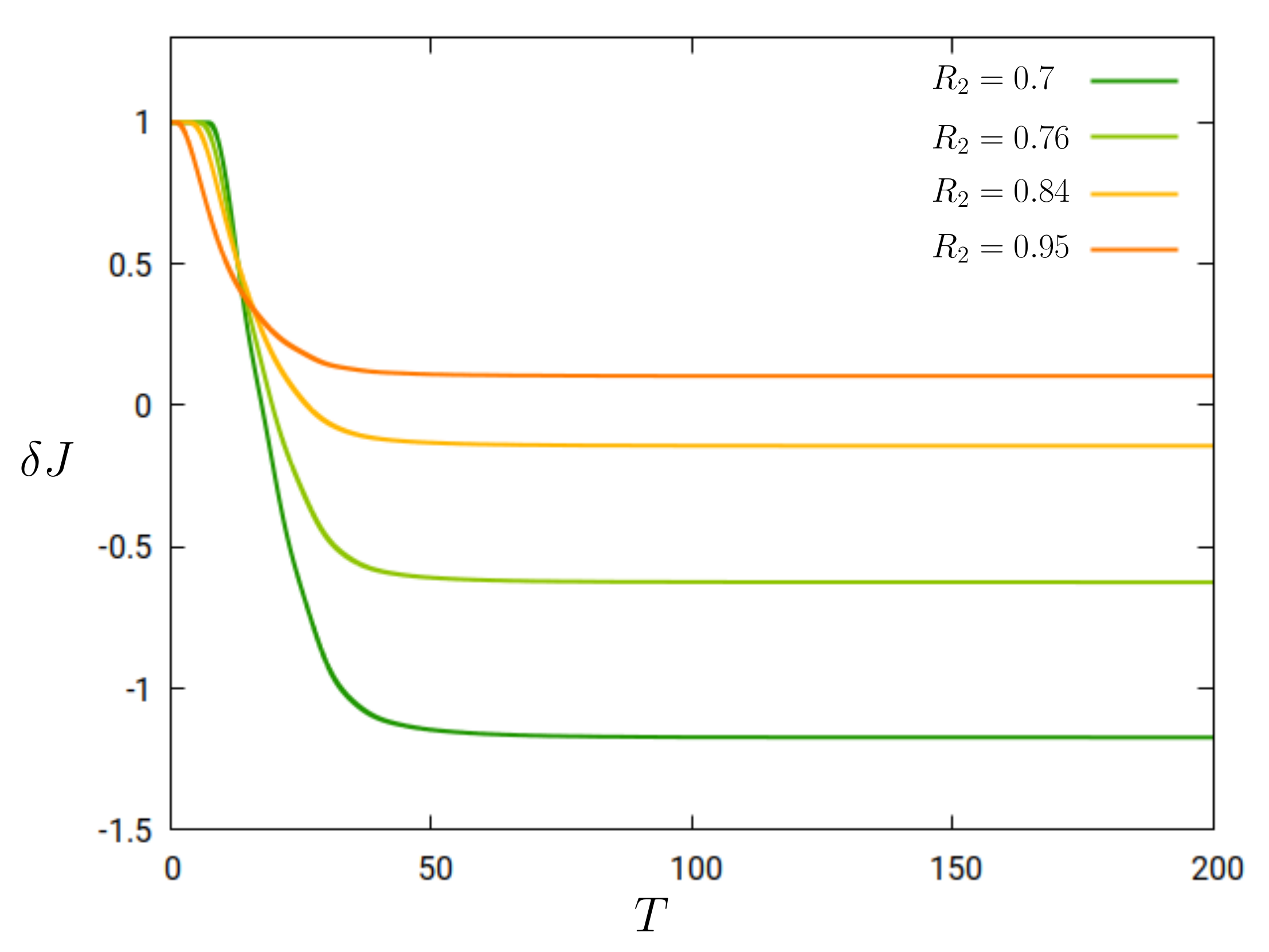}
				\vskip-0.2cm\caption{\footnotesize Angular momentum gains: $s=+1, m=-1$. }
			\end{subfigure}
		}
	\end{centering}\vskip-0.2cm
	\caption{{\footnotesize 	The time dependence of the energy and angular momentum gains, $\delta E$ and $\delta J$, are plotted for counterrotating type-III outgoing  electromagnetic perturbations. 
	}}\label{fig: III-s=+1-super-m=-1} 
\end{figure}
The energy and angular momentum gains of counterrotating type-III outgoing electromagnetic perturbations---with physical parameters $s=+1, \omega_0=0.28, \ell=1, m=-1$ ---are shown on the panels of Fig.\,\ref{fig: III-s=+1-super-m=-1}. There is a slight $\sim 8\%$ energy superradiance for the widest radial profile, with $R_2=0.95$. As opposed to this, no angular momentum gains happens for this profile but stronger and stronger angular momentum gains, $\sim 10\%, \sim 60\%,  \sim 120\%$, can be seen to happen for the other three radial profiles while the domain of the radial profiles is getting narrower and narrower.   

\medskip

\begin{figure}[ht!] 
	\vskip-0.5cm
	\begin{centering}
		{\tiny
			\begin{subfigure}{0.49\textwidth}
				\includegraphics[width=\textwidth]{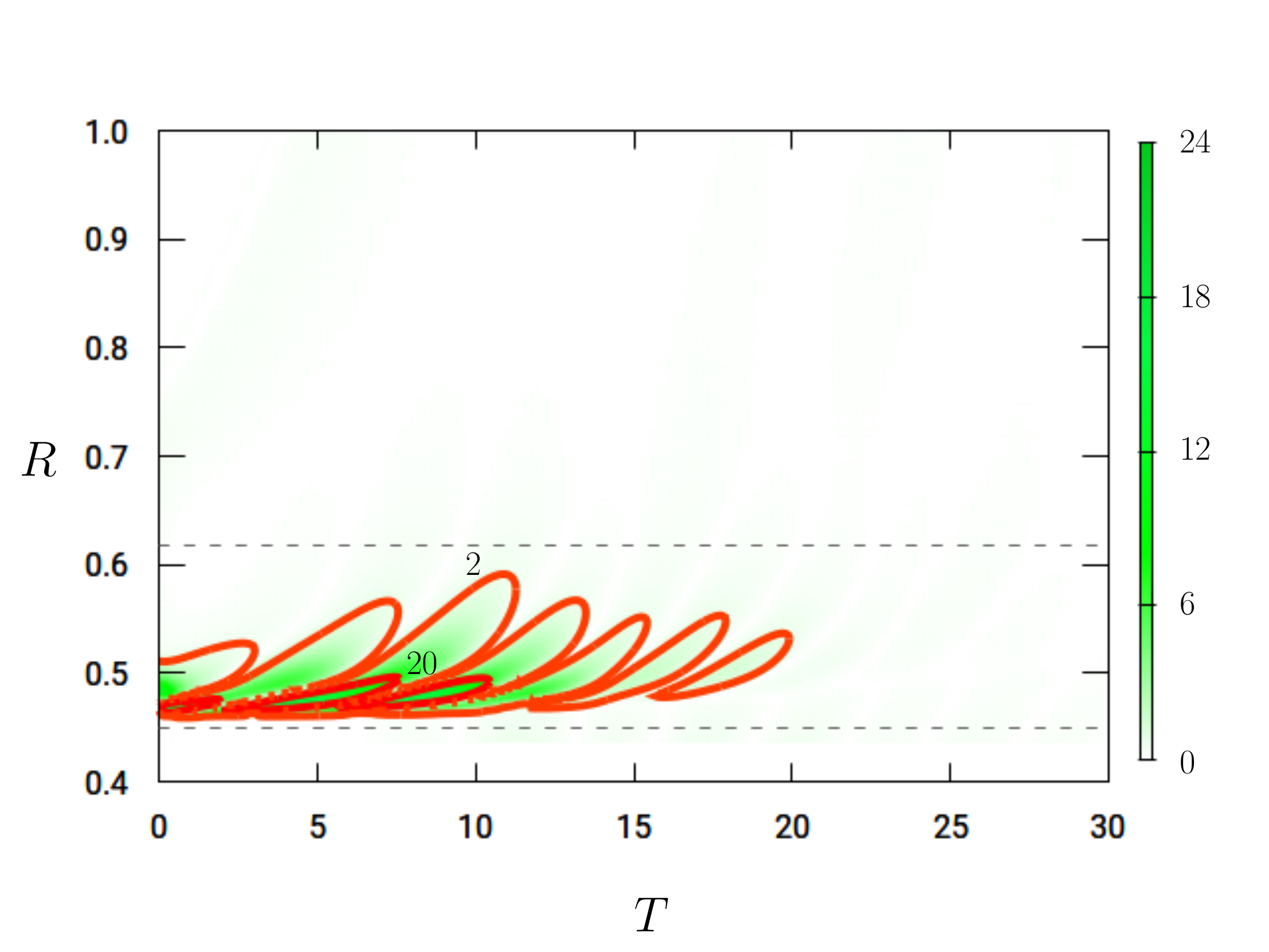}
				\caption{\footnotesize  $\mathscr{E}$ and $\mathscr{J}$ with $s=+2$ and $m=+2$.}
			\end{subfigure}
			\hskip.02\textwidth
			\begin{subfigure}{0.49\textwidth}
				\includegraphics[width=\textwidth]{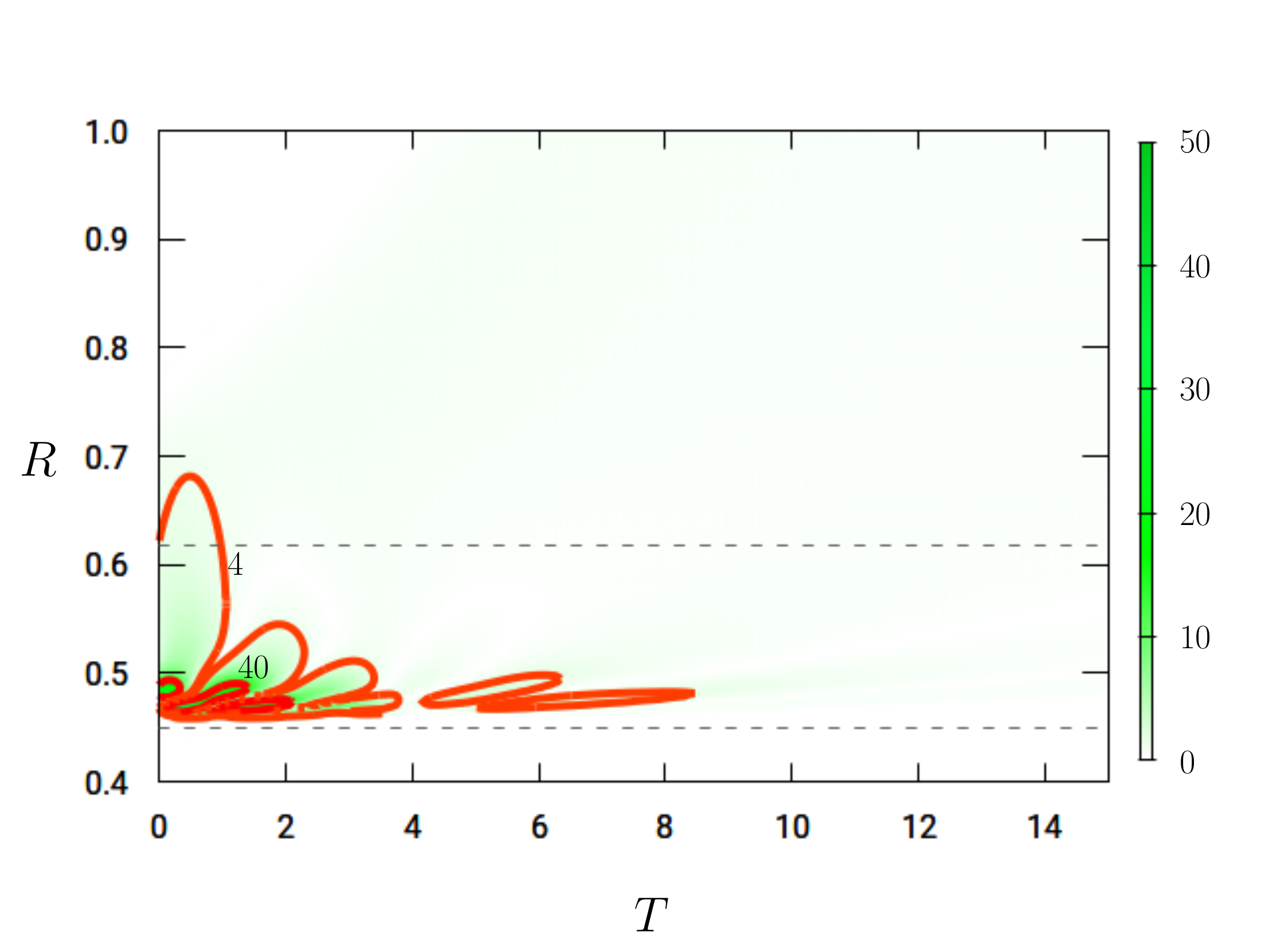}
				\caption{\footnotesize $\mathscr{E}$ and $\mathscr{J}$ with $s=+2$ and $m=-2$.}
			\end{subfigure}
		}
	\end{centering}\vskip-0.2cm
	\caption{{\footnotesize The volume normalized energy and angular momentum current densities, $\mathscr{E}(T,R)$ and  $\mathscr{J}(T,R)$ are plotted for outgoing  co- and counterrotating gravitational perturbations---with physical parameters  $s=+2, \omega_0=0.65,  \ell=2, m=\pm2, R_2= 0.84$---, respectively. 
	}}\label{fig: s=+2-ebr-m=+2} 
\end{figure}

The type-III outgoing gravitational perturbations behave significantly differently. 
The energy and angular momentum transfers of the co- or counterrotating type-III outgoing gravitational perturbations---depicted on Fig. \ref{fig: s=+2-ebr-m=+2} with physical parameters  $s=+2, \omega_0=0.65,  \ell=2, m=+2, R_2= 0.84$---are very close to the ones we saw in case of analogous type-II configurations.  

\begin{figure}[ht!] 
	\begin{centering}
		{\tiny
			\begin{subfigure}{0.49\textwidth}
				\includegraphics[width=\textwidth]{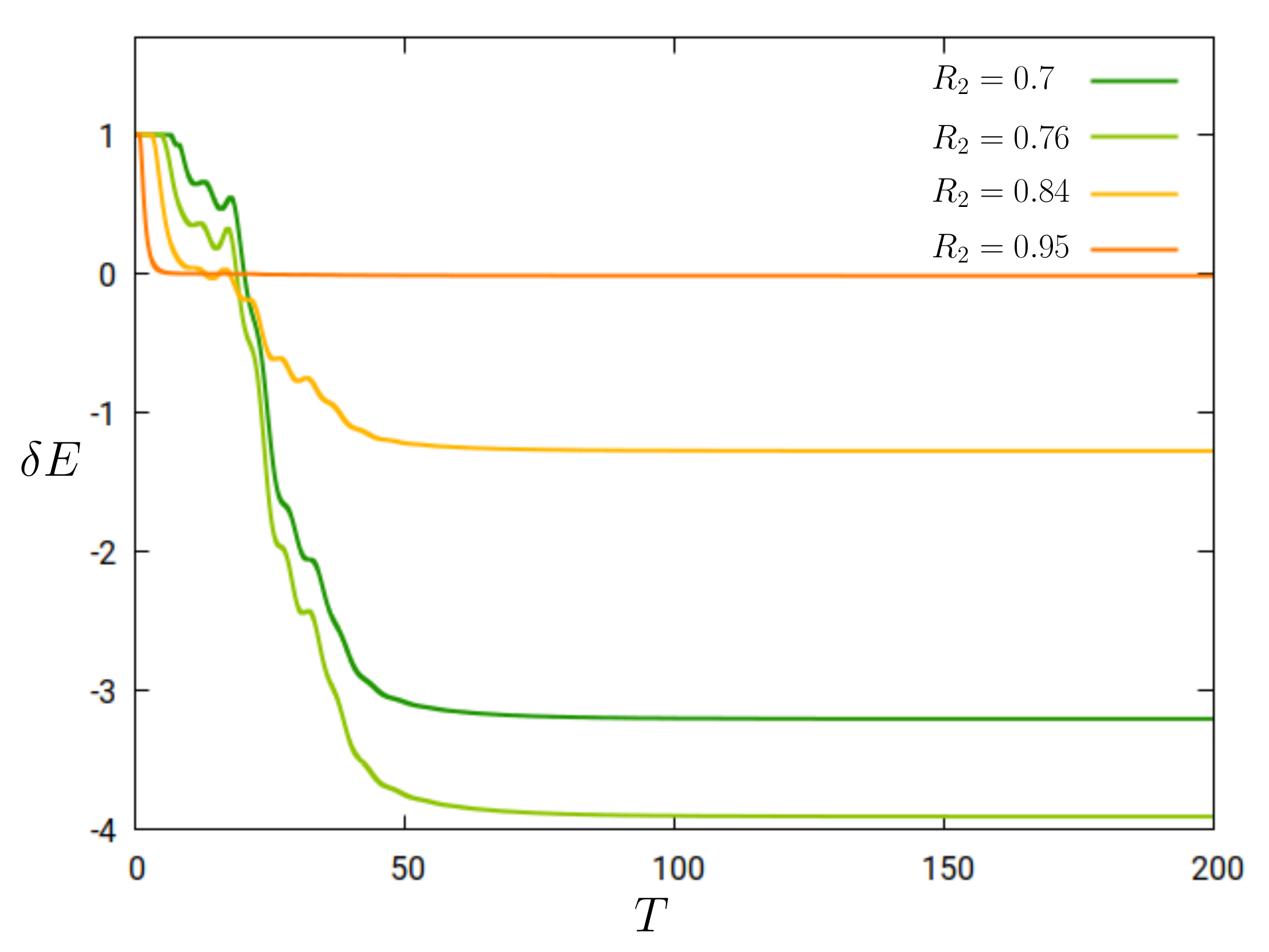}
				\caption{\footnotesize Energy gains: $s=+2, m=+2$. }
			\end{subfigure}
			\hskip.02\textwidth
			\begin{subfigure}{0.49\textwidth}
				\includegraphics[width=\textwidth]{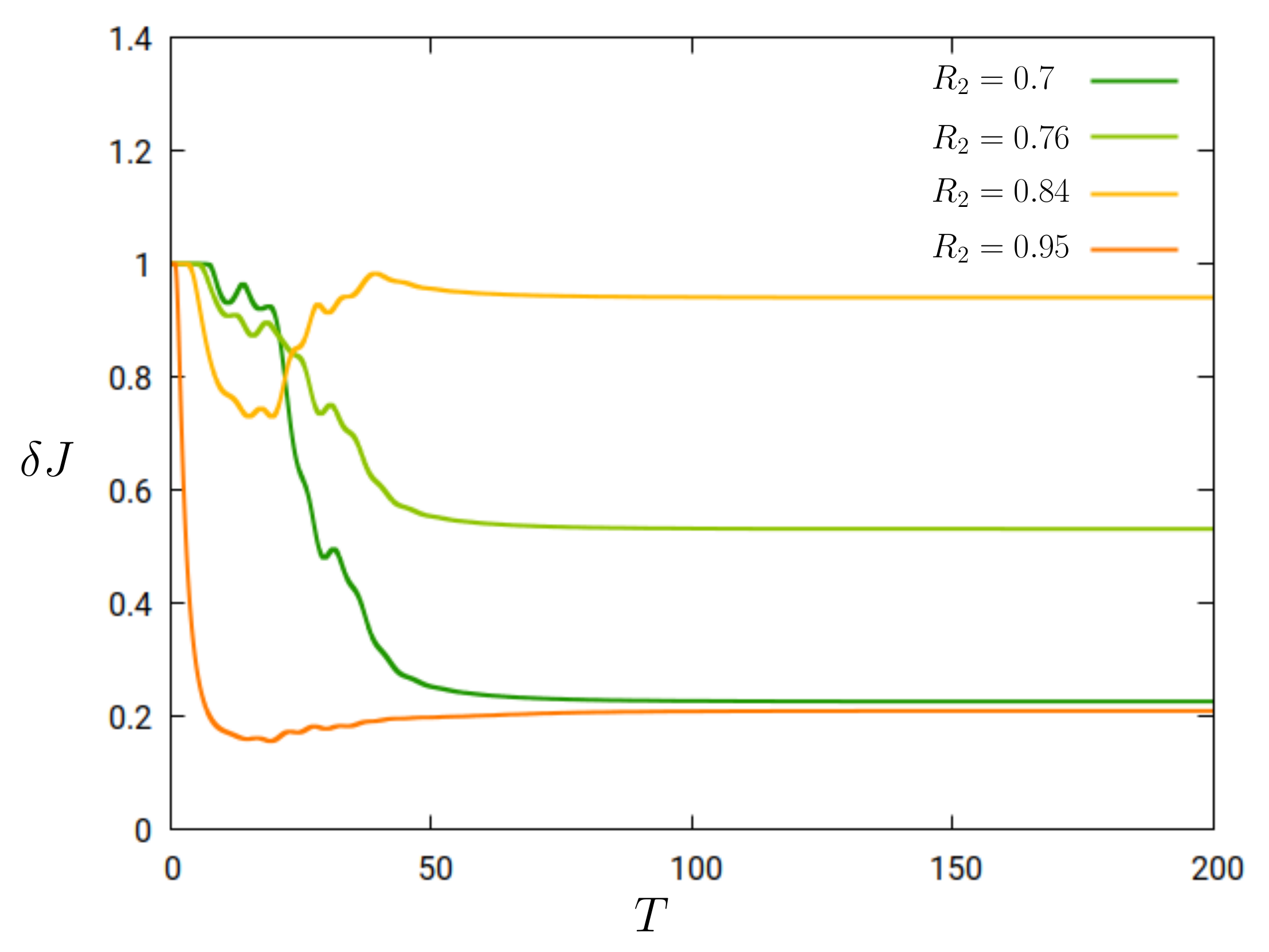}
				\caption{\footnotesize Angular momentum gains: $s=+2, m=+2$. }
			\end{subfigure}
		}
	\end{centering}\vskip-0.2cm
	\caption{{\footnotesize 	The time dependence of the energy and angular momentum gains, $\delta E=(E_0-E_{out})/E_0$ and $\delta J=(J_0-J_{out})/J_0$, are plotted for outgoing corotating  gravitational perturbations. Note that on the left panel the graph of the strongest energy gain $\sim 390\%$, we have found, can also be seen. 
	}}\label{fig: III-s=+2-super-m=+2} 
\end{figure}

Nevertheless, as it is depicted on the panels of Fig.\,\ref{fig: III-s=+2-super-m=+2} the outgoing corotating type-III gravitational perturbations---with physical parameters $s=+2, \omega_0=0.65, \ell=2, m=+2$---produced so far the strongest of energy gains, $ \sim 100\%,\sim 320\%, \sim 390\%$, though the very same configuration did not show the slightest angular momentum gain. 

\medskip

\begin{figure}[ht!] 
	\begin{centering}
		{\tiny
			\begin{subfigure}{0.49\textwidth}
				\includegraphics[width=\textwidth]{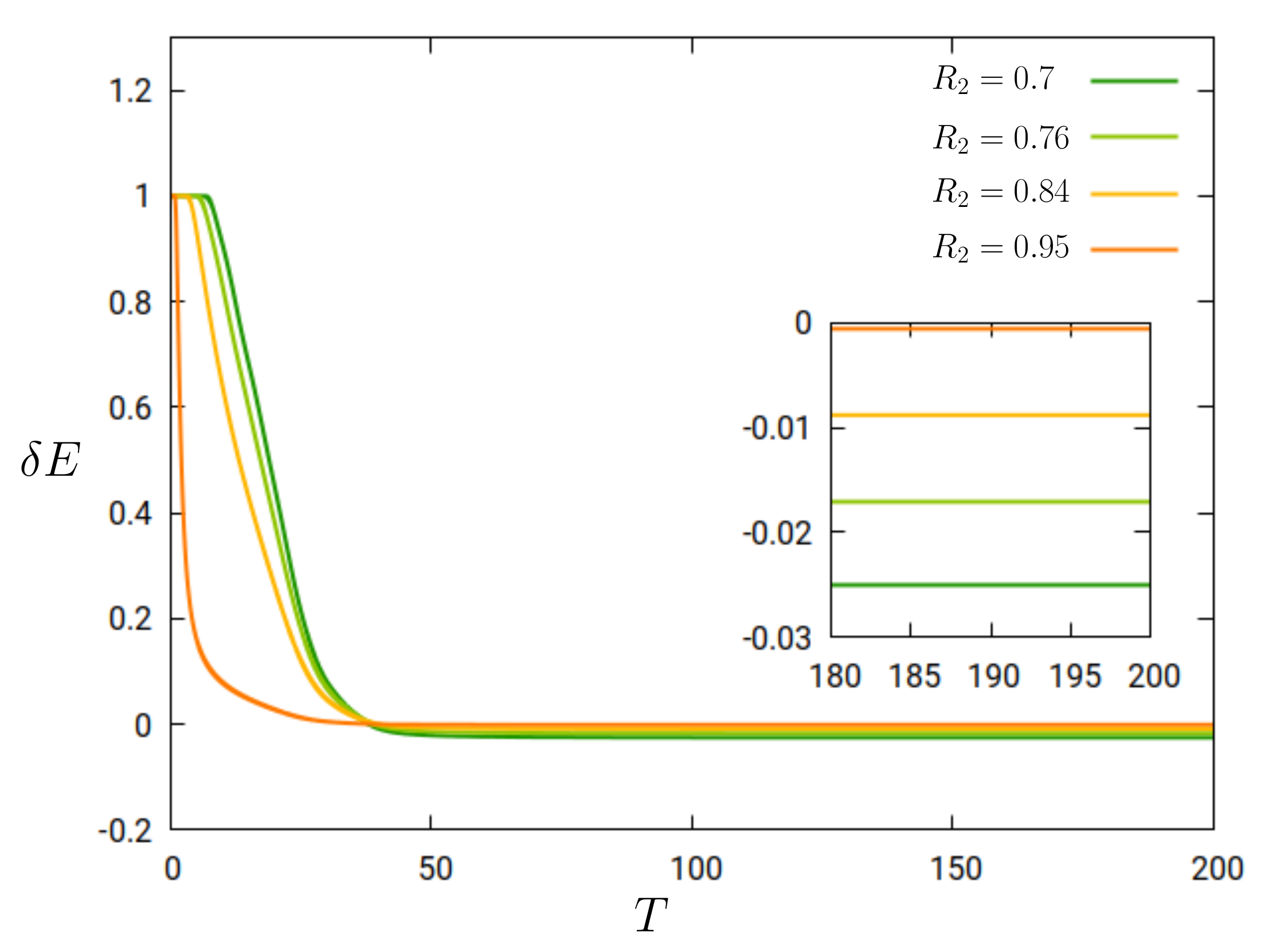}
				\caption{\footnotesize Energy gains: $s=+2, m=-2$.}
			\end{subfigure}
			\hskip.02\textwidth
			\begin{subfigure}{0.49\textwidth}
				\includegraphics[width=\textwidth]{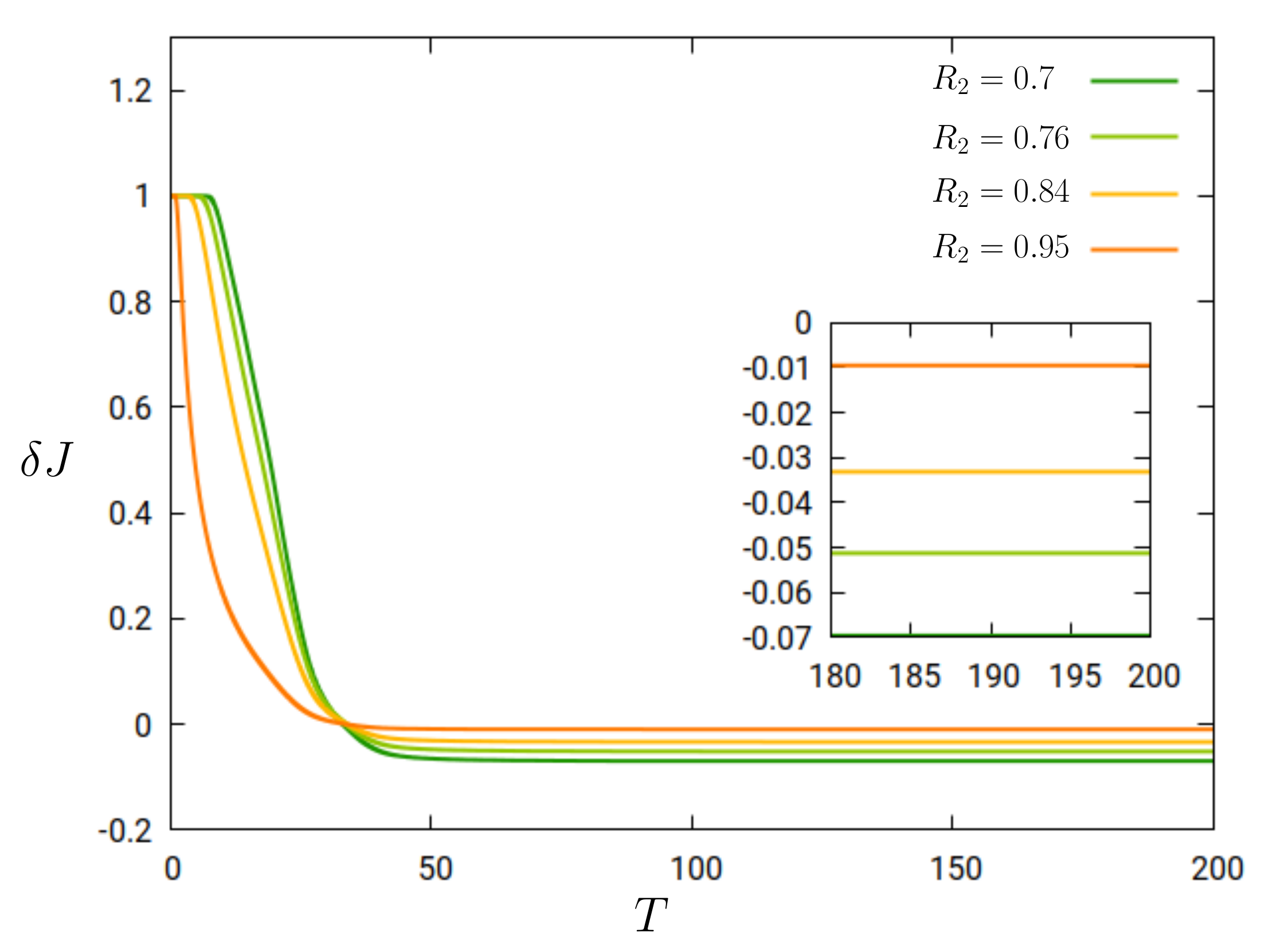}
				\caption{\footnotesize Angular momentum gains: $s=+2, m=-2$. }
			\end{subfigure}
		}
	\end{centering}\vskip-0.2cm
	\caption{{\footnotesize 	The time dependence of the energy and angular momentum gains, $\delta E=(E_0-E_{out})/E_0$ and $\delta J=(J_0-J_{out})/J_0$, are plotted for outgoing counterrotating gravitational perturbations. 
	}}\label{fig: III-s=+2-super-m=-2} 
\end{figure}

Notably,  even the counterrotating outgoing gravitational perturbations---with physical parameters  $s=+2, \omega_0=0.65,  \ell=2, m=-2$, for either of the type-III radial profiles,---produced some energy and angular momentum gains as it is depicted on the panels of Fig.\,\ref{fig: III-s=+2-super-m=-2}. 
Nevertheless,  the scale of superradiance, produced by counterrotating type-III outgoing gravitational perturbations with $R_2=0.7,0.76,0.84,0.95$, is considerably weaker as $\sim 0.01\%, \sim 0.9\%,\sim 1.8\%, \sim 2.5\%$ energy gains and $\sim 1.0\%, \sim 3.3\%,\sim 5.1\%, \sim 7\%$ angular momentum gains showed up in the respective investigations.

\section{Final remarks}\label{sec: final-remarks}

\begin{figure}[ht!] 
	\begin{centering}
		{\tiny
			\begin{centering}
				\begin{subfigure}{0.49\textwidth}
					\hskip1.8cm\includegraphics[width=5.8cm,height=7.6cm]{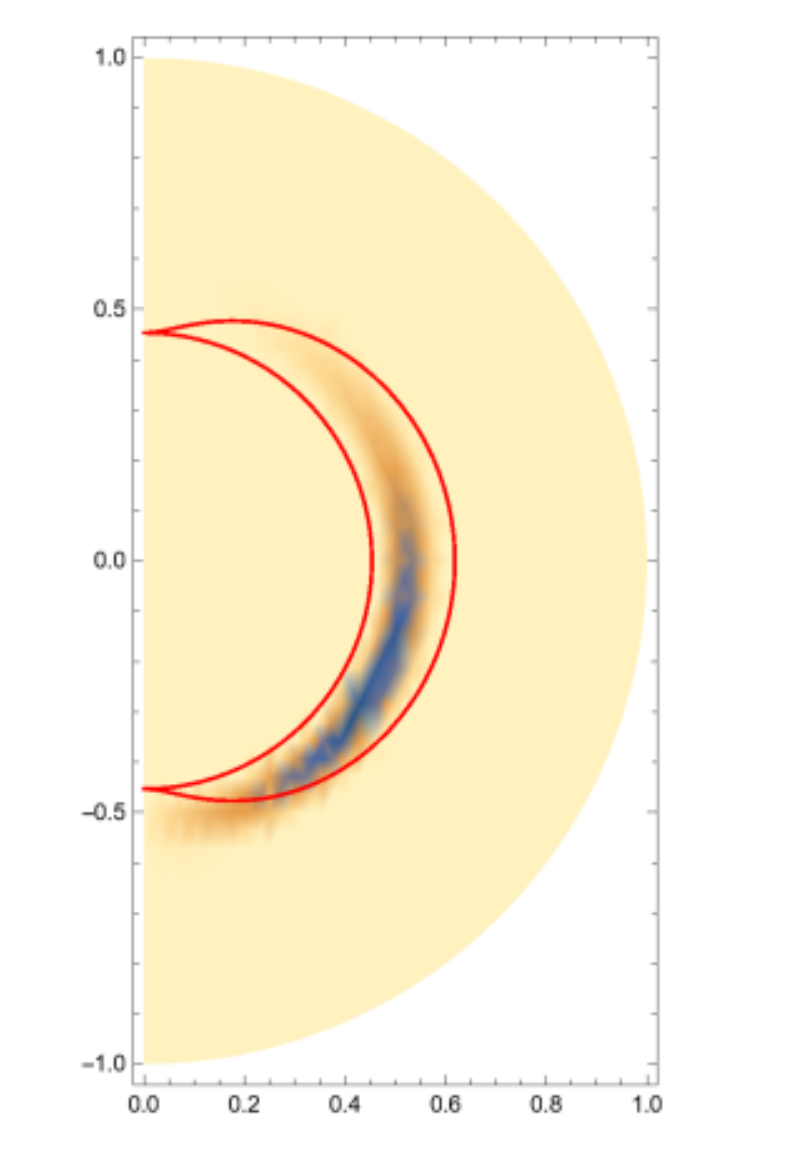}
				\end{subfigure}
			\end{centering}
			\begin{subfigure}{0.49\textwidth}
				\includegraphics[width=\textwidth]{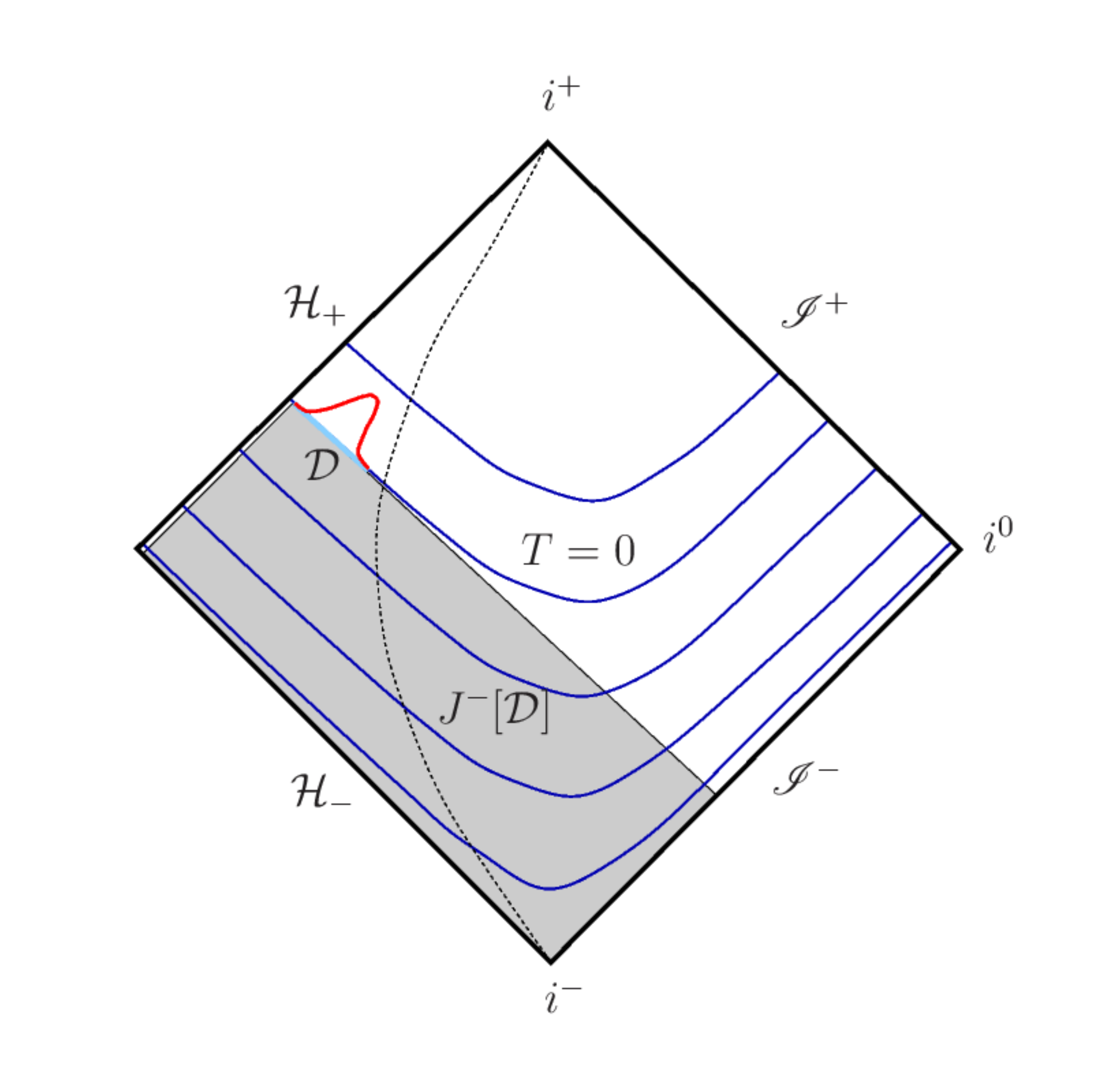}
			\end{subfigure}
		}
	\end{centering}\vskip-0.1cm
	\caption{\footnotesize 
		On the left panel the $\varphi=0$ section of a type-II initial data configuration proportional to ${}_{1}Y_{2}^{2}$, spin $1$ $\ell=m=+2$, spin-weighted spherical harmonic is depicted. The inner colored contour signify the location of the event horizon, the other one does it for the outer boundary ergoregion. Null infinity is located at the $R=1$ semicircle bounding the homogeneously colored region. Note that left vertical boundary denotes the axis of rotation of the Kerr geometry which is axially symmetric, nevertheless, one should keep in mind that the spin-weighted spherical harmonic ${}_{1}Y_{2}^{2}$ is not axially symmetric. \\ 
		On the right panel the $\varphi=const,\vartheta=\pi/2$ slice of the Carter-Penrose diagram of domain of outer communication of the Kerr spacetime is shown. The dotted line is to indicate the location of the outer boundary of the ergoregion. This figure is to demonstrate that the intersection of the causal past, $J^-[\mathcal{D}]$, of the compact support $\mathcal{D}$ of the initial data at $T=0$ and the progenitor time slices will always have a non-negligible part overlapping with the ergoregion if the inner edge of the support is chosen to be close to the black hole horizon.}
	\label{fig: rajz}
\end{figure}

In this paper various aspects of superradiance, on rotating Kerr black hole background, were investigated in the time domain by solving the source free Teukolsky master equation relevant for spin $-2,-1,0,+1,+2$ fields numerically. 
In doing so three different types of initial data were applied. The radial profile of each of these were of compact support. For type-I configurations the support of the radial profile is clearly separated from the ergoregion located in the domain of outer communications. By contrast, the support type-II data is concentrated onto the ergoregion. (Some aspects of the angular dependence of type-II data is depicted on the left panel of Fig.\,\ref{fig: rajz}. See the explanation in the caption of this figure.) For type-III data the corresponding compact support was chosen to start next to the event horizon and to cover, across the ergoregion, larger and larger portions of the domain of outer communication. In addition the initial data---by frequency tuning the incident wave  packets via specifying the $\omega_0$ parameter in \eqref{eq: Initial-data-TR-T-der} properly---was chosen to be optimally superradiant. 

\medskip

One of the main lessons drawn from our investigations is that the scale and various characters of superradiance primarily depends on the type or on the location of the compact support of the radial profile of the initial data. For type-I corotating configurations, regardless how well they are frequency-tuned, considerably large portion of the incident wave  packets endure an almost perfect reflection while approaching the ergoregion. This, in turn, reduces the transmission coefficient, and, thereby, the effect of superradiance is suppressed significantly in the corresponding cases. 
For instance, regardless of the type of the initial data for the scalar field case no energy superradiance occur. 
For this reason it is surprising that angular momentum gain happened in case of time derivative reversed type-II and type-III outgoing scalar fields. The scale of superradiance remained modest regardless of the type of the initial data for electromagnetic perturbations. Impressively strong energy superradiance with notable $\sim116.7\%$ and $\sim320\%, \sim390\%$ energy gains occurred only for co-rotating type-II and type-III gravitational perturbations, respectively. 

\medskip

In \cite{Csizmadia:2012kq,Andras:2014kq}, studying the evolution of scalar perturbations, we experienced an analogous suppression of superradiance while initial data well-separated from the ergoregion was used. Notably, in our present studies, we found a significant superradiance deficit for well-tuned type-I initial data even for spin $s = -1$ and $s = -2$ perturbations. In other words, instead of experiencing superradiance, most of these type-I inbound wave packets get reflected before reaching the ergoregion. This phenomenon seems to be explained by the first and second laws of black hole thermodynamics as follows. On the one hand, according to the first law, the change in the energy, area, and angular momentum of a black hole must be related as
\begin{equation}
dE=\tfrac{\kappa}{8\,\pi}dA + \Omega_{+}\,dJ\,.
\end{equation}
On the other hand, according to the second law, in the examined completely classical scattering processes, the area of the black hole cannot decrease, i.e.
\begin{equation}
dA \geq 0\,.
\end{equation}
Using these two relations, we get
\begin{equation}
0 \leq dE - \Omega_{+}\,dJ\,,
\end{equation}
i.e., any inbound wave packet whose energy and momentum are such that the expression on the right-hand side becomes negative goes against the laws of black hole thermodynamics. Therefore, such a wave packet must undergo a strong reflection. Interestingly, regardless of the spin, the term on the right is negative for all type-I initial data tuned to be superradiant. It is not surprising then that if the support of the initial data is well-separated from the ergoregion, strong reflections occur, and superradiance gets to be suppressed even for spin $s = -1$ and $s = -2$ perturbations.

It is worth noting that the observed strong reflection could also be viewed as a highly indirect justification of Penrose’s weak censorship hypothesis. Assume that a type-I wave packet tuned to the superradiant falls unhindered into a highly spinning Kerr black hole. Provided that backreaction could occur, the wave packet's oversized angular momentum would turn the singularity of the black hole into a naked one.

\medskip
 
To provide a proper measure of the scale of energy superradiance and angular momentum gain we applied the energy and angular momentum balance relations tested previously in \cite{cskritgzs-2019}.  As it was pointed out in \cite{Toth:2018ybm}, conserved currents can be associated with essentially arbitrary pairs of spin $s$ and $-s$ solutions to the Teukolsky master equation. In our investigations the aim was to characterize the superradiance of specific spin $s$ configurations. In order to keep the distinguished position of these specific solutions we decided to use the spin  $-s$ adjoint solutions to the Teukolsky master equation.  This required the use of the constructions offered for electromagnetic perturbations by Cohen-Kegeles in  \cite{Cohen:1974cm,Chrzanowski:1975wv}, and for metric perturbations by Chrzanowski in \cite{Chrzanowski:1975wv} as it is outlined in the appendix of the present paper. It is important to emphasize that when energy and angular momentum gains were reported by referring to via the quantities $\delta E$ and $\delta J$ it is always guaranteed that the energy and angular momentum balances hold at least up to the order $10^{-4}\cdot\delta E$ and $10^{-4}\cdot\delta J$, respectively.

\medskip

\begin{figure}[ht!] 
	\begin{centering}
		{\tiny
			\begin{subfigure}{0.49\textwidth}
				\includegraphics[width=\textwidth]{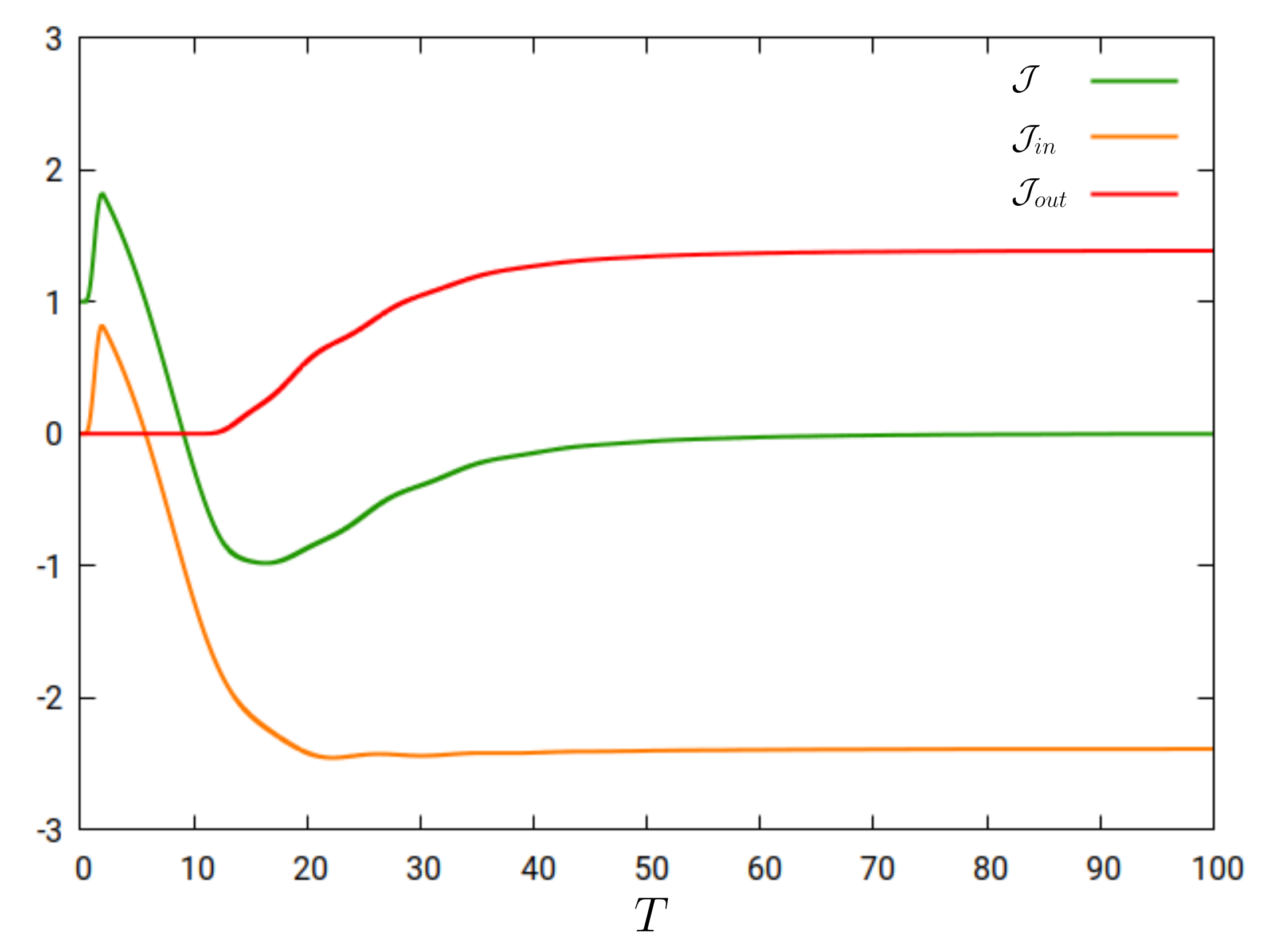}
				\caption{\footnotesize $s=0, m=+1$.}
			\end{subfigure}
			\hskip.02\textwidth
			\begin{subfigure}{0.49\textwidth}
				\includegraphics[width=\textwidth]{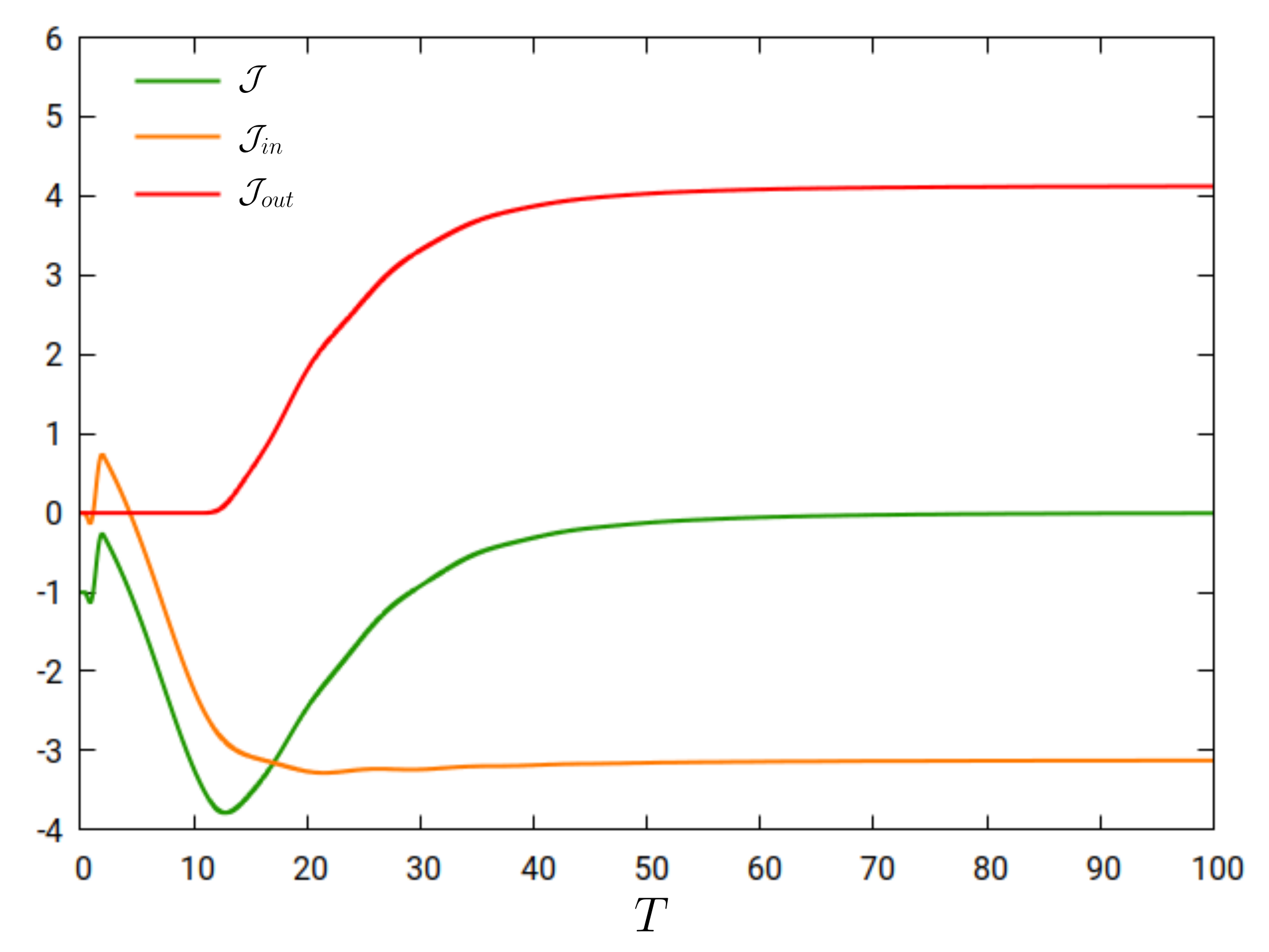}
				\caption{\footnotesize $s=0, m=-1$ }
			\end{subfigure}
		}
	\end{centering}\vskip-0.2cm
	\caption{{\footnotesize The time dependence of the normalized and sign correct angular momentum fluxes $\mathcal{J}_{in}={J}_{in}/|J_0|$ and $\mathcal{J}_{out}={J}_{out}/|J_0|$ at the horizon and at $\scrip$, along with the  normalized and sign correct angular momentum content $\mathcal{J}={J}{(T)}/|J_0|$ of the $T=const$ time slices, are plotted for ingoing co- and counterrotating (time derivative reversed) scalar perturbations, respectively.  
	}}\label{fig: II-s=0-super-m=pm1} 
\end{figure}

When no energy gain but only angular momentum gain occurs---at least in the scalar field case where the conserved energy and angular momentum currents, \eqref{eq:curr1} and \eqref{eq:curr2}, possess immediate physical interpretations---one may suspect that frame dragging enhances the corresponding superradiant behavior. This expectation is supported by the fact that the signs of the angular momentum fluxes through the black hole horizon $J_{in}$ and through null infinity $J_{out}$ were opposite of each other such that, apart from a short initial dynamical period, $J_{in}$ is negative. Notably, as indicated by the $(b)$ panels of Figs.\,\ref{fig: s=0o-super-m=pm1}, \ref{fig: s=+2-super-m=pm2}, \ref{fig: III-s=0-super-m=-1}, and \ref{fig: III-s=+1-super-m=-1}, analogous phenomenon can be seen to apply not only to type-II but to type-III outgoing (i.e.~time derivative reversed) scalar configurations, as well as, to type-II and type-III counterrotating electromagnetic configurations. On the two panels of Fig.\,\ref{fig: II-s=0-super-m=pm1} the  normalized and sign correct angular momentum fluxes through the black hole horizon $\mathcal{J}_{in}={J}_{in}/|J_0|$ and through null infinity $\mathcal{J}_{out}={J}_{out}/|J_0|$, along with the ``normalized'' and ``sign correct'' angular momentum content $\mathcal{J}(T)={J}(T)/|J_0|$ of the $T=const$ time slices are shown separately for the type-II outgoing scalar configurations which were used on Fig.\,\ref{fig: s=0o-super-m=pm1}. Each one of these is normalized by the absolute value $|J_0|=|J(T=0)|$ of the angular momentum,  $J_0$, on the initial slice. Note that the sign correctness, as expected, means here that $J_{0}$ is positive for co-rotating, $m=+1$, configurations, whereas it is negative for  counterrotating, $m=-1$, configurations.

Accordingly, $\mathcal{J}_{out}=J_{out}/|J_{0}|$ and the angular momentum gain $\delta J=(|J_0|-|J_{out}|)/|J_0|$ are, regardless the sign of the $m$ azimuthal parameter, related via the relation $\delta J = 1-|\mathcal{J}_{out}|$. It is clearly visible on the panels of Fig.\,\ref{fig: II-s=0-super-m=pm1} that regardless of the sign of $J_{0}$, apart from a short initial dynamical period, the flux through the event horizon is negative, whereas the flux through $\scrip$ is positive. In addition, as it is anticipated, the frame dragging effect can also be seen (compare the panels of Fig.\,\ref{fig: II-s=0-super-m=pm1}) to be stronger for the counterrotating configuration. 
Note finally that this angular momentum gain, caused by frame dragging, is in entire accordance with the picture we have on mind concerning energy gain in the Penrose process. To see this note that an enhanced negative angular momentum flux through the black hole event horizon ensures the appearance of a scaled up positive flux at null infinity.

\medskip

It is important to mention that in spite of the detailed investigations carried out in the present paper there remained a lot of interesting configurations to be studied. As an immediate example one may think of initial data with radial profile of noncompact support which were left out of the present investigations. This is mainly because in evaluating the energy and angular momentum balance relations we treated the spin $-s$ adjoint solutions,   $\Phi^{(-s)}_{adj}$, as suitable counterparts of the spin $s$ solutions, $\Phi^{(s)}$, to the Teukolsky master equation. There is, however, a substantial technical difficulty coming with this choice. Namely, the use of adjoint solutions in case of gravitational perturbations requires the evaluation of the fourth order derivatives of the $\Phi^{(s)}$ solutions. With the choice of radial profile of compact support we could do this by keeping sufficiently good level of accuracy but this was not possible to be done with initial data the radial support of which  extended to null infinity. In this respect one could also carry out interesting follow up studies, for instance, by applying in the evaluation of the energy and angular momentum balance relation spin $-1,0,+1$  projections of the same electromagnetic perturbations, e.g.~based on the Fackerell-Ipser type approach \cite{Fackerell-1972}, or spin $-2,-1,0,+1,+2$ projections of gravitational perturbations, after applying suitable gauge choice, as suggested in \cite{lars-2011}. This technically could exclude the use of second or fourth order derivatives of spin $s$ solutions to the Teukolsky master equation and, in addition, possibly physically more adequate energy and angular momentum currents could be defined this way in the $s= \pm1,\pm2$ field cases.

\medskip

Let us finish by recalling again that in order to have a sizable superradiance one has to use type-II or type-III initial data. As described above these are configurations where the relevant radial profile of the initial data overlaps considerably with the ergoregion. It is natural to ask then if on the ``progenitor time slices'' of the corresponding configurations could get to be separated from the ergoregion or not. Since the causal past of the compact support of the original initial data intersects all the  earlier hyperbolic time slices near to the black hole horizon the pertinent compact supports of the  progenitor time slices will always remain partly within the ergoregion (as indicated on the right panel of Fig.\,\ref{fig: rajz}). Our numerical experiments---performed by evolving the corresponding initial data configurations backward in time---confirm this expectation. 
All these simple observations do also indicate that sizable superradiant scattering cannot occur unless the initial data is arranged to have considerable overlap with the ergoregion. In other words, if an advanced civilization would try to extract sizable energy or angular momentum from a rotating black hole that cannot be done without being able to engineer, say by some astrophysical processes, suitable scalar, electromagnetic or gravitational perturbations such that nontrivial parts of these perturbation are prepared to have a domain of support ranging essentially from the black hole event horizon, across the ergoregion, to certain location in the domain of outer communication.


\section*{Acknowledgments}

The authors are grateful to Andr\'as L\'aszl\'o and G\'abor T\'oth for helpful comments and suggestions. This research was supported in part by the NKFIH grants K-115434. 

\appendix
\section*{Appendix: Generating adjoint solutions}

This appendix is to recall basics of the methods we used to generate adjoint solutions to Teukolsky's master equation. In doing so the proposals made by Cohen-Kegeles \cite{Cohen:1974cm}, Chrzanowski \cite{Chrzanowski:1975wv}, explained later by Wald in \cite{Wald:1978vm}, were followed.  
In the following subsections the basic ideas and elements of the aforementioned constructions will be outlined first for the generic case, while in the succeeding sections the explicit form of the transformations applicable to electromagnetic perturbations and linear metric perturbations, respectively, are recalled. For the sake of simplicity in the outlined arguments considerations will be restricted to the case of homogeneous TMEs. Note, however, that all the details relevant for the presence of sources can be found in \cite{Cohen:1974cm,Chrzanowski:1975wv,Wald:1978vm}.

\section{Outline of the generic argument}
\renewcommand{\theequation}{GA.\arabic{equation}}
\setcounter{equation}{0}

Consider a physical field $\tf$ that is subject to a linear partial differential equation (PDE) 
\begin{equation}\label{eq: be}
\mathcal{E}\big(\tf\big)=0\,,
\end{equation}
where, in general, $\mathcal{E}$ is a tensor valued linear partial derivative operator and the dependent variable $\tf$ is a tensor field.\,\footnote{Since the  metric $g_{ab}$ is assumed to be defined everywhere
	on $M$ we may, without loss of  generality, restrict considerations to
	tensor fields of type  $(0,m)$, where $m$ is a positive integer.}

\subsection{Reformulations}

It was assumed in \cite{Wald:1978vm} that there exist linear partial derivative operators ${{}^{(s)\hskip-0.5mm}}\mathcal{O}$, ${{}^{(s)\hskip-0.5mm}}\mathcal{T}$ and ${{}^{(s)\hskip-0.5mm}}\mathcal{S}$, in addition to $\mathcal{E}$, such that the followings hold\,\footnote{Note that there is a slight notational difference with respect to the one applied in \cite{Wald:1978vm}. Namely the role of ${{}^{(s)\hskip-0.5mm}}\mathcal{O}$ and ${{}^{(s)\hskip-0.5mm}}\mathcal{T}$ are changed as it appears to be more adequate to denote by ${{}^{(s)\hskip-0.5mm}}\mathcal{T}$ the Teukolsky operator, defined by \eqref{eq: te}.}
\begin{itemize}
	\item  the tensorial variable $\tf$  is mapped by ${{}^{(s)\hskip-0.5mm}}\mathcal{O}$ to a spin-weight-$s$ variable, denoted by $\psi{{}^{(s)\hskip-0.5mm}}$, such that the relation 
		\begin{equation}
			\psi{{}^{(s)\hskip-0.5mm}}={{}^{(s)\hskip-0.5mm}}\mathcal{O}(\tf)
		\end{equation}
		holds.
	\item ${{}^{(s)\hskip-0.5mm}}\mathcal{T}$  is Teukolsky's  linear operator acting on a spin-$s$ variable generated  by ${{}^{(s)\hskip-0.5mm}}\mathcal{O}$ such that 
	\begin{equation}\label{eq: te}
		{{}^{(s)\hskip-0.5mm}}\mathcal{T}{{}^{(s)\hskip-0.5mm}}\mathcal{O}(\tf) = {{}^{(s)\hskip-0.5mm}}\mathcal{T}\psi{{}^{(s)\hskip-0.5mm}}
	\end{equation}
	holds.
	\item finally, the linear operator ${{}^{(s)\hskip-0.5mm}}\mathcal{S}$ is supposed to represent all the manipulations that must be performed to get, by combining it with $\mathcal{E}$, the Teukolsky operator ${{}^{(s)\hskip-0.5mm}}\mathcal{T}$, i.e.
	\begin{equation}\label{eq: be-te}
	{{}^{(s)\hskip-0.5mm}}\mathcal{S}\mathcal{E}\big( \tf \big) = {{}^{(s)\hskip-0.5mm}}\mathcal{T}{{}^{(s)\hskip-0.5mm}}\mathcal{O}\big( \tf \big)= {{}^{(s)\hskip-0.5mm}}\mathcal{T}\psi{{}^{(s)\hskip-0.5mm}}\,.
	\end{equation}
	 This, in particular means that 
	 \begin{itemize}
	 	\item for the linear partial derivative operators $\mathcal{E}$, ${{}^{(s)\hskip-0.5mm}}\mathcal{O}$, ${{}^{(s)\hskip-0.5mm}}\mathcal{T}$ and ${{}^{(s)\hskip-0.5mm}}\mathcal{S}$ the relation
	 	\begin{equation}\label{eq: be-te-abstr}
	 	{{}^{(s)\hskip-0.5mm}}\mathcal{S}\mathcal{E} = {{}^{(s)\hskip-0.5mm}}\mathcal{T}{{}^{(s)\hskip-0.5mm}}\mathcal{O}
	 	\end{equation}
	 	holds, and
	 	\item whenever $\tf$ is a solution to $\mathcal{E}(\tf)=0$ then $\psi{{}^{(s)\hskip-0.5mm}}={{}^{(s)\hskip-0.5mm}}\mathcal{O}(\tf)$  is a solution to  ${{}^{(s)\hskip-0.5mm}}\mathcal{T}\psi{{}^{(s)\hskip-0.5mm}}=0$. 
	\end{itemize}
\end{itemize}

\subsection{The adjoint operator} 

	Consider a linear operator  $\mathcal{P}$ that maps $n$-index tensor fields to $m$-index tensor fields. The linear operator $\mathcal{P}^\dagger$ is called the adjoint of $\mathcal{P}$ if it maps $m$-index tensor fields to $n$-index tensor fields, and it is determined, up to the total divergence $\nabla_a\sigma^a$, where $\sigma^a$ is a smooth vector field on $\mathbb{R}^4$, via the relation 	
	\begin{equation}
		\psi^{a_1\ldots a_m}(\mathcal{P}\tf)_{a_1\ldots a_m}-(\mathcal{P}^\dagger\tf)^{b_1\ldots b_n}\phi_{b_1\ldots b_n}=\nabla_a\sigma^a\,.
	\end{equation}
	For the composition of the linear operators $\mathcal{P}$ and $\mathcal{Q}$ the familiar relation $(\mathcal{P}\mathcal{Q})^\dagger=\mathcal{Q}^\dagger\mathcal{P}^\dagger$ can be seen to hold. A linear operator $\mathcal{P}$ is called to be self-adjoint if $\mathcal{P}^\dagger=\mathcal{P}$ holds, which can happen only if $m=n$.

\subsection{Wald's $1^{st}$ theorem}

The first claim in \cite{Wald:1978vm} was that if $\psi{{}^{(-s)\hskip-0.5mm}}$ is a solution to the adjoint of Teukolsky's equation then $\tf={{}^{(-s)\hskip-0.5mm}}\mathcal{S}^\dagger\psi{{}^{(-s)\hskip-0.5mm}}$ is a solution to the original linear equation \eqref{eq: be}. More precisely, we have the following. 

\begin{theorem}\label{th: wald1} 
Assume that $\psi{{}^{(-s)\hskip-0.5mm}}$ is the solution to the adjoint of the Teukolsky equation
		\begin{equation}\label{eq: te-adj}
			{{}^{(-s)\hskip-0.5mm}}\mathcal{T}^\dagger\psi{{}^{(-s)\hskip-0.5mm}}=0 \,,
		\end{equation}	 
where ${{}^{(-s)\hskip-0.5mm}}\mathcal{T}^\dagger=\big[{{}^{(s)\hskip-0.5mm}}\mathcal{T}\big]^\dagger$, and assume that $\mathcal{E}$ is self-adjoint, i.e.\,$\mathcal{E}^\dagger=\mathcal{E}$. 
Then, in virtue of \eqref{eq: be-te-abstr}, the tensor field  $\tf={{}^{(-s)\hskip-0.5mm}}\mathcal{S}^\dagger\psi{{}^{(-s)\hskip-0.5mm}}$ is a solution to \eqref{eq: be} as the following relations hold
		\begin{align}\label{eq: er1}
		   	\mathcal{E}({{}^{(-s)\hskip-0.5mm}}\mathcal{S}^\dagger\psi{{}^{(-s)\hskip-0.5mm}})  & =  \mathcal{E}^\dagger\big({{}^{(-s)\hskip-0.5mm}}\mathcal{S}^\dagger\psi{{}^{(-s)\hskip-0.5mm}}\big)=({{}^{(s)\hskip-0.5mm}}\mathcal{S}\mathcal{E})^\dagger\psi{{}^{(-s)\hskip-0.5mm}} =  ({{}^{(s)\hskip-0.5mm}}\mathcal{T}{{}^{(s)\hskip-0.5mm}}\mathcal{O})^\dagger \psi{{}^{(-s)\hskip-0.5mm}} \nonumber \\ &  ={{}^{(-s)\hskip-0.5mm}}\mathcal{O}^\dagger\big({{}^{(-s)\hskip-0.5mm}}\mathcal{T}^\dagger\psi{{}^{(-s)\hskip-0.5mm}} \big)=0\,.
		\end{align}
\end{theorem}

\subsection{Wald's $2^{nd}$ theorem} 

In \cite{Wald:1978vm} it was also shown that whenever $\psi{{}^{(-s)\hskip-0.5mm}}$ is a solution to the adjoint of the Teukolsky equation then $\psi{{}^{(s)\hskip-0.5mm}}={{}^{(s)\hskip-0.5mm}}\mathcal{O}\big({{}^{(-s)\hskip-0.5mm}}\mathcal{S}^\dagger\psi{{}^{(-s)\hskip-0.5mm}}\,\big)$ is a solution to the opposite spin Teukolsky equation, i.e., ${{}^{(s)\hskip-0.5mm}}\mathcal{T}\psi{{}^{(s)\hskip-0.5mm}}=0$. 

\begin{theorem}\label{th: wald2}
	Assume that $\psi{{}^{(-s)\hskip-0.5mm}}$ is the solution to the adjoint of the Teukolsky equation 
	\begin{equation}\label{eq: te-adj2}
	{{}^{(-s)\hskip-0.5mm}}\mathcal{T}^\dagger\psi{{}^{(-s)\hskip-0.5mm}}=0 \,.
	\end{equation}	 
	Then, in virtue of \eqref{eq: er1},  $\mathcal{E}\big({{}^{(-s)}\hskip-0.5mm}\mathcal{S}^\dagger{{}^{(-s)\!}}\psi\big)=0$, which, along with \eqref{eq: be-te-abstr} and the self-adjointness of $\mathcal{E}$, implies that $\psi{{}^{(s)\hskip-0.5mm}}={{}^{(s)}\hskip-0.5mm}\mathcal{O}\big({{}^{(-s)}\hskip-0.5mm}\mathcal{S}^\dagger\psi{{}^{(-s)}\hskip-0.5mm}\big)$ is indeed a solution to the Teukolsky equation as the  following relations hold
	\begin{align}
	{{}^{(s)\hskip-0.5mm}}\mathcal{T}\big({{}^{(s)\hskip-0.5mm}}\mathcal{O}\big({{}^{(-s)\hskip-0.5mm}}\mathcal{S}^\dagger\psi{{}^{(-s)\hskip-0.5mm}}\big)\big) & =\big({{}^{(s)\hskip-0.5mm}}\mathcal{T}{{}^{(s)\hskip-0.5mm}}\mathcal{O}\big)\big({{}^{(-s)\hskip-0.5mm}}\mathcal{S}^\dagger\psi{{}^{(-s)\hskip-0.5mm}}\big) =	\big({{}^{(s)\hskip-0.5mm}}\mathcal{S}\mathcal{E}^\dagger\big)\big({{}^{(-s)\hskip-0.5mm}}\mathcal{S}^\dagger\psi{{}^{(-s)\hskip-0.5mm}}\big) \nonumber \\ & = {{}^{(s)\hskip-0.5mm}}\mathcal{S}\big(\mathcal{E}\big({{}^{(-s)\hskip-0.5mm}}\mathcal{S}^\dagger\psi{{}^{(-s)\hskip-0.5mm}}\big)\big)=0\,.
	\end{align}
\end{theorem}

\section{The electromagnetic and metric perturbations}

As it was observed first by Teukolsky \cite{Teukolsky-I} the electromagnetic and metric perturbations of a fixed Kerr background, respectively, whenever they are written out by making use of the Newman-Penrose formalism \cite{newman:penrose}, the equations relevant for the extreme spin values decouple from the other equations. What makes these extreme spin modes to be distinguished is that essential physical information---such as ingoing and outgoing radiation---is carried by them.\,\footnote{The basic notions and notation of the Newman-Penrose formalism \cite{newman:penrose}, along with the gauge fixing applied by Kinnersley \cite{Kinnersley}, is assumed to be known and they will be cited without explanations.} 

\medskip

\subsection{Electromagnetic perturbations}
\renewcommand{\theequation}{ED.\arabic{equation}}
\setcounter{equation}{0}

The symbolic form of \eqref{eq: be} describing the electromagnetic perturbations of algebraically special background reads as
\begin{equation}\label{eq: base-EM}
	\mathcal{E}_E(A_b)=0\,,
\end{equation}
where now the vector potential $A_a$ plays the role of the $(0,1)$ type tensor field $\tf$, and the linear operator $\mathcal{E}_E$ acts on the vector potential as 
\begin{equation}\label{eq: op-maxwell}
\big[\mathcal{E}_E(A_b)\big]_a=\big[\,g^b{}_a\,\Box-g^{bc}\nabla_c\nabla_a\,\big]\,A_b\,.
\end{equation}
It is straightforward to check that the linear operator $\mathcal{E}_E$ is self-adjoint. 

\medskip

The additional linear operators ${{}^{(s)\hskip-0.5mm}}\mathcal{O}$, ${{}^{(s)\hskip-0.5mm}}\mathcal{T}$ and ${{}^{(s)\hskip-0.5mm}}\mathcal{S}$, relevant for the extreme spin values $s=+1$ and $s=-1$, along with the pertinent form of the two theorems by Wald, are given in the following subsections. 

\subsubsection{The ${{}^{(\pm 1)\hskip-0.5mm}}\mathcal{O}$ operators}

Recall first that the Maxwell spinor components $\phi_0$ and $\phi_1$  are defined by the contractions
\begin{align}
\phi_0 & =\ell^am^bF_{ab}=[m^aD-\ell^a\delta]\,A_a\,,\\
\phi_2 & =\overline{m}^an^bF_{ab}=[n^a\overline{\delta}-\overline{m}^a\Delta]\,A_a\,.
\end{align}
In terms of these the Teukolsky variables read as \cite{Teukolsky-I}
\begin{align}
\psi^{(+1)} & = \phi_0\,,\\
\psi^{(-1)} & = (\Psi_2)^{-2/3} \,\phi_2 \,,
\end{align}
where the only nonvanishing and gauge invariant Weyl-scalar component on Kerr background is given by the contraction of the Weyl tensor $C_{abcd}$ and the tetrad vectors as \cite{Teukolsky-I}
\begin{equation}
\Psi_2= -C_{abcd}\,l^a{m}{}^{\,b}n^c\overline{m}{}^{\,d}\,.
\end{equation}

The operators ${{}^{(\pm 1)\hskip-0.5mm}}\mathcal{O}$, mapping the field $A_a$ to the extreme spin Teukolsky variables $\psi^{(\pm 1)}$, can be read off the above relations. They are 
\begin{align}
\psi^{(+1)} = [{{}^{(+1)\hskip-0.5mm}}\mathcal{O}]^aA_a  & = \big[m^a\,D-\ell^a\,\delta\big]\,A_a\nonumber\\
&=\big[(D-\epsilon+\overline{\epsilon}-\nprv)\,m^{a}-(\delta+\overline{\varpi}-\beta-\overline{\alpha})\,\ell^{a}\big]\,A_a,\\
\psi^{(-1)}  = [{{}^{(-1)\hskip-0.5mm}}\mathcal{O}]^aA_a  & =(\Psi_2)^{-2/3}\big[n^a\,\overline{\delta}-\overline{m}^a\,\Delta\big]\,A_a \nonumber\\
&=(\Psi_2)^{-2/3} \big[(\overline{\delta}+\alpha+\overline{\beta}-\overline{\tau})\,n^{a}-(\Delta+\overline{\mu}-\overline{\gamma}+\gamma)\,\overline{m}^{a}\big]\,A_a\,.
\end{align}

\subsubsection{The ${{}^{(\pm 1)\hskip-0.5mm}}\mathcal{T}$ and ${{}^{(\pm 1)\hskip-0.5mm}}\mathcal{T}^\dagger$ operators}

Recall that by applying the operators ${{}^{(\pm 1)\hskip-0.5mm}}\mathcal{T}$ to the fields $\psi^{(\pm 1)}$ we get the homogeneous TME.  The relevant form of the operators ${{}^{(\pm 1)\hskip-0.5mm}}\mathcal{T}$ are as \cite{Chrzanowski:1975wv,Wald:1978vm}
\begin{align}
{{}^{(+1)\hskip-0.5mm}}\mathcal{T} = \big[\,(\ld-\npe{} &+\npev-2\npr-\nprv) \,(\nd+\npm-2\npg) \nonumber\\ & -(\md-\npb-\npav-2\npt+\nppv)\,(\mvd+\npp-2\npa)\,\big]\,,\\
{{}^{(-1)\hskip-0.5mm}}\mathcal{T} = \big[\,(\Delta+\npr{}  & -\nprv+\npmv)\,(D+\npr+2\npe) \nonumber\\ & -(\mvd+\npa+\npbv-\nptv)\,(\md+\npt+2\npb)\,\big]\,.
\end{align}

Since the following relations hold for the adjoint of the Newman--Penrose derivative operators 
\begin{align}
D^\dagger&=-(D+\epsilon+\overline{\epsilon}-\npr-\nprv),\\
\Delta^\dagger&=-(\Delta-\gamma-\overline{\gamma}+\mu+\overline{\mu}),\\
\delta^\dagger&=-(\delta+\beta-\overline{\alpha}-\tau+\overline{\varpi}).
\end{align}
the relations 
\begin{align}
{{}^{(+1)\hskip-0.5mm}}\mathcal{T}^\dagger  = \big[{{}^{(-1)\hskip-0.5mm}}\mathcal{T}\big]^\dagger \,,\\
{{}^{(-1)\hskip-0.5mm}}\mathcal{T}^\dagger =  \big[{{}^{(+1)\hskip-0.5mm}}\mathcal{T}\big]^\dagger\,,
\end{align}
 applied in Theorem\,\ref{th: wald1} also follow.

\subsubsection{The ${{}^{(\pm1)\hskip-0.5mm}}\mathcal{S}$ and ${{}^{(\mp1)\hskip-0.5mm}}\mathcal{S}^\dagger$ operators}

The operators ${{}^{(\pm1)\hskip-0.5mm}}\mathcal{S}$ can be seen to take the form \cite{Cohen:1974cm,Chrzanowski:1975wv}
\begin{align}
\label{eq:S1}
{{}^{(+1)\hskip-0.5mm}}\mathcal{S}{}^{a}&=(\md-\npb-\npav-2\npt+\nppv)\,\ell^a -(D-\npe+\npev-2\npr-\nprv)\,m^a\,,\\
\label{eq:Sm1}
{{}^{(-1)\hskip-0.5mm}}\mathcal{S}{}^{a}&=(\Psi_2)^{-2/3}[(\nd+\npr-\nprv+2\npm+\npmv)\,\overline{m}^a -(\mvd+\npa+\npbv+2\npp-\nptv)\,n^a]\,,
\end{align}
whereas their adjoints are 
\begin{align}
\big[{{}^{(+1)\hskip-0.5mm}}\mathcal{S}{}^\dagger\big]^a & = [\,m^{a}\,(D+2\epsilon+\npr)-\ell^{a}\,(\delta+2\beta+\tau)\,]\,,\\
\big[{{}^{(-1)\hskip-0.5mm}}\mathcal{S}{}^\dagger\big]^a & = [\,n^{a}\,(\overline{\delta}-2\alpha-\varpi)-\overline{m}^{a}\,(\Delta-2\gamma-\mu)\,]\,(\Psi_2)^{-2/3}\,.
\end{align}

\subsection{Wald's $1^{st}$ theorem in case of electrodynamics} 

By applying these adjoint operators  ${{}^{(\pm1)\hskip-0.5mm}}\mathcal{S}{}^\dagger$ 
to solutions $\psi^{(+1)}$ and $\psi^{(-1)}$ to the Teukolsky equations we can generate solutions ${{}^{(+1)\hskip-0.5mm}}{\widehat A}_a$ and ${{}^{(-1)\hskip-0.5mm}}{\widehat A}_a$  to \eqref{eq: base-EM}, respectively, given as 
\begin{align}
{{}^{(-1)\hskip-0.5mm}}{\widehat A}_a =  {} &  \mathfrak{Re}\,\big[\,{{}^{(-1)\hskip-0.5mm}}\mathcal{S}{}^\dagger \psi^{(-1)}\,\big]\nonumber\\
= {} & \tfrac{1}{2}\big[\,m_a\,(D+2\epsilon+\npr)\,\psi^{(-1)}-\ell_a\,(\delta+2\beta+\tau)\,\psi^{(-1)}\nonumber\\
&+\overline{m}_a\,(D+2\overline{\epsilon}+\nprv)\,\overline{\psi^{(-1)}}-\ell_a\,(\overline{\delta}+2\overline{\beta}+\overline{\tau})\,\overline{\psi^{(-1)}}\,\big],
\end{align}
and
\begin{align}
{{}^{(+1)\hskip-0.5mm}}{\widehat A}_a = {} & \mathfrak{Re}\,\big[{{}^{(+1)\hskip-0.5mm}}\mathcal{S}{}^\dagger \psi^{(+1)}\,\big]\nonumber\\
=&\tfrac{1}{2}\,\big[n_a\,(\overline{\delta}-2\alpha-\varpi)\,(\Psi_2)^{-2/3}\psi^{(+1)}-
\overline{m}_a\,(\Delta-2\gamma-\mu)\,(\Psi_2)^{-2/3}\psi^{(+1)}\nonumber\\
&+n_a\,(\delta-2\overline{\alpha}-\overline{\varpi})\,\overline{(\Psi_2)^{-2/3}\psi^{(+1)}}-
m_a\,(\Delta-2\overline{\gamma}-\overline{\mu})\,\overline{(\Psi_2)^{-2/3}\psi^{(+1)}}\,\big]\,,
\end{align}
where $\mathfrak{Re}$ denotes the operation taking the real value of the pertinent complex variable.
Note that neither ${{}^{(+1)\hskip-0.5mm}}{\widehat A}_a$ nor ${{}^{(-1)\hskip-0.5mm}}{\widehat A}_a$ coincides with the vector potential $A_a$ from which the fields $\psi^{(+1)}$ and $\psi^{(-1)}$ are generated by the relation $\psi^{(\pm1)} = [{{}^{(\pm1)\hskip-0.5mm}}\mathcal{O}]^aA_a$.

\subsection{Wald's $2^{nd}$ theorem in case of electrodynamics} 

In virtue of Theorem\,\ref{th: wald2} while acting on the above yielded vector potentials ${{}^{(\pm1)\hskip-0.5mm}}{\widehat A}_a$ by the operators ${{}^{(\pm1)\hskip-0.5mm}}\mathcal{O}$ from the $\psi^{(\pm1)}$ fields new spin $\pm 1$ solutions $\widehat{\psi}^{(\pm1)}$ can be generated which can be given as 
\begin{align}
\widehat{\psi}^{(+1)} & = \tfrac{1}{2}\, (D-\epsilon+\overline{\epsilon}-\nprv)\,(D+2\overline{\epsilon}+\nprv)\,\overline{\psi^{(-1)}}\,, \label{eq: hatpsip1}\\
\widehat{\psi}^{(-1)} & = \tfrac{1}{2}\,(\Psi_2)^{-2/3}\, (\Delta+\overline{\mu}-\overline{\gamma}+\gamma) (\Delta-2\overline{\gamma}-\overline{\mu}) [\,(\overline{\Psi_2})^{-2/3}\overline{\psi^{(+1)}}\,]\,.
\end{align}
By applying the gauge fixing of Kinnersley \cite{Kinnersley}, implying in particular $\epsilon=0$, along with the equation $D\npr=\npr^2$ [which is guaranteed by the Petrov D type of the background], \eqref{eq: hatpsip1} simplifies to
\begin{align}
\widehat{\psi}^{(+1)} & = \tfrac{1}{2}\,DD\,\overline{\psi^{(-1)}}\,. 
\end{align}

\subsection{Linear metric perturbations}
\renewcommand{\theequation}{MP.\arabic{equation}}
\setcounter{equation}{0}

The linearized Einstein's equations 
\begin{equation}\label{eq: base-GR}
 \mathcal{E}_G(h_{ef})=0\,,
\end{equation}
where the metric perturbation $h_{ab}\,(=g_{ab}-\eta_{ab})$ plays the role of the $(0,2)$ type tensor field $f$, whereas the linear operator $\mathcal{E}_G$ acts on the $h_{ab}$ perturbations as  
\begin{multline}\label{eq: op-lin-pert}
[\mathcal{E}_G(h_{ef})]_{ab}=\tfrac{1}{2}        \Big[g^{f}{}_{a}g^{eg}\nabla_{g}\nabla_{b}+g^{f}{}_{b}g^{eg}\nabla_{g}\nabla_{a}\\
-g^{e}{}_{a}g^{f}{}_{b}\Box-g^{ef}\nabla_{a}\nabla_{b}-g_{ab}g^{eg}g^{fh}\nabla_{g}\nabla_{h}+g_{ab}g^{ef}\Box\,\Big]\, h_{ef}\,.
\end{multline}

The operator $\mathcal{E}_G$ is self-adjoint and the additional linear operators ${{}^{(s)\hskip-0.5mm}}\mathcal{O}$, ${{}^{(s)\hskip-0.5mm}}\mathcal{T}$ and ${{}^{(s)\hskip-0.5mm}}\mathcal{S}$, relevant for the spin $s=+2$ and $s=-2$ fields, along with the pertinent form of the theorems by Wald, are given in the succeeding subsections. 

\subsubsection{The ${{}^{(\pm 2)\hskip-0.5mm}}\mathcal{O}$ operators}

Recall first that the Weyl spinor components $\Psi_0$ and $\Psi_4$  are defined by the contractions
\begin{align}
\Psi_0 & =-C_{abcd}\,l^am^bl^cm^d \,,\\
\Psi_4 & =-C_{abcd}\,n^a\overline{m}{}^{\,b}n^c\overline{m}{}^{\,d} \,,
\end{align}
whereas the Teukolsky variables, satisfying the homogeneous Teukolsky master equations, respectively, read as \cite{Teukolsky-I} 
\begin{align}
\psi^{(+2)}=& \ \Psi_0 = [{{}^{(+2)\hskip-0.5mm}}\mathcal{O}]^{ab} h_{ab}\,,\\
\psi^{(-2)}=&\ (\Psi_2)^{-4/3}\cdot\Psi_4 = [{{}^{(-2)\hskip-0.5mm}}\mathcal{O}]^{ab} h_{ab}\,, 
\end{align}
where the explicit form of the operators ${{}^{(\pm 2)\hskip-0.5mm}}\mathcal{O}$, relating the metric perturbations $h_{ab}$ to the extreme spin value Teukolsky variables $\psi^{(\pm 2)}$, can be deduced from equations (B11) and (B12) of \cite{Chrzanowski:1975wv} and they read as 
\begin{align}
\label{eq:O2}
[{{}^{(+2)\hskip-0.5mm}}\mathcal{O}]^{ab} h_{ab} =  {} & \Psi_0 = -\tfrac12 \{(\md+\nppv-3\npb-\npav)(\md+\nppv-2\npb-2\npav)\,\ell^a\ell^b\nonumber\\
&+(D-\nprv-3\npe+\npev)(D-\nprv-2\npe+2\npev)\,m^am^b\nonumber\\
&-[(D-\nprv-3\npe+\npev)(\md+2\nppv-2\npb)\nonumber\\
&+(\md+\nppv-3\npb-\nprv)(D-2\nprv-2\npe)]\,n^{(a}\overline{m}^{b)}\}\,h_{ab}\,,
\\
\label{eq:Om2}
\hskip-0.4cm 
 [{{}^{(-2)\hskip-0.5mm}}\mathcal{O}]^{ab} h_{ab} = {} & (\Psi_2)^{-4/3}\cdot\Psi_4 \nonumber\\
 = {} & -\tfrac12 (\Psi_2)^{-4/3}
 \{(\mvd-\nptv+3\npa+\npbv)(\mvd-\nptv+2\npa+2\npbv)\,n^an^b\nonumber\\
&+(\nd+\npmv+3\npr-\nprv)(\nd+\npmv+2\npr-2\nprv)\,\overline{m}^a\overline{m}^b\nonumber\\
&-[(\nd+\npmv+3\npr-\nprv)(\mvd-2\nptv+2\npa)\nonumber\\
&+(\mvd-\nptv+3\npa+\npbv)(\nd+2\npmv+2\npr)]\,n^{(a}\overline{m}^{b)}\}\,h_{ab}.
\end{align}

\subsubsection{The ${{}^{(\pm 2)\hskip-0.5mm}}\mathcal{T}$ operators}

By applying the operators ${{}^{(\pm 2)\hskip-0.5mm}}\mathcal{T}$ to the fields $\psi^{(\pm 2)}$ we get the homogeneous TME.  These operators ${{}^{(\pm 2)\hskip-0.5mm}}\mathcal{T}$ read as \cite{Chrzanowski:1975wv}
\begin{align}
{{}^{(+2)\hskip-0.5mm}}\mathcal{T} = {} & [(\ld-3\npe+\npev-4\npr-\nprv)(\nd-4\npr+\npm) \nonumber\\ & -(\md+\nppv-\npav-3\npb-4\npt)(\mvd-4\npa+\npp)-3\Psi_2]\,, \\
{{}^{(-2)\hskip-0.5mm}}\mathcal{T} = {} & [(\Delta+3\npr-\nprv+\npmv)(D+4\npe+3\npr) \nonumber\\ &- (\mvd-\nptv+\npbv+3\npa)(\md+4\npb+3\npt)-3\Psi_2]\,.
\end{align}
A straightforward but heavy calculation verifies then that the relations  
\begin{align}
{{}^{(+2)\hskip-0.5mm}}\mathcal{T}^\dagger =  \big[{{}^{(-2)\hskip-0.5mm}}\mathcal{T}\big]^\dagger \,,\\
{{}^{(-2)\hskip-0.5mm}}\mathcal{T}^\dagger =  \big[{{}^{(+2)\hskip-0.5mm}}\mathcal{T}\big]^\dagger\,,
\end{align}
applied in Theorem\,\ref{th: wald1}, also hold.

\subsubsection{The ${{}^{(\pm 2)\hskip-0.5mm}}\mathcal{S}$ operators}

The operators ${{}^{(\pm2)\hskip-0.5mm}}\mathcal{S}$ can be seen to take the form  \cite{Chrzanowski:1975wv}
\begin{align}
\label{eq:S2}
&{{}^{(+2)\hskip-0.5mm}}\mathcal{S}{}^{ab}\nonumber\\
&\hskip0.4cm =(\md+\nppv-\npav-3\npb-4\npt)\big[(D-2\npe-2\nprv)\,\ell^{(a}m^{b)}-(\md+\nppv-2\npav-2\npb)\,\ell^a\ell^b\big]\nonumber\\
&\hskip0.8cm +(D-3\npe+\npev-4\npr-\nprv)\big[(\md+2\nppv-2\npb)\,\ell^{(a}m^{b)}-(D-2\npe+2\npev-\nprv)\,m^am^b\big]\,, \\
\label{eq:Sm2} 
& {{}^{(-2)\hskip-0.5mm}}\mathcal{S}{}^{ab}=(\Psi_2)^{-4/3}\nonumber\\
& \hskip0.4cm \cdot\Big\{(\nd+3\npr-\nprv+4\npm+\npmv)\big[(\mvd-2\nptv+2\npa)\,n^{(a}\overline{m}^{b)}-(\nd+2\npr-2\nprv+\npmv)\,\overline{m}^a\overline{m}^b\big]\nonumber\\
&\hskip0.6cm+(\mvd-\nptv+\npbv+3\npa+4\npp)\big[(\nd+2\npr+2\npmv)\,n^{(a}\overline{m}^{b)}-(\mvd-\nptv+2\npbv+2\npa)\,n^an^b\big]\Big\}\,.
\end{align}

\medskip

The adjoints of operators ${{}^{(\pm2)\hskip-0.5mm}}\mathcal{S}$ can be given as  \cite{Chrzanowski:1975wv}
\begin{align}
\label{eq:S2adj}
& [{{}^{(+2)\hskip-0.5mm}}\mathcal{S}{}^\dagger]_{ab} \nonumber \\
& =\big[\ell_{(a}m_{b)}\,(D+3\npe+\npev-\npr+\nprv)-\ell_a\ell_b\,(\md+3\npb+\npav-\npt)\big] (\md+4\npb+3\npt) \nonumber\\
& +\big[\ell_{(a}m_{b)}\,(\md+3\npb-\npav-\npt-\nppv)-m_am_b\,(D+3\npe-\npev-\npr)\big] \, (D+4\npe+3\npr)  \\
\label{eq:Sm2adj}
& [{{}^{(-2)\hskip-0.5mm}}\mathcal{S}{}^\dagger]_{ab} \nonumber \\
& =\Big\{\big[n_{(a}\overline{m}_{b)}\,(\mvd-3\npa+\npbv+\npp+\nptv)-\overline{m}_a\overline{m}_b\,(\nd-3\npr+\nprv+\npm)\big](\nd-4\npr-3\npm) \nonumber\\
& +\big[n_{(a}\overline{m}_{b)}\,(\nd-3\npr-\nprv+\npm-\npmv)-n_an_b\,(\mvd-3\npa-\npbv+\npp)\big](\mvd-4\npa-3\npp)\Big\}\nonumber\\
& \hskip 10.7cm \cdot (\Psi_2)^{-4/3}
\end{align}

\subsection{Transformations of linear metric perturbations} 

In virtue of Theorem\,\ref{th: wald1}  by applying these adjoint operators  ${{}^{(\pm2)\hskip-0.5mm}}\mathcal{S}{}^\dagger$ 
to solutions $\psi^{(+2)}$ and $\psi^{(-2)}$ to the Teukolsky equations solutions ${{}^{(+2)\hskip-0.5mm}}{\widehat h}_{ab}$ and ${{}^{(-2)\hskip-0.5mm}}{\widehat h}_{ab}$  to \eqref{eq: base-GR}, respectively, can be generated, given formally as 
\begin{align}
{{}^{(+2)}}{\widehat h}_{ab} & =  \mathfrak{Re}\big[[{{}^{(+2)\hskip-0.5mm}}\mathcal{S}{}^\dagger]_{ab}\,\psi^{(+2)}\big] \\
{{}^{(-2)}}{\widehat h}_{ab} & =  \mathfrak{Re}\big[[{{}^{(-2)\hskip-0.5mm}}\mathcal{S}{}^\dagger]_{ab}\,\psi^{(-2)}\big]
\end{align}
In analogy to the electromagnetic case neither ${{}^{(+2)}}{\widehat h}_{ab}$ nor ${{}^{(-2)}}{\widehat h}_{ab}$ has anything to do with the metric perturbation $h_{ab}$ from which the fields $\psi^{(+2)}$ and $\psi^{(-2)}$ are generated as $\psi^{(\pm2)} = [{{}^{(\pm2)\hskip-0.5mm}}\mathcal{O}]^{ab}h_{ab}$. 

\medskip

By appealing now to Theorem\,\ref{th: wald1} we get that by acting on the metric perturbations ${{}^{(+2)}}{\widehat h}_{ab}$ and ${{}^{(-2)}}{\widehat h}_{ab}$ by the operators ${{}^{(\pm2)\hskip-0.5mm}}\mathcal{O}$ we can generate, from the $\psi^{(\pm2)}$  new fields spin $\pm 2$ solutions $\widehat{\psi}^{(\pm2)}$  which can be given as \cite{Chrzanowski:1975wv}
\begin{equation}\label{eq: hatpsip2}
\widehat{\psi}^{(+2)}=\tfrac{1}{4}(D-\nprv-3\npe+\npev)(D-\nprv-2\npe+2\npev)(D-3\npev+\npe-4\nprv-\npr)(D-2\npev+2\npe-\npr)\,\overline{\psi^{(-2)}}.
\end{equation}
\begin{align}
\widehat{\psi}^{(-2)} = {} & \tfrac{1}{4}(\Psi_2)^{-4/3} \nonumber \\ & \cdot(\nd+\npmv+3\npr-\nprv)(\nd+\npmv+2\npr-2\nprv)(\nd-3\nprv+\npr+\npmv)(\nd-4\nprv-3\npmv)\nonumber \\ & \cdot(\overline{\Psi_2})^{-4/3}\,\overline{\psi^{(2)}}
\end{align}

As in the electromagnetic case, by applying the gauge fixing of Kinnersley \cite{Kinnersley}, with $\epsilon=0$, along with the equation $D\npr=\npr^2$ \eqref{eq: hatpsip2} simplifies to
\begin{align}
\widehat{\psi}^{(2)}=&\tfrac{1}{4}DDDD\overline{\psi^{(-2)}}.
\end{align}

\end{document}